\documentclass[times,twocolumn,final]{elsarticle}
\usepackage{medima}
\usepackage{framed,multirow}

\usepackage{amssymb}
\usepackage{latexsym}
\usepackage{float}
\usepackage{subcaption}
\usepackage{amsmath}
\usepackage{amsfonts}
\usepackage[ruled,vlined]{algorithm2e}
\usepackage{url}
\usepackage{xcolor}
\usepackage{bm}

\newcommand{\m}{{\bm{\mu}}(\bm{\theta})}
\newcommand{\jac}{\bm{J}}

\newcommand{\ttil}{\tilde{ \bm{\theta}}}
\newcommand{\mt}{{\bm{\mu}}}
\newcommand{\R}{\mathbb{R}}
\newcommand{\z}{\bm{Z}}
\usepackage{hyperref}
\definecolor{newcolor}{rgb}{.8,.349,.1}

\journal{Medical Image Analysis}

\begin{document}

\verso{Riwaj Byanju  \textit{et~al.}}

\begin{frontmatter}

\title{Time efficiency analysis for undersampled quantitative MRI acquisitions}%
%\tnotetext[tnote1]{This is an example for title footnote coding.}

\author[1]{Riwaj \snm{Byanju}\corref{cor1}}
\cortext[cor1]{Corresponding author:}
\ead{r.byanju@erasmusmc.nl}
\author[1]{Stefan \snm{Klein}}
\author[1]{Alexandra \snm{Cristobal-Huerta}}
\author[1]{Juan A. \snm{Hernandez-Tamames}}
\author[1]{Dirk H. J. \snm{Poot}}

\address[1]{Department of Radiology and Nuclear Medicine, Erasmus MC,  Rotterdam, 3015 GE, The Netherlands}
%\address[2]{Affiliation 2, Address, City and Postal Code, Country}

\received{27 May 2021}
\finalform{}
\accepted{}
\availableonline{}
\communicated{}

\begin{abstract}
%%%
To realize Quantitative MRI (QMRI) with clinically acceptable scan time, acceleration factors achieved by conventional parallel imaging techniques are often inadequate. Further acceleration is possible using model-based reconstruction. We propose a theoretical metric called TEUSQA: Time Efficiency for UnderSampled QMRI Acquisitions to inform sequence design and sample pattern optimisation. TEUSQA is designed for a particular class of reconstruction techniques that directly estimate tissue parameters, possibly using prior information to regularize the estimation. TEUSQA can be used to evaluate undersampling patterns for multi-contrast QMRI sequences targeting any tissue parameter. To verify the time efficiency predicted by TEUSQA, we performed Monte Carlo simulations and an accelerated parameter mapping with two sequences (Inversion prepared fast spin echo for $T_1$ and $T_2$ mapping and 3D GRASE for $T_2$ and B0 inhomogeneity mapping). Using TEUSQA, we assessed several ways to generate undersampling patterns \textit{in silico}, providing insight into the relation between sample distribution and time efficiency for different acceleration factors. The time efficiency predicted by TEUSQA was within 15\% of that observed in the Monte Carlo simulations and the prospective acquisition experiment. The assessment of undersampling patterns showed that a class of good patterns could be obtained by low-discrepancy sampling. We believe that TEUSQA offers a valuable instrument for developers of novel QMRI sequences pushing the boundaries of acceleration to achieve clinically feasible protocols. Finally, we applied a time-efficient undersampling pattern selected using TEUSQA for a 32-fold accelerated scan to map  $T_1$ \& $T_2$ mapping of a healthy volunteer.

\end{abstract}


\end{frontmatter}

%\linenumbers

%\section{Note}
%\label{sec1}
%Please use \verb+elsarticle.cls+ for typesetting your paper.
%Additionally load the package \verb+medima.sty+ in the preamble using
%the following command: 
%\begin{verbatim} 
%  \usepackage{medima}
%\end{verbatim}

%Following commands are defined for this journal which are not in
%\verb+elsarticle.cls+. 
%\begin{verbatim}
%  \received{}
%  \finalform{}
%  \accepted{}
%  \availableonline{}
%  \communicated{}
%\end{verbatim}
%%
%Any instructions relavant to the \verb+elsarticle.cls+ are applicable
%here as well. See the online instruction available on:
%\makeatletter
%\if@twocolumn
%\begin{verbatim}
% http://support.stmdocs.com/wiki/
% index.php?title=Elsarticle.cls
%\end{verbatim}
%\else
%\begin{verbatim}
% http://support.stmdocs.com/wiki/index.php?title=Elsarticle.cls
%\end{verbatim}
%\fi

%\subsection{Entering text}
%\textcolor{newcolor}{\bf There is no page limit.}

\section{Introduction}
\label{sec:introduction}
Traditional MR images are weighted and do not actually measure tissue properties \cite{ma_magnetic_2013,tofts_quantitative_2005}. This may complicate the diagnosis of subtle changes in these tissue properties \cite{warntjes_rapid_2008}. By measuring tissue properties, Quantitative MR imaging (QMRI) promises to reduce the sensitivity to the exact acquisition protocol and improve the reproducibility and comparability of the results \cite{weiskopf_quantitative_2013}. 

In conventional MR imaging, weighted images are obtained by pulse sequences that enhance contrasts between tissues but do not quantitatively measure any specific tissue property. 
On the contrary, in QMRI, the pulse sequence acquires images of multiple spin states, followed by a model fitting step to infer the tissue properties quantitatively. Naive implementations of this approach, where multiple fully sampled images are acquired, require considerably more scan time than conventional MR imaging. Long scan time makes the acquisition more sensitive to patient motion and other system imperfections, rendering it impractical for clinical use \cite{altbach_extending_2013,hilbert_accelerated_2018}.

Scan time can be reduced by undersampling the k-space and exploiting prior information and/or complementary information. In parallel imaging, the complementary information is provided by the different sensitivity profiles of channels in a multi-channel coil  \cite{hamilton_recent_2017}. However, the achievable acceleration is limited \cite{vasanawala_improved_2010}. Conventional MR images have been accelerated further using prior information about sparsity by compressed sensing reconstructions \cite{lustig_sparse_2007,murphy_fast_2012} and prior information learned from deep neural networks \cite{knoll_deep-learning_2020}. As QMRI has multiple contrast data, reconstruction techniques that exploit the relationship across contrast images as part of estimation can accommodate even higher acceleration. Recently developed MR fingerprinting approaches rely on  information about signal evolution, obtained by Bloch simulations and encoded by a dictionary \cite{ma_magnetic_2013}. Similarly, \cite{tamir_t_2017} add the temporal signal relaxation in the parallel imaging forward model to improve the trade-off between the image resolution and scan time. These results have shown that reconstruction techniques leveraging redundancy across contrasts have the potential to accelerate QMRI acquisitions sufficiently to make them clinically feasible. 

Model-based reconstruction uses models of the relations among the contrast images during reconstruction to allow parameter estimation from undersampled data \cite{block_model-based_2009,sumpf_model-based_2011,majumdar_accelerating_2011,zhao_model-based_2014,sumpf_fast_2014,ben-eliezer_accelerated_2016,tamir_t_2017,hilbert_accelerated_2018,zimmermann_accelerated_2018,hu_super_2019,zhao_direct_2016,zhang_accelerating_2015,zhao_accelerated_2015,mandava_accelerated_2018,chenxi_hu_trust_2015,knopp_iterative_2009,tran-gia_model-based_2015}. A overview of such techniques is provided by Zhao et al. \cite{zhao_model-based_2014}. An implicit assumption in model-based reconstruction methods is that the information from each contrast complements those from other contrasts to allow joint reconstruction or estimation.  The spatial information provided by each contrast depends on the distribution of k-space samples, i.e., the undersampling pattern, which could be different for each contrast \cite{knoll_adapted_2011, cristobal-huerta_k-space_2018, tsao_k-t_2003,haldar_oedipus_2019,haldar_super-resolution_2009,bahadir_adaptive_2019}. However, identifying good undersampling patterns has received only limited attention \cite{levine_--fly_2018}. 

An exhaustive empirical search for the most efficient undersampling pattern is impossible due to the excessively large number of possible patterns. Therefore, there is a need for a theoretical technique  for designing good undersampling patterns. For specific MRI modalities various frameworks have been developed for optimisation of sequence settings such as echo time (TE), inversion time (TI), echo spacing (ESP) and repetition time (TR) \cite{leitao_efficiency_2021,deoni_rapid_2003,crawley_comparison_1988,asslander_optimized_2019,zhao_optimal_2019,poot_optimal_2010,jones_optimal_1996,brihuega-moreno_optimization_2003}. A number of them use the Cram\'er Rao lower bound (CRLB) as metric, for instance: for $T_2$ imaging by Jones et al. \cite{jones_optimal_1996}, for diffusion measurements by Brihuega-Moreno et al. \cite{brihuega-moreno_optimization_2003}, for diffusion kurtosis imaging by Poot et al. \cite{poot_optimal_2010}, for MRF by Zhao et al. \cite{zhao_optimal_2019}. Such metrics have been used to evaluate undersampling patterns as well \cite{haldar_super-resolution_2009,haldar_oedipus_2019}. 
Levine et al. \cite{levine_--fly_2018} proposed a metric for evaluation of undersampling patterns for a class of techniques that uses a linear subspace of the model to reconstruct dynamic image series. For model-based reconstruction that directly estimates tissue parameter maps from undersampled k-space, Zhao et al. \cite{zhao_model-based_2014} derived an expression for the CRLB that is applicable with and without sparsity constraints. However, to the best of our knowledge, there are no studies dedicated to evaluating  undersampling patterns for this class of model-based reconstruction techniques.

Our aim is to develop a framework for theoretical evaluation of undersampling patterns that can take any tissue parameters or acquisition protocol into account. There are two challenges for using a CRLB based framework. First, as calculation of the CRLB requires the inversion of a large information matrix, it is computationally expensive.  Second, there are degeneracies: in voxels with (almost) zero proton density the other parameters are not identifiable and this may impact other voxels since fitting is performed in k-space domain. One of the ways to get around the degeneracies is by the inclusion of prior information. However, this makes the estimator biased to the prior information while the CRLB assumes an unbiased estimator.

In this work, we propose a theoretical metric called TEUSQA: Time Efficiency for UnderSampled QMRI Acquisitions.  TEUSQA  can be used to evaluate undersampling patterns for multi-contrast QMRI sequences targeting any tissue parameter.  It is based on the CRLB and takes into account sequence-related settings as well as the undersampling pattern. TEUSQA overcomes computational complexity by using a central patch of k-space for evaluation. To make the estimation free from degeneracies and keep the metric's generalisability, we propose a `weak' prior that only comes into play when the information on a voxel is not available from the measurements. TEUSQA accounts for this prior information by computing the posterior distribution using Bayes theorem similarly to Bayesian CRLB \cite{van_trees_detection_2013}. We evaluate TEUSQA with two sequences: Inversion prepared fast spin echo (3D IP-FSE) and Gradient recalled echo sequence (3D GRASE) to verify its generalisability. We show with Monte Carlo simulations and prospective acquisitions that it can accurately predict the variance observed in actual scans with full-sized k-space. Using TEUSQA, we evaluate several undersampling pattern generation techniques and identify a key property called discrepancy, which can aid in the generation of time efficient undersampling patterns. We show with a prospectively undersampled in-vivo scan, that such patterns can be used to obtain $T_1$ and $T_2$ maps.

\section{Theory}
\subsection{Derivation of time efficiency}
\label{sec: derrivation and down}
\subsubsection{Signal model for an undersampled QMRI acquisition}
%Let $\bm{\theta}$ be a column vector composed of parameter vectors ($T_1$, $T_2$, etc) $\bm{\theta}_{\bm{x}} \in \R^{P}$ for all voxels in image domain $\Omega^x$  where  $\bm{x}$ is positional index and $P = |\bm{\theta}_{\bm{x}}|$. 
%Let $\bm{\theta}$ be a map of parameter vectors ($T_1$, $T_2$, etc) of length P, with elements $\bm{\theta}_{\bm{x}} \in R^P$ for all voxels in the image domain $\Omega^x$.
Let $\bm{\theta_{\bm{x}}} \in \mathbb{R}^P$ be a column vector of $P$ tissue parameters ($T_1$, $T_2$, etc.) at position $\bm{x}$ of the (cartesian) voxel grid of the image $\Omega^x \subset \mathbb{R}^3$. Let $\bm{\theta} \in \mathbb{R}^L$ be the concatenation of $ \bm{\theta}_{\bm{x}} \forall \bm{x} $, with length $L=P|\Omega^x|$. Let $f_q( \bm{\theta}_{ \bm{x}})$ be a function that predicts the signal of a contrast state in an acquisition scheme that acquires $Q$ different contrast states indexed by $q \in [1, Q] $. Let $C_{ \bm{x}, c}\in \mathbb{C}$ indexed as $c \in [1, C]$ be the coil sensitivity map where $C$ is the number of coils used in the acquisition. Let $F_{ \bm{k}, \bm{x}} =  e^{ - i \bm{x}^{T} \bm{k}/|\Omega^x|}$ be the Fourier transform operator between the image domain $\Omega^x$ and the k-space domain $\Omega^k$ where $ \bm{k}$ represents the multi-dimensional k-space index.
%\begin{equation}
%	\label{eq: fourier_operator}
%	F_{ \bm{k}, \bm{x}} =  e^{ - i \bm{x}^{T} \bm{k}/|\Omega^x|}.
%\end{equation}
We define the domain of sampled k-space as  $\Omega^{k, S}_{q} \subseteq \Omega^{k}$, which may be different for each contrast $q$.
Then the expected value for a k-space measurement  $\mu_{q, \bm{k},c}( \bm{\theta})$, is given by
\begin{equation}
	\label{eq: model}
	\mu_{q, \bm{k},c}( \bm{\theta}) = {\sum_{ \bm{x} \in  \Omega^x }} F_{ \bm{k}, \bm{x}} C_{\bm{x}, c} f_q( \bm{\theta}_{ \bm{x}})  .
\end{equation}

The noise in the acquired k-space can be assumed to be of complex Gaussian distribution having independent real and imaginary parts \cite{henkelman_measurement_1985,brown_magnetic_2014}.
The modeled signal shown in Equation \ref{eq: model} can be represented in complex notation as
$\mu_{q, \bm{k},c} = \Re\left\{\mu_{q, \bm{k},c}\right\} + i \Im\left\{\mu_{q, \bm{k},c}\right\} $ 
%\begin{equation}
%\bm{\mu} = \left( \begin{array}{c} \Re\{\bm{\mu}\} \\ 
%     \Im\{\bm{\mu}\} \end{array} \right) \in \mathbb{R}^{2N}.
%\end{equation}
% where $N  =  C \sum_{ q }|\Omega^{k, S}_{q}|$ 
and the measured complex-valued signal $Z_{q,\bm{k},c}$ can also be represented in the same way. For notational convenience, we define $\grave{\mu}_{q,\bm{k},c} = \left( \begin{array}{c} \Re\{ \mu_{q,\bm{k},c} \}\\ \Im\{ \mu_{q,\bm{k},c} \}\end{array} \right)$ and $\grave{Z}_{q,\bm{k},c} = \left( \begin{array}{c} \Re\{ Z_{q,\bm{k},c} \}\\ \Im\{ Z_{q,\bm{k},c} \}\end{array} \right)$. 
Let $vec_*( A_*)$  make a vector out of $A$, iterating over all indices $ *$, then $\bm{Z} = vec_{q, \bm{k}, c}\left( \grave{Z}_{q,\bm{k},c}\right)   \in \mathbb{R}^{2N},$
%and including the real and imaginary part of the complex values as separate elements
%$\bm{Z} = vec_{q, \bm{k}, c}\left( Z_{q,\bm{k},c}\right)  \in  \mathbb{C}^N$
%\begin{equation}
%	\bm{Z} = vec_{q, \bm{k}, c}\left( \grave{Z}_{q,\bm{k},c}\right)   \in \mathbb{R}^{2N},
%\end{equation}
where $N  =  C \sum_{ q }|\Omega^{k, S}_{q}|$ and $\bm{\mu} = vec_{q, \bm{k}, c}\left( \grave{\mu}_{q,\bm{k},c}\right)  $ indexed as  ${{Z}}_{n}$, ${\mu}_{n}$ with $n \in \{1, \dots 2N\}$. Assuming independent noise in the measurements, the joint probability density function (PDF) of all measurements across coils and contrasts  is given by:
\begin{equation}
	\label{eq: pdf2a}
	p(\bm{Z}  |  \bm{\theta}, \sigma ) = \prod^{Q}_{q=1} \prod_{\bm{k} \in \Omega^{k, S}_{q}}\prod^{C}_{c=1} p \left( \grave{Z}_{q, \bm{k}, c} |  \bm{\theta} ,\sigma \right) = \prod_{n=1}^{2N} p({Z}_{n}|\bm{\theta},\sigma),
\end{equation}
where $\sigma$ is the standard deviation of noise. Then $p({Z}_{n}|\bm{\theta},\sigma)$ is a Gaussian distribution given by:
\begin{equation}
	\label{eq: pdfformula0a}
	p({Z}_{n}|  \bm{\theta} ,\sigma)   =    \tfrac{1}{{{\sigma}}\sqrt{2\pi}}   \left(e^{-\tfrac{1}{2 \sigma}  \left({Z}_{n} - {\mu}_{n}( \bm{\theta})  \right)^2  }\right).
\end{equation}
%\subsubsection{Fisher information}
%We begin derivation of TEUSQA without including the impact of the prior information. In this case, the estimator would be a maximum likelihood estimator. For such estimator, CRLB quantifies the lower bound of the variance of $\bm{\theta}$ given that the estimator is unbiased \cite{cavassila_cramer-rao_2001}. CRLB has been used to predict variance for MR sequences earlier; however, most are based on  evaluations performed on the spatial domain. In this work, we derive it to include  the effects of undersampling by deriving it based on measurements in k-space. 

\subsubsection{Estimator and prior information}
\label{sec: estimator}

To estimate $\bm{\theta}$ given the measurements $\bm{Z}$, various estimators such as least square \cite{hilbert_accelerated_2018,zimmermann_accelerated_2018,hu_super_2019} or maximum likelihood \cite{block_model-based_2009,sumpf_model-based_2011,zhao_model-based_2014} have often been used in combination with prior information such as sparsity among the neighbouring voxels \cite{zhao_model-based_2014,zimmermann_accelerated_2018}.
In the TEUSQA framework, we  assume a weak prior on the parameters to avoid degeneracies. We consider a spatially independent prior with a normal distribution with mean $\overline{{ \bm{\theta}}}_{1} \in \R^{P}$ and covariance matrix $\bm{\Gamma_{1}} \in \R^{P\times P}$. %This is similar to a Gaussian prior without spatial correlation. 
Let  $\overline{{ \bm{\theta}}}_{1}$ and $\bm{\Gamma_{1}}$ be replicated $ |\Omega^{ {x}}|$ times to form $\overline{{ \bm{\theta}}}$ and $\bm{\Gamma} $ respectively. Then, the prior distribution over all the voxels is given by: 
\begin{equation}  
	\label{eq: prior}
	p(  \bm{\theta} |  \overline{{ \bm{\theta}}}, \bm{\Gamma} )   =     \prod_{ \bm{x} \in \Omega^{ {x}}  }    \mathcal{N}(   \bm{\theta_{ \bm{x}} } | \overline{ \bm{\theta}}_{1} ,\bm{\Gamma_{1}})  =   \frac{e^{-\frac{1}{2}(\bm{\theta} - \overline{\bm{\theta}})^{T} \bm{\Gamma}^{-1} (\bm{\theta} - \overline{\bm{\theta}}) }}{\sqrt{(2\pi )^{L} \det{\bm{\Gamma}}}}.
\end{equation}

To estimate $\bm{\theta}$, we use a maximum a-posteriori (MAP) estimate, which in the limit of infinitely weak prior converges to a maximum likelihood estimate:
%The stopping criteria was based on change in residue to ensure, best possible estimation is made.
\begin{equation}  \label{eq: MAP}
	\hat{\bm{\theta}}  = \arg \max_{\bm{\theta}}  \left[\log \left(p(\bm{Z} | {\bm{\theta}} ,\sigma) \right) + \log\left(p(  \bm{\theta} |  \overline{{ \bm{\theta}}}, \bm{\Gamma} )   \right) \right].
\end{equation}

\subsubsection{Prediction of parameter variance maps}

To predict a theoretical lower bound on the variance of $\bm{\theta}$ for the case of model-based reconstruction of an undersampled QMRI acquisition, Zhao et al. \cite{zhao_model-based_2014} derived an expression for the CRLB.  
It is defined by the inverse of the Fisher information matrix $\bm{I}( \bm{\theta}, \sigma ) \in \mathbb{R}^{L \times L}$,  given by:
\begin{align}
	\label{fisher2}
	\bm{I}(\bm{\theta}, \sigma) &  =  \mathbb{E}_{\bm{Z}}  \left[   \left(   \frac{\partial  \ln p({\bm{Z}} |  \bm{\theta} ,\sigma)}{\partial \bm{\theta}}  \right)    \left(   \frac{\partial  \ln p({\bm{Z}} |  \bm{\theta} ,\sigma)}{\partial \bm{\theta}}  \right)^{ T} \right] \\& = \sum\limits_{n=1}^{2N} \sum\limits_{m=1}^{2N}     \tfrac{\partial {{\mu}}_n }{\partial \bm{\theta}} \tfrac{\partial {{\mu}}_m}{\partial\bm{\theta}^T} \mathbb{E}\left[  \tfrac{\partial \ln p({{Z}}_{n} |   \bm{\theta} ,\sigma)}{\partial {{\mu}}_n} \tfrac{\partial \ln p({{Z}}_{m} |  \bm{\theta} ,\sigma)}{\partial {{\mu}}_m} \right]  \\&  =\frac{1}{\sigma^2} \jac(\bm{\theta})^{T} \jac(\bm{\theta})
\end{align}
where $ \jac(\bm{\theta})    =   \tfrac{\partial \bm{{\mu}}}{\partial \bm{\theta}} \in \mathbb{R}^{2N \times L}$. Although this expression nicely takes into account the effect of undersampling, it is only valid for pure maximum likelihood estimators as it neglects the effect of the prior information.

To account for the prior in the variance estimate for $\bm{\theta}$, we derive an approximate posterior distribution $p( \bm{\theta} | \bm{Z}, \sigma) $ considering the Bayes theorem similarly to the derivation of  Bayesian CRLB \cite{van_trees_detection_2013}.  In this derivation, we assume that the signal decay along the contrast given by  $f_q( \bm{\theta}_{ \bm{x}})$ can be considered locally linear  around the ground truth parameter values $\ttil$. 
The remaining terms in ${\mu}_{q, \bm{k},c}( \bm{\theta}) $ are linear, hence, $
\bm{\mu}( \bm{\theta}) \approx \bm{\mu}(\ttil) + \jac(\bm{\theta}) \{  \bm{\theta} - \ttil \}$.

In \ref{section: S1} we show that the posterior distribution of $\bm{\theta}$  is given by: $p( \bm{\theta} | \bm{Z}, \sigma) \propto \mathcal{N}(   \bm{\theta} | \breve{ \bm{\theta}} , \breve{ \bm{\Gamma} })$
with mean and covariance given by: 
\begin{equation}
	\breve{ \bm{\theta}}  =    \tfrac{1}{\sigma^2}  \breve{\bm{\Gamma}}   \left[  {  \jac( \ttil )^{T} \left\{  \bm{Z} -    {\bm{\mu}}( \ttil ) + \jac( \ttil ) \ttil \right\}  +  \bm{\Gamma}^{-1} \overline{ \bm{\theta}}   }   \right]  
\end{equation}

\begin{equation}  
	\label{eq prior_last1}
	\breve{\bm{\Gamma}} = \left[ \bm{\Gamma}^{-1} + \bm{I}(\ttil, \sigma) \right]^{-1}.
\end{equation}

The diagonal $\bm{d} \in \R^{L}$ of $\bm{\breve{\Gamma}} $ gives the individual posterior variance of each parameter at each voxel.
This vector can  be remapped  to $P$ variance maps over $\Omega^{x}$,
\begin{equation}
	\label{eq: vp1}
	{V_{\bm{x}, p}} =  \mathit{vec}_{\bm{x}, p}^{-1}\left(\bm{d}\right).
\end{equation}
%Note that with a weak prior ${V_{\bm{x}, p}}$ is much smaller than the prior variance (diagonal of $\bm{\Gamma_1}$) for voxels where the parameters can be recovered from the measurements. For voxels with e.g. zero proton density ${V_{\bm{x}, p}}$ converges to  the prior variance (diagonal of $\bm{\Gamma_1}$).
For voxels with low information content, e.g. with zero proton density, the posterior variance  ${V_{\bm{x}, p}}$ equals the prior variance (specified by the diagonal of $\bm{\Gamma_1}$). For voxels with high information content where the parameters can be reliably recovered from the measurements, the posterior variance converges to the inverse Fisher information matrix.

%Still the performance should be similar, hence the evaluations of experiments \ref{sec: mc} show accelerated computations can predict precision of full scale acquisitions.
\subsubsection{Accelerated computation of variance maps}

Evaluation of TEUSQA for a map with 6 parameters per voxel and size of $256 \times 256$ will require inversion of matrix of size $(6 \times256 \times 256)^{2}$. Such matrix inversions take long computation time and are inconvenient for repetitive use, often needed when designing undersampling patterns and scan protocols. Thus, accelerated computation is desirable.  We propose the evaluations to be performed in a downsampled parameter map, which leads to a smaller patch of k-space  $\Omega^{k, D}$. The evaluation of  $V_{\bm{x}, p}$ is dominated by computation of  $\bm{I}(\ttil, \sigma)$ where the number of computations is given by $ 2N L^{2} = 2 C \sum_{ q }|\Omega^{k, S}_{q}| (P|\Omega^{x}|)^2$. Assuming $|\Omega^{x}| = |\Omega^{k}|$ a smaller k-space would save computations by a factor  $\frac{|\Omega^{k}|^3}{|\Omega^{k, D}|^3}$ as well as avoid large matrix inversion by decreasing the size of $\bm{I}(\ttil, \sigma)$.

We hypothesize that using relatively small patches of k-space is sufficient to capture all the relevant aspects involved in model-based reconstruction. Even though a small k-space patch leads to a lower resolution image,  the model and the acquisition settings remain pertinent in the computation. Moreover, small scale features in k-space translate into large-scale image features, and coil sensitivity maps vary smoothly in the spatial domain. Therefore, a relatively small patch of k-space  is adequate to capture the variations in the coil sensitivity maps, thus capturing the parallel imaging induced acceleration. However, the downsampling ratio  should be limited to a factor that retains the variations in the coil sensitivity maps.

Due to the downsampling, the undersampling pattern would differ between the actual scan and the k-space considered by TEUSQA. Assuming that the full-sized and downsized undersampling patterns are generated using the same pattern generation technique, we provide an appropriate compensation factor that can be used to compensate for the difference. In the Equation \ref{fisher2}, summation is over the number of observations ($2N =   2 C \sum_{ q }|\Omega^{k, S}_{q}|$). The reduction in size of k-space $|\Omega^{k}|$ would decrease the number of observations.  
Assuming all observations have same SNR, the factor representing noise $\frac{1}{\sigma^2}$ should scale linearly with the number of observations; therefore, the required scaling factor to be applied to $\bm{I}(\ttil, \sigma)$ is $ \frac{|\Omega^{k}|}{|\Omega^{k, D}|}$.

%Assuming that the fullsized and downsized undersampling patterns are generated in the same way, we provide appropriate compensation factor that can be used to compensate for the difference. The summation in Equation \ref{fisher2}, which indicates number of observations, would be from $1$ to $2 N_d = 2C|\Omega^{k, D}|$ instead of $1$ to $2N =  2C|\Omega^{k}|$. By downsizing the k-space we decrease the number of observations, thus, assuming the noise scales linearly with the number of observations we compute the required scaling factor to be $\frac{|\Omega^{k}|}{|\Omega^{k, D}|}$.

\subsubsection{ From posterior variance to time efficiency}
The variance maps predicted by CRLB can be used to quantify noise amplification using metrics such g-factor \cite{velikina_accelerating_2013} or d-factor \cite{hu_super_2019}. 
%Hu et al. \cite{hu_super_2019} take the standard deviation predicted by CRLB to compute d-factor, similar metrics have been used to quantify noise amplification due to undersampling \cite{velikina_accelerating_2013}.  
Such metrics consider the ratio between square root of variance predicted for an undersampled acquisition with that of a fully sampled acquisition.  
For verification of these metrics with an actual scan, a fully sampled scan is needed. However, that is impractical  due to long scan times. Therefore, to facilitate comparison with actual scans, we propose to predict the coefficient of variation instead:
\begin{equation}
	\label{eq: normalised_Cvp}
	CV_{\bm{x}, p} =  \frac{\sqrt{ V_{\bm{x}, p}}}{{\tilde{\theta}}_{{\bm{x}, p}}}
\end{equation}
where ${{\tilde{\theta}}_{{\bm{x}, p}}}$ represents ground truth values of the parameters in the corresponding voxel. We take the average $CV_{\bm{x}, p}$ over a region of interest (ROI) $\Omega^{x, ROI} \subseteq \Omega^x $ to aggregate the result for parameter $p$: 
\begin{equation}
	\label{eq: CV}
	CV_{p} =  \sum_{ \bm{x} \in  \Omega^{x, ROI} } \frac{CV_{\bm{x}, p} }{ |\Omega^{x, ROI}|}.
\end{equation}  This $CV_{p}$ depends on sequence settings as well as the undersampling pattern. To quantify the information gained per unit scan time $T$, the final time efficiency metric TEUSQA is defined as 
\begin{equation}
	\label{eq: Time_eff}
	\eta_{p} =  \frac{1}{CV_{p}^2 \times T}.
\end{equation}
The inclusion of scan time in the metric allows comparison across different acceleration factors, and facilitates analysing the trade-off between scan time and precision.

\subsection{Undersampling patterns}
Scan time can be reduced by undersampling the two phase encoding dimensions in 3D Cartesian acquisitions. So, we consider the frequency encoding dimension to be fully sampled. The number of possible undersampling patterns for the two phase encoding dimensions and the contrast encoding dimension of  a 3D cartesian acquisition is excessively large ($2^{{P_1} \times P_2 \times Q}$, where $P_1$ and $P_2$ are size of the two phase encoding dimensions). Consequently, instead of trying to find an optimal pattern using TEUSQA, we propose an alternative approach. We compare several undersampling pattern generation techniques using TEUSQA and using a geometric property called Discrepancy. We correlate the results of TEUSQA and Discrepancy to gain insights that could help design time efficient patterns.

\begin{figure}[!t]
	\centering
	\begin{subfigure}[t]{0.22\textwidth}
		\includegraphics[width=\linewidth,trim={3cm 4cm 3cm 2cm},clip]{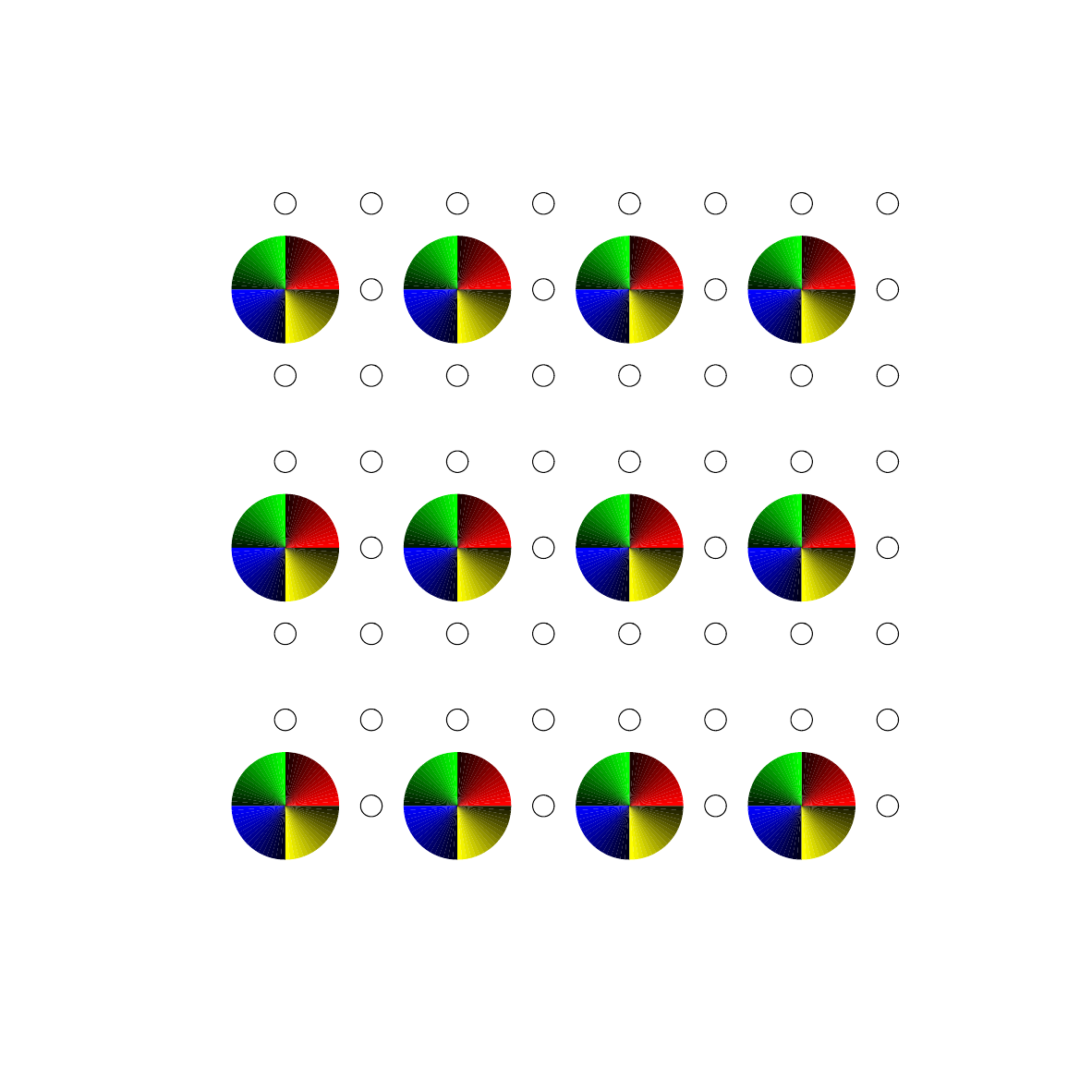}
		\caption{\centering Regular}
		\label{subfig: regular}
	\end{subfigure}
	\begin{subfigure}[t]{0.22\textwidth}
		\includegraphics[width=\linewidth,trim={3cm 4cm 3cm 2cm},clip]{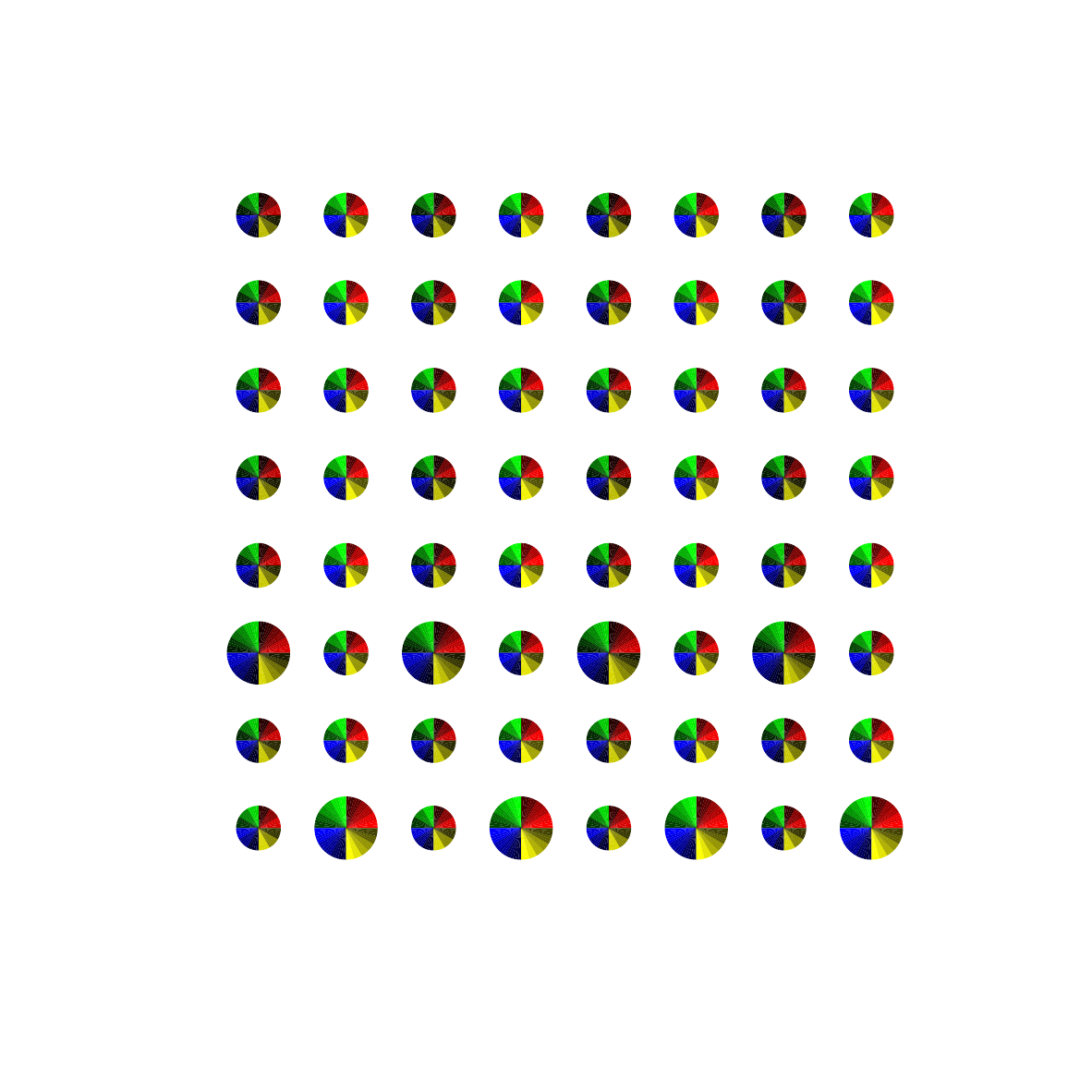}
		\caption{\centering Treg}
		\label{subfig: MR-I}
	\end{subfigure}
	\begin{subfigure}[t]{ 0.22\textwidth}
		\includegraphics[width=\linewidth,trim={3cm 4cm 3cm 2cm},clip]{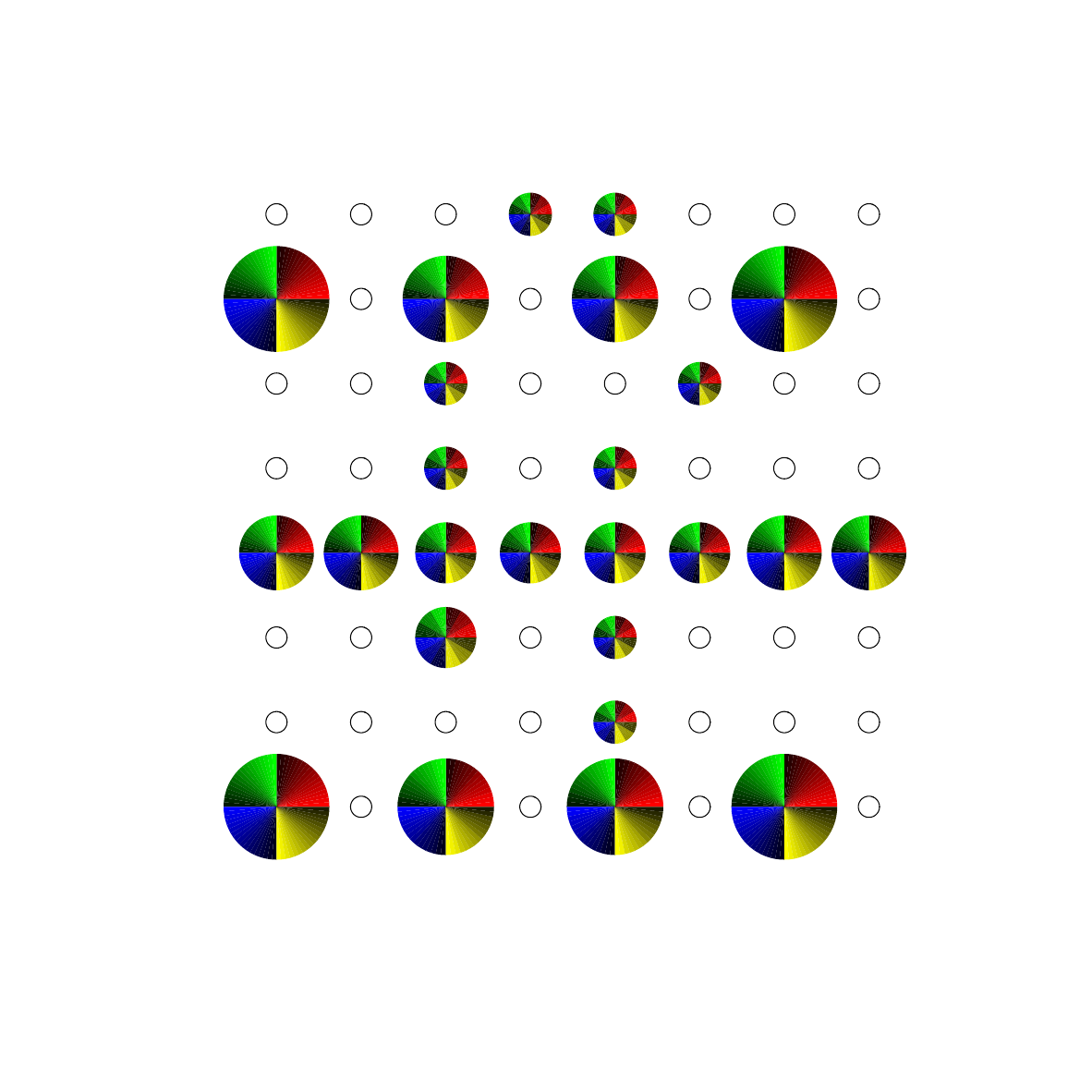}
		\caption{\centering Sreg}
		\label{subfig: MR-II}
	\end{subfigure}
	\begin{subfigure}[t]{0.22\textwidth}
		\includegraphics[width=\linewidth,trim={3cm 4cm 3cm 2cm},clip]{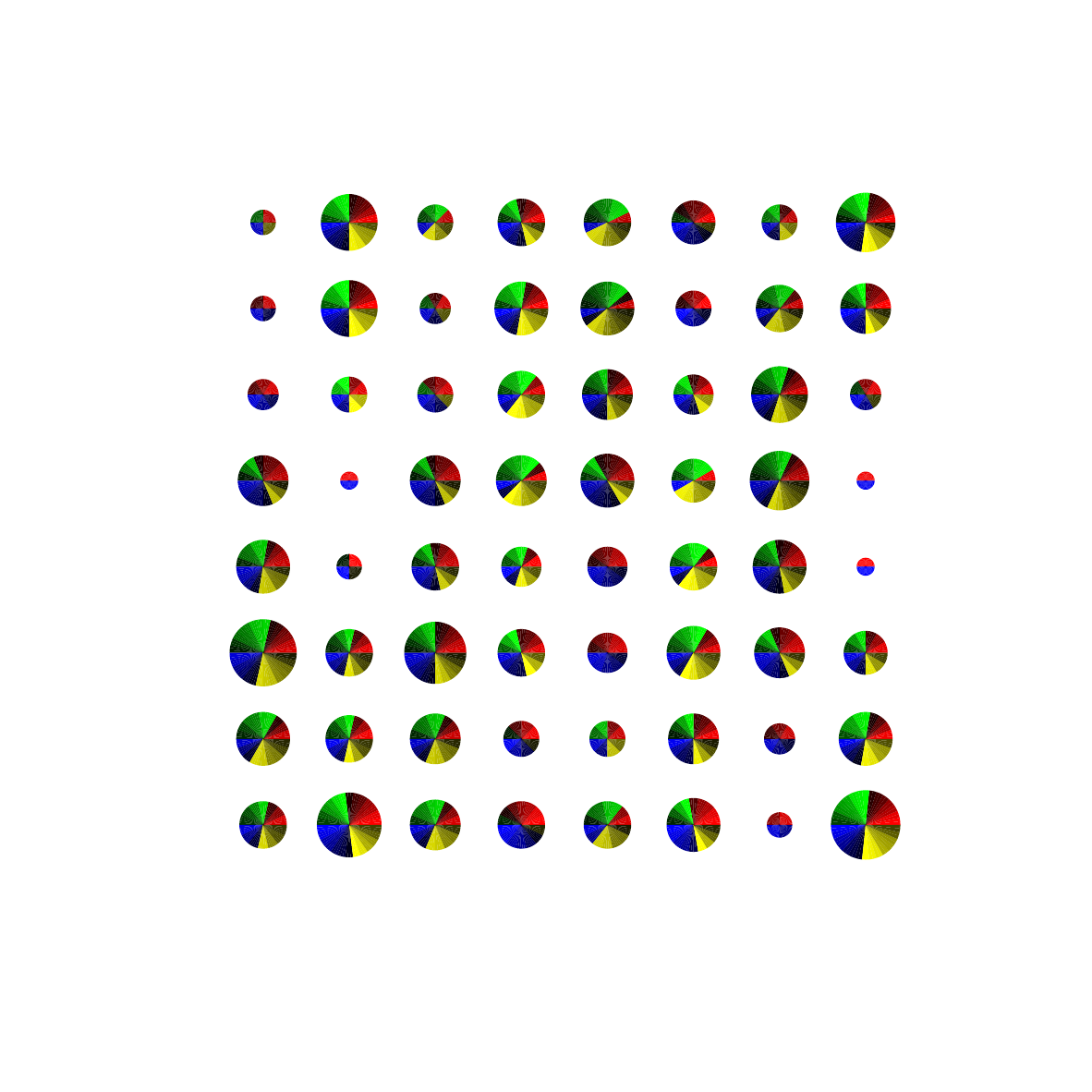}
		\caption{\centering TSreg}
		\label{subfig: MR-III}
	\end{subfigure}
	\begin{subfigure}[t]{0.22\textwidth}
		\includegraphics[width=\linewidth,trim={3cm 4cm 3cm 2cm},clip]{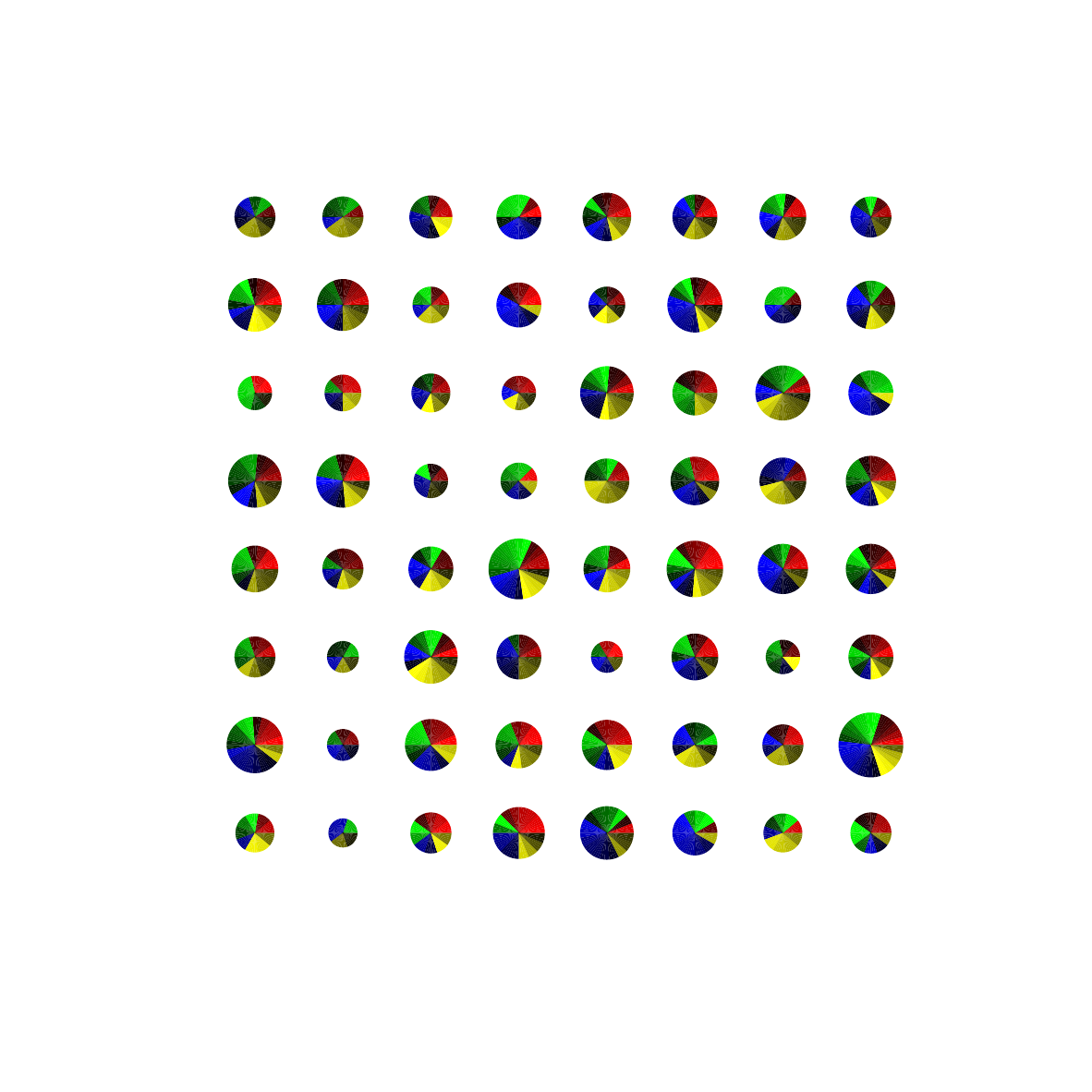}
		\caption{\centering Random}
		\label{subfig: random}
	\end{subfigure}
	\begin{subfigure}[t]{0.22\textwidth}
		\includegraphics[width=\linewidth,trim={3cm 4cm 3cm 2cm},clip]{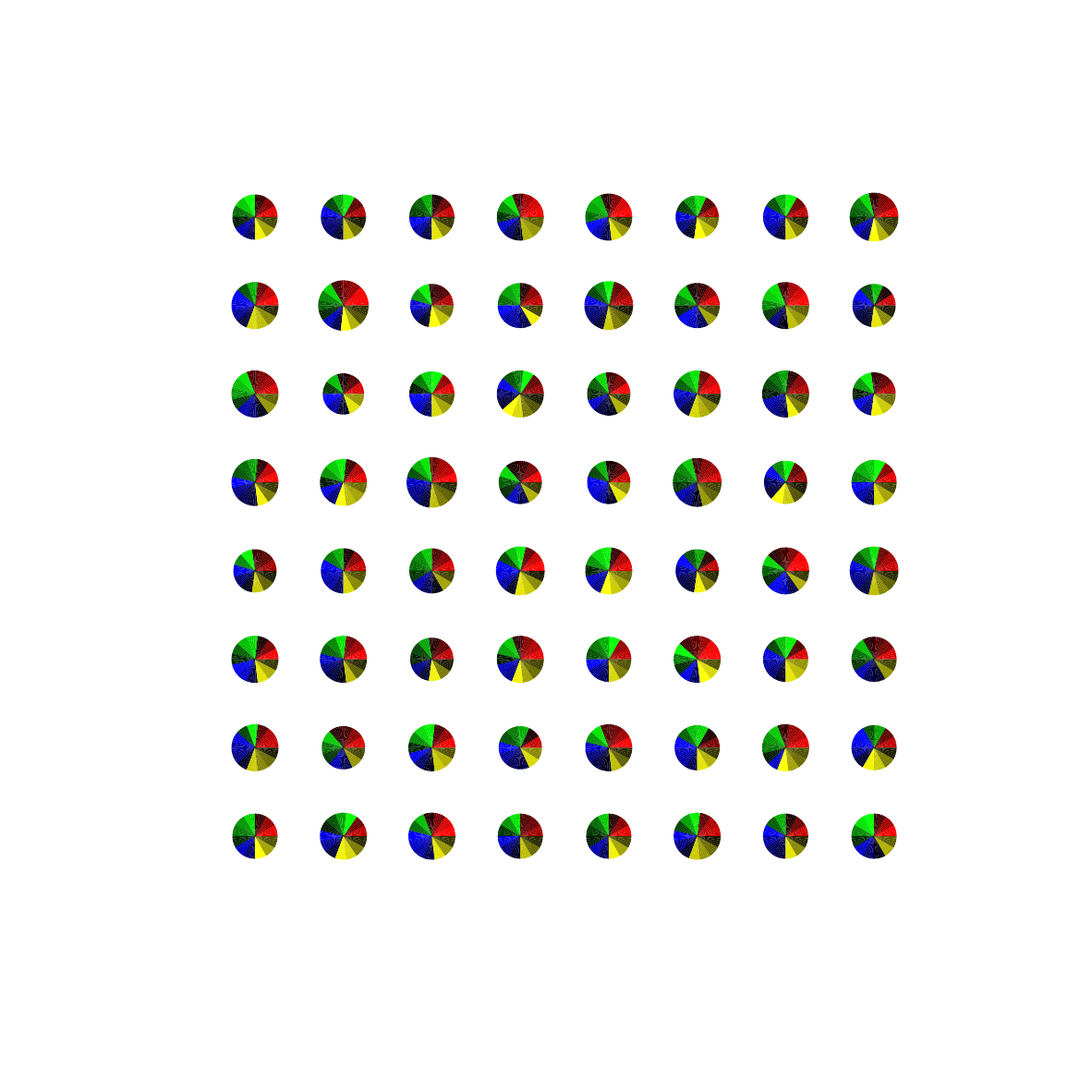}
		\caption{\centering Halton}
		\label{subfig: halton}
	\end{subfigure}
	\begin{subfigure}[t]{0.49\textwidth}
		\includegraphics[width = 1\linewidth]{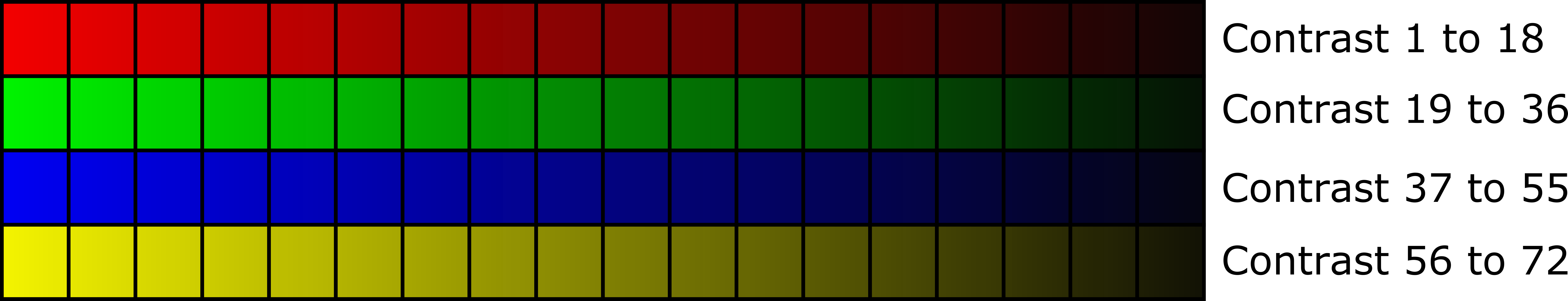}
		\caption{\centering Legend}
		\label{subfig: legend}
	\end{subfigure}
	\caption{{Graphical  representation of  undersampling patterns for $\bm{R} = [ 2, 3 ]$, k-space size of $ 8 \times 8 $ and number of contrasts $Q = 72$. Legend: each row represents a set of contrasts, grouped together with according to contrast property such as inversion times. }}
	\label{Fig: Und}
\end{figure}

%Therefore, instead of trying to find optimal pattern using TEUSQA, we propose an alternative approach. We compare several undersampling pattern generation techniques using TEUSQA along with their geometric property. The relation between their geometric property and time efficiency computed from the TEUSQA framework could provide insights into designing efficient undersampling patterns.
%\ref{Fig: Und}
Figure \ref{Fig: Und} shows a $8 \times 8$ patch of k-space generated by the patterns with acceleration factor $\bm{R} = [2, 3]$ and $Q = 72$. In this visualisation, each pie slice corresponds to one readout and has a constant area. Hence, the area of each pie corresponds to the number of times with which that k-space position is sampled in the entire acquisition. 

\subsubsection{Undersampling pattern generation techniques}
\label{sec: pattern}
The following notation is used to define all k-space positions for  undersampling patterns: 
\begin{align}
	\Omega^{k, S}_{q} = \left\{ \left[\begin{matrix}
		R_1 & S_1 \\
		S_2 & R_2 \\
	\end{matrix}  \right] \left[\begin{matrix}  x \\ y \\ \end{matrix}\right] +  \bm{\delta}  \forall  x, y  \in  \mathbb{N}  \right\}
\end{align}
where $\bm{R} = [R_1, R_2] $ are acceleration factors with associated total acceleration factor $R=R_1 \times R_2$, $\bm{\delta} = [\delta_1, \delta_2]$ are translational shifts,  $\bm{S} = [S_1, S_2]$ are the shears applied to the k-space pattern in the phase encoding 1 and phase encoding 2 dimensions respectively. $\bm{R} $, $\bm{S}$, and $\bm{\delta}$ could be constants, or functions of $q$.
\paragraph{Regular} In this most basic undersampling pattern,  $\bm{\delta} = [0, 0]$  and $ \bm{S}   =   [0, 0] : \Omega^{k, S}_{q}   =  \left\{\left[xR_1,  yR_2\right]  \forall  x,y \in  \mathbb{N} \right\}.$ Note that the same k-space positions are sampled for each contrast $q$. \newline
\paragraph{Translated Regular (Treg)}	 

To obtain complementary spatial encoding the regular undersampling patterns can be translated with respect to each other for the different contrasts $q$. The translations we investigate are the patterns: $\Omega^{k, S}_{q}      = \left\{\left[xR_1,  yR_2\right] + \bm{\delta}  \forall  x,y \in  \mathbb{N} \right\},$    $\delta_2 = {q} \bmod {R_2}$, and  $\delta_1 =  {\left\lfloor\frac{q}{R_2} \right\rfloor} \bmod {R_1}.$\newline
\paragraph{Sheared Regular (Sreg)}
To vary aliasing patterns for the different contrasts, the shearing rate can be varied: $ \Omega^{k, S}_{q} = \left\{\left[xR_1 + S_2,  yR_2 + S_2 \right] \forall x, y \in  \mathbb{N} \right\}$, $S_2 = {q} \bmod  {R_2}$, and $S_1 =   {\left\lfloor \frac{q}{R_2} \right\rfloor} \bmod {R_1}.$ \newline
\paragraph{Translated and Sheared Regular (TSreg)}	  	 
Both Treg and Sreg provide limited number of variation along the contrast dimension and patterns get repeated along $q$. To increase the number of variations, both translations and sharing rates are varied along the contrast dimension: $\Omega^{k, S}_{q} = \left\{\left[xR_1 + S_2 , yR_2 + S_1 \right] + \bm{\delta}    \forall  x, y  \in  \mathbb{N} \right\}$, $\delta_2 = q \bmod {R_2}$, $\delta_1 =   {\left\lfloor \frac{q}{R_2} \right\rfloor} \bmod {R_1}$, $ S_2 = {q} \bmod {R} $,  and  $S_1 =  {\left\lfloor \frac{q}{R} \right\rfloor} \bmod {R}$.\newline
\paragraph{Random}
Random undersampling is know to produce incoherent undersampling artifacts useful for reconstruction using compressed sensing \cite{knoll_adapted_2011}. In this pattern, the k-space positions are sampled randomly for each contrast. For each contrast independently the required number of samples $ |\Omega_{q}^{k, S}|  = |\Omega^{k}| / (R_1 R_2) $   is picked randomly without replacement from the pool of all k-space positions ($\Omega^{k}$). 

\paragraph{Random sampling with Halton sequence (Halton)}
The Random sampling technique described above considers each contrast independently and does not distribute the samples considering the contrast dimension. Consequently, it generates areas with an uneven density of sampled k-space positions within each contrast.  To address these issues, we propose another random pattern generation technique based on Halton sampling, a well-known low-discrepancy sampling technique \cite{wang_randomized_2000}.  Speidel et al. \cite{speidel_quasi-random_2018} used a similar low-discrepancy sequence to generate undersampling patterns for single-point imaging. The details of the implementation are presented as pseudo code in \ref{A: algo}.

\subsubsection{Discrepancy} 

To study the patterns purely on the basis of their geometry, we introduce a measure called Discrepancy. Such measure can provide us insights that are useful for pattern generation without being specific to an acquisition model. Discrepancy has been used to test whether a set of points is equidistributed in the field of integration theory \cite{shirley_discrepancy_1991}. From the several ways to quantify Discrepancy we use the $L_2$ form, which quantifies the $L_2$ error when using the set of points when integrating some class of smooth functions \cite{heinrich_efficient_1996}, and is given by:
%Among various ways to quantify it, we do it using the $L_2$ form of Discrepancy, which quantifies the L2 error of using the set of points to integrate some class of smooth functions, which can be evaluated by \cite{heinrich_efficient_1996}:

\begin{equation}
	D^2 = 3^{-d} - 2^{1-d} \sum_{i=1}^{m} v_{i} \prod_{u=1}^{d}(1 - x^{2}_{i,u}) + \sum_{i,j = 1}^{m} v_{i} v_{j} \prod_{u=1}^{d} (1  - \max(x_{i,k}, x_{j,u}))
\end{equation}
where $d$ is the dimensionality of the pattern, including the two phase encoding dimensions and the contrast dimension, $\bm{x_{i}} \in [0, 1]^d$ represents a point, $u$ indexes elements of $\bm{x_{i}}$ and ${v_{i}} \in \R$ defines a weight of each sampled point which we consider to be same for all the points.

\section{Methods}
%\subsection{Experiments}
We used two sequences for verification of TEUSQA: 1) 3D IP-FSE; 2) 3D GRASE. In the main manuscript, we describe all experiments and results for 3D IP-FSE. The same information is presented for 3D GRASE in \ref{A: 3D GRASE}.
	
\subsection{Sequence and estimator details}
\label{sec: sequence used} 
%as it naturally limits T1 and T2 to positive values
3D IP-FSE is used for joint $T_1$ and $T_2$ mapping QMRI protocol where  each echo is considered as a different contrast. The parameter vector  $\bm{\theta_x} = \left[ \Re(M_0), \Im(M_0), \ln(T_1), \ln(T_2) \right]$, where   $\Re(M_0)$, $\Im(M_0)$ are real and imaginary component of the complex valued apparent proton density $M_0$. The logarithm of $T_1$ and $T_2$ were taken, as it naturally limits the $T_1$ and $T_2$ to positive values and lets the Gaussian prior select the order of magnitude. We used sequence settings in Table \ref{tab: scan set} where the TIs were selected according to  \cite{barral_robust_2010} which targets brain $T_1$. The prediction function  $f_q( \bm{\theta}_{ \bm{x}})$ performs extended phase graph (EPG) simulation  \cite{busse_fast_2006}. In post processing the $\ln(T_1)$ and $\ln(T_2)$ maps were converted to $T_1$ and $T_2$ using principles of propagation of uncertainty. This conversion was also applied in the time efficiency analysis.  As prior we used: $\overline{ \bm{\theta}}_1 =[0,0, \ln(1000), \ln(70)]$ and
$\bm{\Gamma_1} =
\left[ {\begin{array}{cccc}
		20^2 & 0 & 0 & 0\\
		0 & 20^2 & 0 & 0\\
		0 & 0 & \ln(10)^2 & 0\\
		0 & 0 & 0 & \ln(7)^2\\
\end{array} } \right]$  which corresponds to an a-priori $1 \text{-} \sigma$ interval for $T_1$ of $[100, \ldots, 10000]$ ms centered at $T_1$ = $1000$ ms and for $T_2$ of $[10, \ldots, 490]$ ms centered at $T_2 = 70 $ms. The variance  on $M_0$ depends on the intensities of the images which scale arbitrarily between different acquisitions and have to be adjusted accordingly. The $1 \text{-} \sigma$ intervals for   $\Re{\{M_0\}}$ and $\Im{\{M_0\}}$  of  $[-20, \ldots, 20]$ centered at $\Re{\{M_0\}} = \Im{\{M_0\}}=0$ were about $5$ times the root mean square value present in the ground truth map used.

%\begin{table}[h]
%	\centering
%	\caption{Acquisition settings for 3D IP-FSE.}
%	\begin{tabular}{|c|c|}
%		\hline
%		\textbf{Sequence settings} & \textbf{Values} \\ \hline
%		Inversion delay (TI) & \begin{tabular}[c]{@{}c@{}}2400, 1100, 50, 400   (ms)\end{tabular} \\ 
%		Repetition time (TR) & \begin{tabular}[c]{@{}c@{}}2552  (ms)\end{tabular} \\ 
%		\begin{tabular}[c]{@{}c@{}}Echo train length   (ETL)\end{tabular} & 18 \\ 
%		Echo spacing (ESP) & 6 (ms) \\ 
%		Flip angles (FA) & $180^{\circ}$ \\
%		Contrasts (${Q}$)	& 72 \\ \hline
%	\end{tabular}
%	\label{table: tabs1}
%\end{table}

\begin{table*}
	\centering
	\caption{Acquisitions settings and scan protocols used in this work. Note that ground truth experiments use same settings accept a smaller Acquisition matrix and Field of View in $PE_2$ incase of 3D IP-FSE.}
	\begin{tabular}{|c|c|c|c|} 
		\hline
		\textbf{Scans}                                                                          & \begin{tabular}[c]{@{}c@{}}\textbf{3D IP-FSE}\\\textbf{~phantom scans}\end{tabular} & \begin{tabular}[c]{@{}c@{}}\textbf{In-vivo scan }\\\textbf{3D IP-FSE}\end{tabular} & \begin{tabular}[c]{@{}c@{}}\textbf{~3D GRASE }\\\textbf{ scans}\end{tabular}  \\ 
		\hline
		\multicolumn{1}{|c}{}                                                                   & \multicolumn{1}{c}{}                                                             & \multicolumn{1}{c}{}                                                            &                                                                                      \\ 
		\hline
		\textbf{\textit{Acquisition settings }}                                                 & \multicolumn{1}{c}{}                                                             & \multicolumn{1}{c}{}                                                            &                                                                                      \\ 
		\hline
		Acquisition matrix                                                                      & \begin{tabular}[c]{@{}c@{}}AP: 64~ \\LR: 128 \\SI: 128\end{tabular}              & \begin{tabular}[c]{@{}c@{}}AP: 96\\LR: 76\\SI: 128\end{tabular}                 & \begin{tabular}[c]{@{}c@{}}AP: 128\\LR: 128\\SI: 84\end{tabular}                     \\ 
		\hline
		Field of view (mm)                                                                      & \begin{tabular}[c]{@{}c@{}}$PE_1$: 128\\$PE_2$: 128\\FE: 128\end{tabular}              & \begin{tabular}[c]{@{}c@{}}$PE_1$: 230\\$PE_2$: 182.4\\FE: 307\end{tabular}           & \begin{tabular}[c]{@{}c@{}}FE: 204.8\\$PE_1$: 204.8\\$PE_2$: 134.4\end{tabular}            \\ 
		\hline
		Number of coils (C)                                                                     & 8                                                                                & 8                                                                               & 32                                                                                   \\ 
		\hline
		Acceleration factor (R)                                                                 & 32                                                                               & 32                                                                              & 16  (including calibration region)                                                                                 \\ 
		\hline
		Scan time                                                                               &       $\!\!\!(64\! \times\! 128 \! \times \sum{\!TR} ) /R \!\approx \!\!$  45  min. $\!\!\!$                                                                           &    $\!\!\!(96\! \times\! 76 \! \times \sum{\!TR} ) /R \!\approx$  43  min. $\!\!\!$                                                                              &    $\!\!\!(128\! \times\! 84 \! \times \!TR ) /R \!\approx$  20  min. $\!\!\!$                                                                                   \\ 
		\hline
		Scan for calibration region                                                             & $\!\!\! 12 \!\times 12 \!\times\! 4  \times\! \sum{\!TR}\! \approx\!\! $ 24 min.                                                                                 & NA                                                                              & $ \!\!\! 12 \!\times 12 \!\times\! \!TR\! \approx\!\! $ 5 min.                                                                                    \\ 
		\hline
		\multicolumn{1}{|c}{}                                                                   & \multicolumn{1}{c}{}                                                             & \multicolumn{1}{c}{}                                                            &                                                                                      \\ 
		\hline
		\textbf{\textit{Sequence settings}}                                                     & \multicolumn{1}{c}{}                                                             & \multicolumn{1}{c}{}                                                            &                                                                                      \\ 
		\hline
		Inversion delay (TI)                                                                    & 2400, 1100, 50, 400 (ms)                                                         & 2400, 1100, 50, 400 (ms)                                                        & NA                                                                                   \\ 
		\hline
		Repetition time (TR)                                                                    & 2552 (ms)                                                                        & 2552 (ms)                                                                       & 1800 (ms)                                                                            \\ 
		\hline
		Echo train length (ETL)                                                                 & 18                                                                               & 32                                                                              & 32                                                                                   \\ 
		\hline
		Echo spacing (ESP)                                                                      & 6 (ms)                                                                           & 6 (ms)                                                                          & 10 (ms)                                                                              \\ 
		\hline
		Flip angles (FA)                                                                        & $180^\circ$                                                                              & $180^\circ$                                                                             & $180^\circ$                                                                                  \\ 
		\hline
		Contrasts (Q)                                                                           & 72                                                                               & 128                                                                             & 96                                                                                   \\ 
		\hline
		\begin{tabular}[c]{@{}l@{}}Delay between Spin echo \\and Gradient echo ($\Delta_t$)\end{tabular} & NA                                                                               & NA                                                                              & 2 (ms)                                                                               \\
		\hline
	\end{tabular}
\label{tab: scan set}
\end{table*}
\subsection{Verification of TEUSQA with numerical simulation}
\label{sec: MC}

\subsubsection{Acquisition of ground truth map} 
\label{sec: MC_1}
To obtain realistic ground truth parameter maps we performed a fully sampled scan of the Eurospin II T05 (Diagnostic Sonar LTD, Livingston, Scotland) with sequence settings described in Section \ref{sec: sequence used} and a $3.0$ T  clinical scanner (Discovery MR750, GE Healthcare, Waukesha, WI) using an 8-channel head coil. As performing a fully sampled acquisition on both  phase encoding directions takes an impractically long  scan time, the acquisitions were performed with a reduced acquisition matrix of size  $ 8 \times 64 \times 128$ in $PE_2 (SI) \times PE_1 (LR) \times FE (AP)$. Only the central slice of the reduced $PE_2$ dimension was selected for further processing. This slice will be used as if it was acquired along the $PE_1$ and $PE_2$ dimensions for the experiments in the following sections. The eight coil sensitivity maps $C^{GT}_{\bm{x}, c}$ were computed from the  first contrast of the fully sampled scans using the \textit{ESPIRIT} technique \cite{uecker_espirit-eigenvalue_2014} and the \textit{BART} toolbox \cite{uecker_generalized_2016}. Subsequently, the parameter maps used as ground truth, $\bm{\theta}^{GT}_{\bm{x}}$, were estimated by least squares fitting of  $f_q(\theta_{ \bm{x}})$ to each voxel of the contrast images.

%\paragraph{Procedure for Monte Carlo simulation}

\subsubsection{Time efficiency based on Monte Carlo simulation} 
\label{sec: mc}
As validation of the time efficiency metric $\eta_{p}$ we compared it to results from a Monte Carlo experiment.  The forward model in Equation \ref{eq: model} was used to generate MR signals in the k-space domain using the ground truth parameter map $\bm{\theta}^{GT}_{\bm{x}}$ and coil sensitivity maps $C^{GT}_{\bm{x}, c}$. All the undersampling patterns described in Section \ref{sec: pattern} with set of $\bm{R}\!\!:\!\! U \!\!=\!\! \{ [1,1] , [1,2] , [2,2] , [2,4] , [3,3] , [3,4] , [4,4] , [4,6] , [8,4] , [6,6] ,$ \\$ [8,6] \} $ were used. 
A complex Gaussian noise ($\sigma$) equivalent to the SNR of 50 was added in the 100 Monte Carlo iterations, where `signal' was taken as the root mean square of the predicted full k-space of all contrasts. The parameters were recovered using Equation \ref{eq: MAP} for each noise realisation. A ROI was manualy drawn, selecting voxels inside the spheres for which nominal values were available. For each voxel inside the ROI the CV over the Monte Carlo iterations was evaluated using Equation \ref{eq: normalised_Cvp} taking nominal value as ground truth. Subsequently Equation \ref{eq: Time_eff} was used to compute $\eta^{MC}_p$. Note that due to the limit of 100 Monte Carlo iterations, the 95$\%$ confidence interval is $[0.74, 1.29] \eta^{MC}_p$.

%An ROI was manually created, selecting voxels inside the spheres for which nominal values were provided in the phantom's datasheet. The CV over the Monte Carlo iterations was evaluated for all voxels in the ROI for each noise realisation. The nominal value of the parameters from the datasheet was used as the mean value in the CV.  The $\eta^{MC}_{p}$ was computed from the CV for each voxel according to Equation \ref{eq: Time_eff}.
%subsubsection{Ratio of $\eta^{MC}_{p}$ to  $\eta_{p}$} The resulting $\eta^{MC}_{p}$ maps  were downsampled to $\Omega^{x, D}$ with nearest-neighbor interpolation. The ratio of $\eta^{MC}_{p}$ to  $\eta_{p}$ was computed  voxel-wise and results were  plotted as box plot for comparison.

\begin{figure}
	\centering
	\begin{subfigure}[t]{ 0.2\textwidth}
		\includegraphics[width=\linewidth]{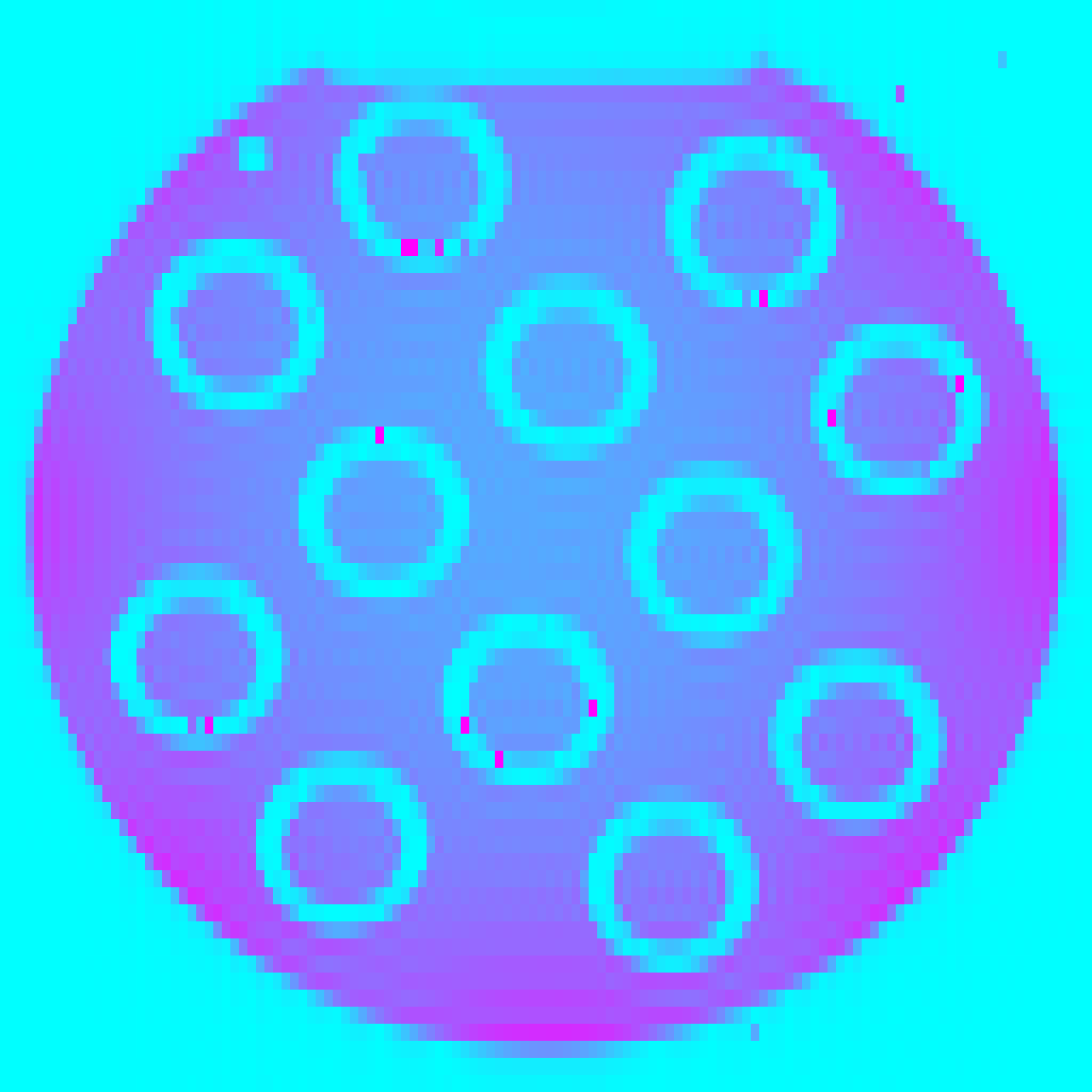}
		\caption{\centering {Full-size magnitude of $M_0$ map}}
		\label{subfig: M0rf}
	\end{subfigure}
	\begin{subfigure}[t]{0.2\textwidth}
		\includegraphics[width=\linewidth]{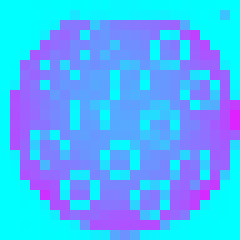}
		\caption{\centering { Downsampled magnitude of $M_0$ map}}
		\label{subfig: M0rd}
	\end{subfigure}
	\begin{subfigure}[t]{0.05\textwidth}
		\centering
		\includegraphics[scale=0.65]{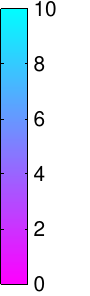}
	\end{subfigure}
	\begin{subfigure}[t]{ 0.2\textwidth}
		\includegraphics[width=\linewidth]{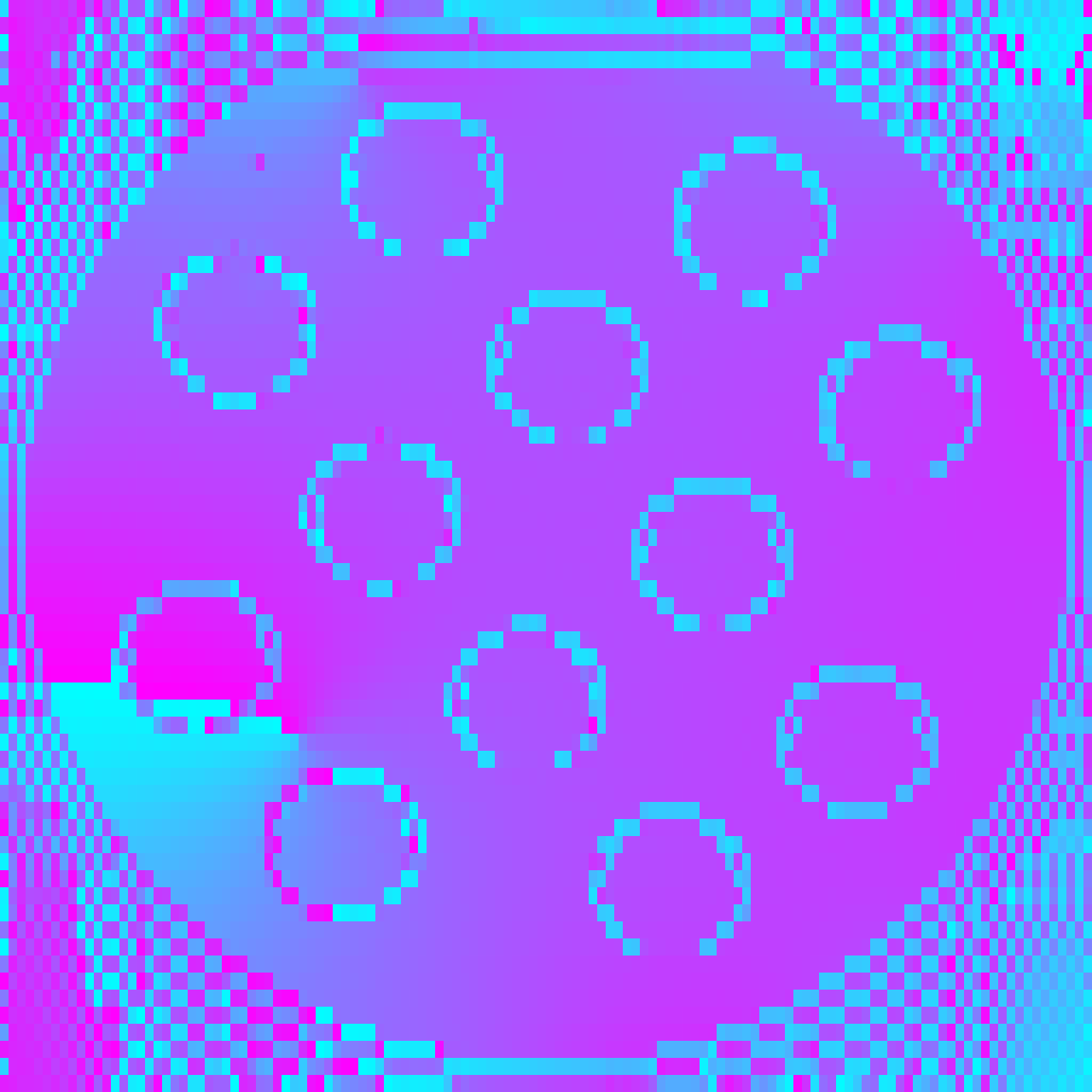}
		\caption{\centering {Full-size phase map of $M_0$}}
		\label{subfig: M0if}
	\end{subfigure}
	\begin{subfigure}[t]{0.2\textwidth}
		\includegraphics[width=\linewidth]{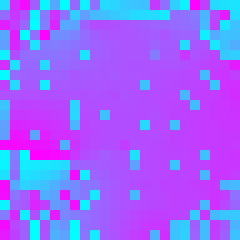}
		\caption{\centering { Downsampled phase map of $M_0$}}
		\label{subfig: M0id}
	\end{subfigure}
	\begin{subfigure}[t]{0.05\textwidth}
		\centering
		\includegraphics[scale=0.1]{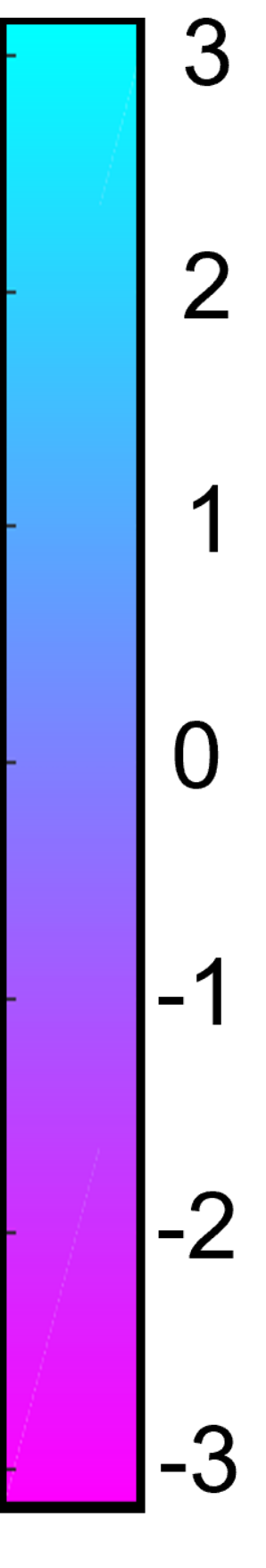}
	\end{subfigure}
	\begin{subfigure}[t]{ 0.2\textwidth}
		\includegraphics[width=\linewidth]{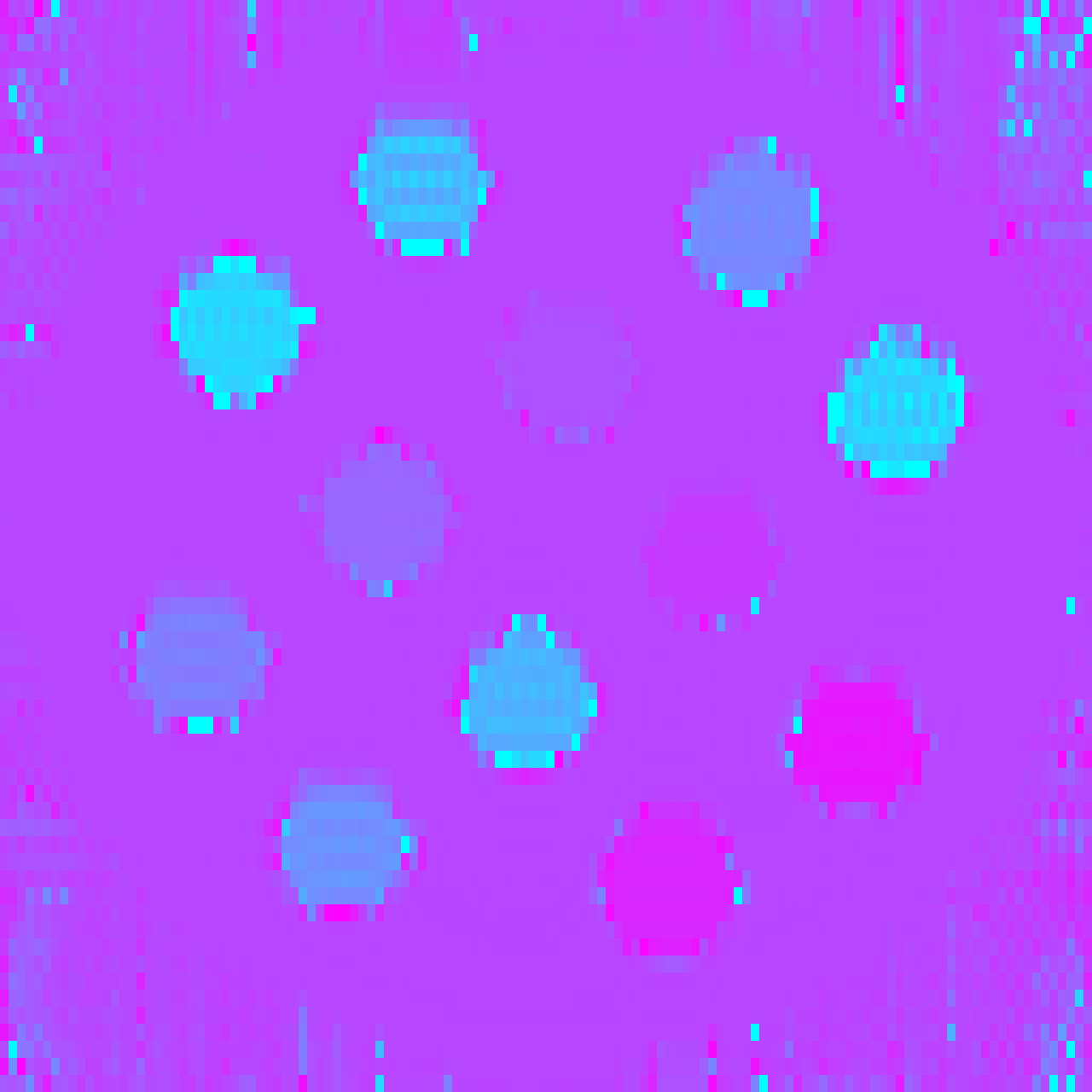}
		\caption{\centering {Full-size $T_1$ map (ms)}}
		\label{subfig: T1f}
	\end{subfigure}
	\begin{subfigure}[t]{0.2\textwidth}
		\includegraphics[width=\linewidth]{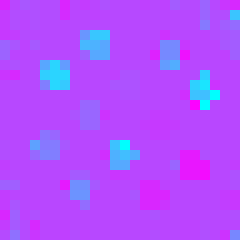}
		\caption{\centering { Downsampled $T_1$ map (ms)}}
		\label{subfig: T1d}    \end{subfigure}
	\begin{subfigure}[t]{0.05\textwidth}
		\centering
		\includegraphics[scale=0.65]{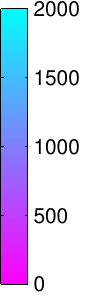}
	\end{subfigure}
	\begin{subfigure}[t]{ 0.2\textwidth}
		\includegraphics[height = \linewidth, width=\linewidth]{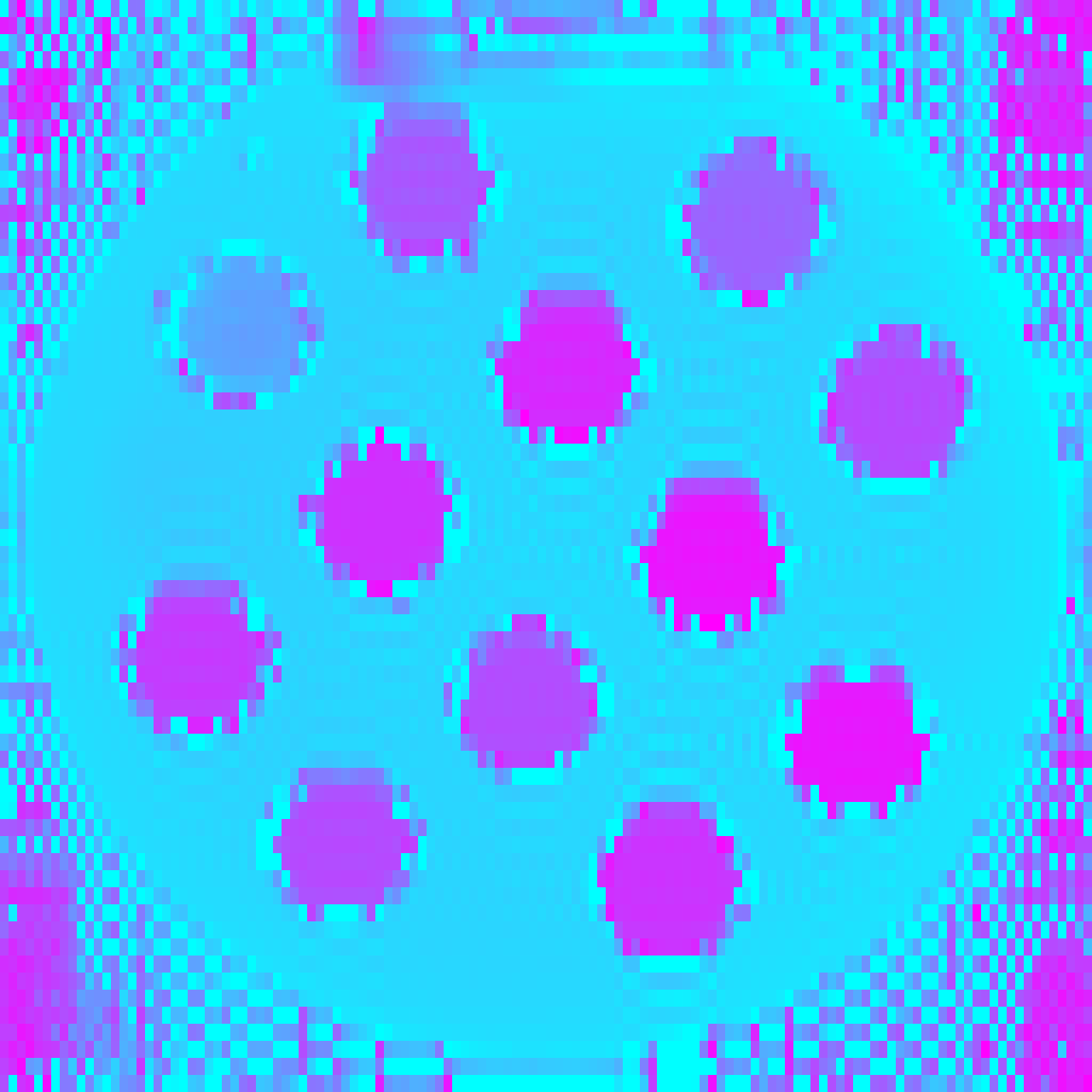}
		\caption{\centering {Full-size $T_2$ map (ms)}}
		\label{subfig: T2f}       \end{subfigure}
	\begin{subfigure}[t]{0.2\textwidth}
		\includegraphics[height = \linewidth, width=\linewidth]{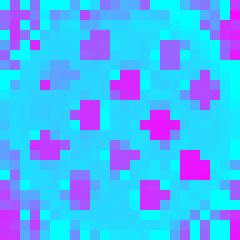}
		\caption{\centering { Downsampled $T_2$ map (ms)}}
		\label{subfig: T2d}       \end{subfigure}
	\begin{subfigure}[t]{0.05\textwidth}
		\centering
		\includegraphics[scale=0.65]{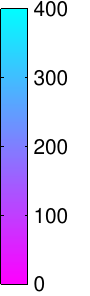}
	\end{subfigure}
	
	\caption{ The ground truth maps acquired from a fully sampled acquisition of the Eurospin II TO5 phantom used in the Monte Carlo simulations are shown on the left, and downsampled version of the map for computation of TEUSQA is shown on the right.  }
	\label{Fig: GTs}
\end{figure}

\subsubsection{Computation of $\eta_{p}$} The $\bm{\theta}_{\bm{x}}^{GT}$ and  $C^{GT}_{\bm{x}, c}$ were downsampled to $|\Omega^{x, D}| = 24 \times 24$ by applying nearest-neighbor interpolation on each parameter map separately while undersampling patterns were generated specifically for $\Omega^{x, D}$. Both the original and downsampled ground truth maps are shown in  Figure \ref{Fig: GTs}. The ROI mask was obtained by applying the same downsampling operation to  $\bm{\theta}_{\bm{x}}^{GT}$, to the ROI mask used for Monte Carlo simulation. The $ \eta_{p} $ for each voxel corresponding to voxels in the full-size map where nominal values are available was computed according to Equation \ref{eq: Time_eff}.
%\ref{Fig: GTs}
\subsubsection{Ratio of $\eta^{MC}_{p}$ to  $\eta_{p}$} The resulting $\eta^{MC}_{p}$ maps  were downsampled to $\Omega^{x, D}$ with nearest-neighbor interpolation. The ratio of $\eta^{MC}_{p}$ to  $\eta_{p}$ was computed  voxel-wise and results were  plotted as box plot for comparison.

\subsection{Comparison of undersampling patterns}
For the selection of an appropriate pattern for a prospective acquisition  and to gain insight into the relation between $\eta_p$ and Discrepancy,  $\eta_{p}$ was computed for all  patterns described in Section \ref{sec: pattern} and acceleration factors in set $U$. The downsampled parameter map $\bm{\theta}^{D}_{\bm{x}}$ and coil sensitivity maps $C^{D}_{\bm{x}, c}$ were used along with the noise level equivalent to SNR of 50 for computation of TEUSQA. The Discrepancy of the patterns was also computed for comparison.

\subsection{Verification of TEUSQA with prospective acquisition}
To compare TEUSQA on actual acquisitions, we performed test-retest variability of a prospectively undersampled acquisition.

\subsubsection{Prospective acquisition}	
The Halton  pattern with the  acceleration factor  $R = 32$ was selected for a prospective acquisition of the ISMRM model 130 phantom \cite{jiang_repeatability_2017} using the same system as the ground truth map acquisition. Sequence settings equal to Table \ref{tab: scan set} were used with the acquisition settings given in Table \ref{tab: scan set}. An additional patch of k-space of size $12 \times 12$ was acquired in the center of k-space as calibration region for generating coil sensitivity maps. This region was acquired for all echoes even though only one is needed. We use only the first contrast ($ q = 1 $) for computing coil sensitivity maps, we acquired additional patches for all echoes. These were used for estimation, however they were ignored in the computation of $\eta_{p}$  as we do not consider it as part of the undersampling pattern and hypothesize that its inclusion in the estimation brings negligible difference between  $\eta_{p}$ and observed time efficiency. 
%
%\begin{table}[h]
%	\caption{{Acquisition settings for the prospectively  undersampled phantom scan}}
%	\begin{tabular}{|c|c|}
%		\hline
%		\textbf{Acquisition settings} & \textbf{Values} \\ \hline
%		Acquisition matrix & $AP \!\! : \!\! 64, LR \!\! : \!\! 128, SI \!\! : \!\! 128$ \\ 
%		Field of view ($mm^3$) & 	 $\!\!\!	\!\!\! PE_1 \!\! : \!\! 128, PE_2 \!\! : \!\! 128, FE \!\! : \!\! 128 	\!\!\! 	\!\!\!$ \\
%		Number of coils (${C}$)	& 8 \\ 
%		Acceleration factor ($R$)	& 32 \\ 
%		Scan time & 	 $\!\!\!(64\! \times\! 128 \! \times \sum{\!TR} ) /R \!\approx \!\!$  45  min. $\!\!\!$ \\ 
%		$\!\!\! $Scan time for calibration region	$\!\!\! $	&  $\!\!\! 12 \!\times 12 \!\times\! 4  \times\! \sum{\!TR}\! \approx\!\! $ 24 min.  \\ \hline
%	\end{tabular}
%	\label{table: tabs3}
%\end{table}
Two scans were performed in the same scan session to measure the CV in the acquisitions. The coil sensitivity maps were computed using the calibration region for each slice in the FE direction. 
The noise level was computed from a patch of $10 \times 20 \times 20$ at the edge of k-space, across all coil channels and image contrasts. Specifically, the noise level was set to the standard deviation of the difference between the two acquisitions across all sampled positions of k-space in that patch divided by $\sqrt{2}$. 
The parameter estimation was performed separately for each slice in the FE dimension to allow the parallel processing of slices.

\subsubsection{Time efficiency based on prospective acquisition}	The phantom has multiple layers of sphere arrays with a range of specific $T_1$, $T_2$, and proton density values \cite{jiang_repeatability_2017}. 
Using the given sequence settings we cannot accurately map spheres having  $T_1$ and $T_2$ values that lie outside the range:  first TI $  \leq T_1 \leq \text{last TI}$ and second TE $  \leq T_2 \leq \text{last TE}  $ respectively. These spheres were excluded during the manual ROI selection. For each parameter, a map of the difference between the two acquisitions was computed. The voxels within each ROI of the difference map were divided by their nominal value and divided by $\sqrt{2}$. The standard deviation with its $95\%$ confidence bounds assuming a normal distribution was computed for all voxels (in all ROIs) of the resulting map to obtain the CV of the prospective scan.  The CV, along with lower and upper bounds, was used to compute the time efficiency denoted as $\eta^{ACQ}_{p}$ using Equation \ref{eq: Time_eff}.

\subsubsection{Computation of $\eta_{p}$}
The $\eta_{p} $ was computed for each slice independently in the FE direction using the estimated map from the first acquisition. Taking $\bm{\tilde{\theta}_{ \bm{x}}}$ one slice at a time, the parameter maps and coil sensitivity maps were downsampled using nearest-neighbor interpolation to $32 \times 32$. The Halton pattern was generated for these downsampled dimensions.

\subsection{In-vivo scan}
The Halton pattern was used to perform a healthy volunteer scan targetting the brain. Informed consent was obtained after review by our Institutional Review Board. The same sequence settings were used as phantom; however, the ETL was increased to 32 to account for higher $T_2$ values observed in human brain. (Gray matter: 104 - 134 ms  white matter: 70 - 84 ms ) \cite{wansapura_nmr_1999}. The acquisition settings are shown in Table \ref{tab: scan set} resulting in $Q = 128$. In order to get coil sensitivity maps from the scan, one of the contrasts was used to sample only a small patch of k-space from the center. We selected $q = 78$ based on an experiment in which, for each $q$, TEUSQA was evaluated without samples in echo $q$.
%\begin{table}[h]
%	\caption{{Acquisition settings for the prospectively  undersampled in-vivo scan}}
%	\begin{tabular}{|c|c|}
%		\hline
%		\textbf{Acquisition settings} & \textbf{Values} \\ \hline
%		Acquisition matrix & $AP \!\! : \!\! 96, LR \!\! : \!\! 76, SI \!\! : \!\! 128$ \\ 
%		Field of view ($mm$) & 	 $\!\!\!	\!\!\! PE_1 \!\! : \!\! 230, PE_2 \!\! : \!\! 182.4, FE \!\! : \!\! 307 	\!\!\! 	\!\!\!$ \\
%		Number of coils (${C}$)	& 8 \\ 
%		Acceleration factor ($R$)	& 32 \\ 
%		Scan time & 	 $\!\!\!(96\! \times\! 76 \! \times \sum{\!TR} ) /R \!\approx$  43  min. $\!\!\!$ \\ 
%		 \hline
%	\end{tabular}
%	\label{table: tabs4}
%\end{table}

\section{Results}
\label{sec:results}
\subsection{Verification of   TEUSQA  with numerical  simulations}
\begin{figure*}
	\begin{subfigure}{\textwidth}
		\centering
		\includegraphics[width = 0.5\linewidth]{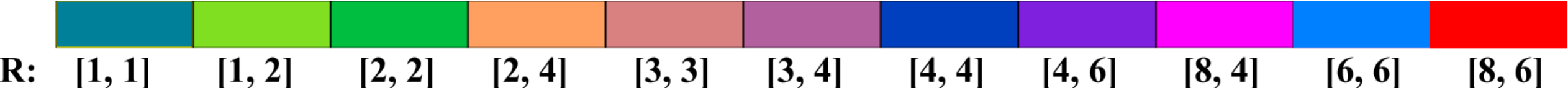}
		\label{Fig: expt_1_legend_t1a}
	\end{subfigure}
	\begin{subfigure}{\textwidth}
		\centering
		\includegraphics[width = \linewidth]{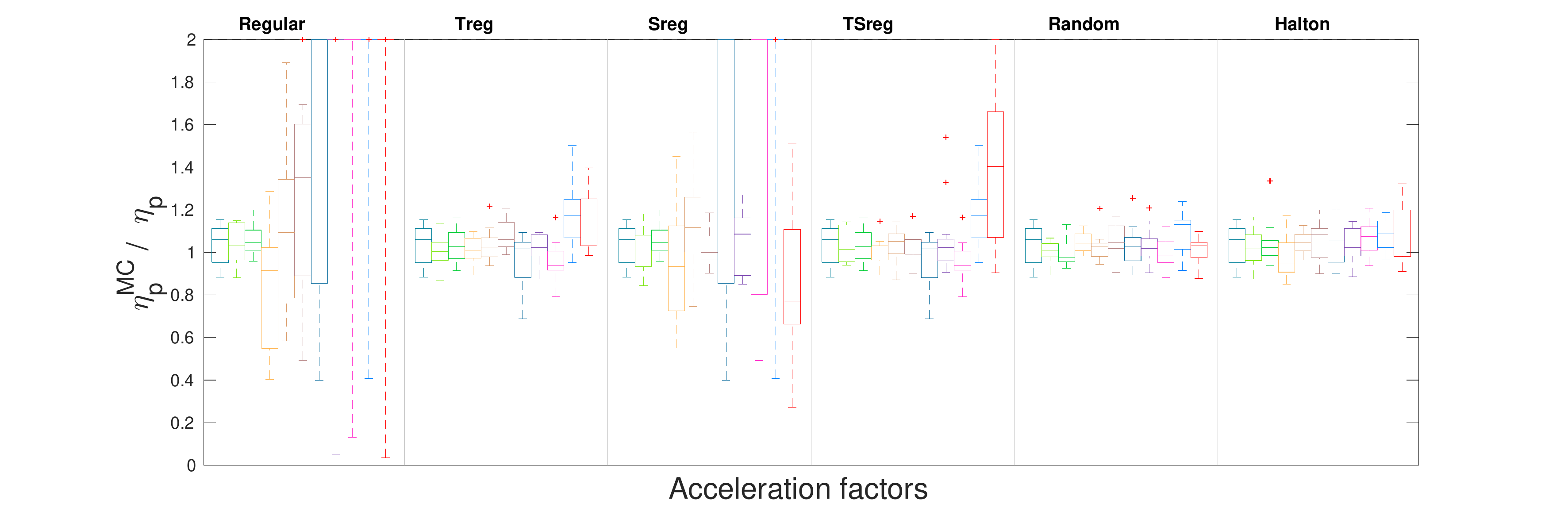}
%		\subcaption{ 	\centering  Result for $T_1$}
%		\label{Fig: expt_1_t1s}
	\end{subfigure}
	\caption{ Results from Monte Carlo simulation for evaluation of TEUSQA for  $T_1$. The  box plots are grouped together according to the undersampling patterns shown by the vertical line separating the figure. The  box plots are colored according to the respective acceleration factor shown in the legend. Each  box plot represents the distribution of $\eta^{MC}_{p}/\eta_{p}$ over voxels within the ROI of the phantom for a particular acceleration factor. The ratio is shown for range $[0, 2]$.} 
	\label{Fig: expt_1_t1}
\end{figure*}

\begin{figure*}
	\begin{subfigure}{\textwidth}
		\centering
		\includegraphics[width = 0.5\linewidth]{legend_expt1_new}
		\label{Fig: expt_1_legend_t1a}
	\end{subfigure}
	\begin{subfigure}{\textwidth}
		\centering
		\includegraphics[width = \linewidth]{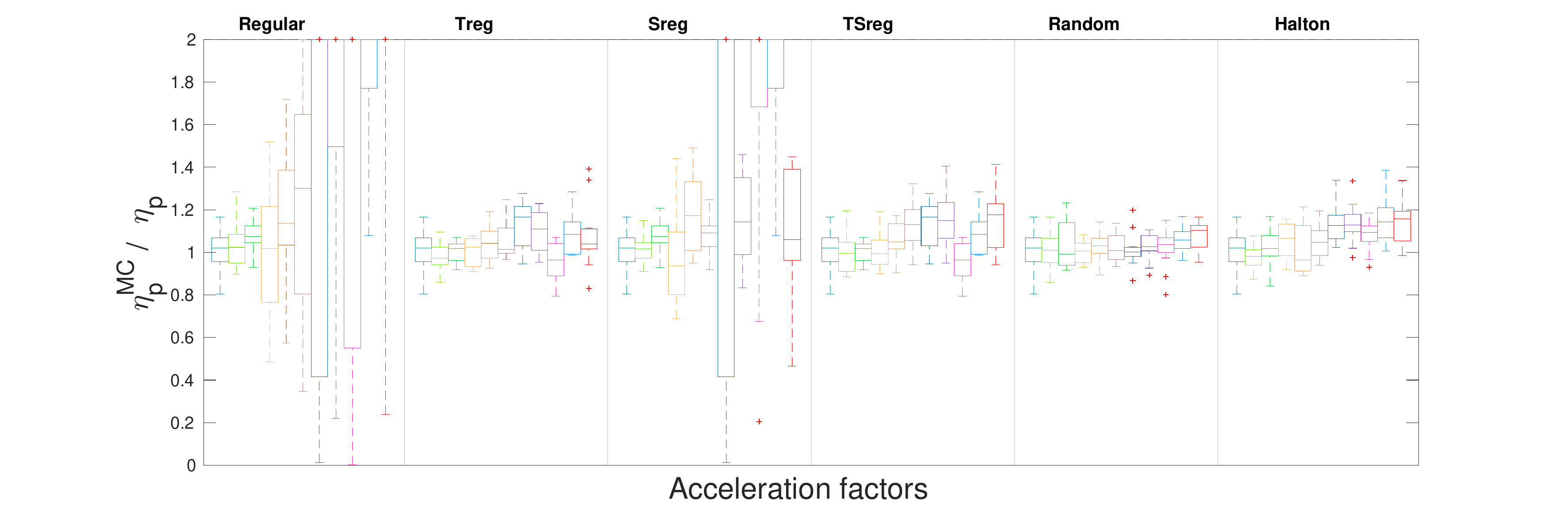}
%		\subcaption{	\centering Result for $T_2$}
%		\label{Fig: expt_1_t2s}
	\end{subfigure}
	\caption{ Results from Monte Carlo simulation for evaluation of TEUSQA for $T_2$. The  box plots are grouped together according to the undersampling patterns shown by the vertical line separating the figure. The  box plots are colored according to the respective acceleration factor shown in the legend. Each  box plot represents the distribution of $\eta^{MC}_{p}/\eta_{p}$ over voxels within the ROI of the phantom for a particular acceleration factor. The ratio is shown for range $[0, 2]$.} 
	\label{Fig: expt_1_t2}
\end{figure*}

% \ref{Fig: expt_1_t1} and \ref{Fig: expt_1_t2}
Figures \ref{Fig: expt_1_t1} and \ref{Fig: expt_1_t2} show the  box plots of ratio between $\eta^{MC}_{p}$ and $\eta_{p}$ for $T_1$ and $T_2$ respectively  for all undersampling patterns and acceleration factors.  Each box represents the $25^{th}$ and $75^{th}$ percentile of the distribution of all the voxels for a particular acceleration factor and undersampling pattern. 
Observe that for both $T_1$ and $T_2$, with Treg, TSreg, Random and Halton  undersampling patterns, the boxes lie within 0.85 and 1.15 with the exception of two cases where $\bm{R} = [8, 6]$. This was not the case for Regular and Sreg particularly for high acceleration factors $\bm{R} \geq [2, 4]$. Note that due to the finite number of Monte Carlo iterations the box size is expected to be $0.9$ to $1.09$.

The computation time of $\eta_p$ for the fully sampled scan was 1.4 minutes for one slice and always lower for undersampled scans on a workstation with a 3.8 GHz Intel Core i7-1065G7 Processor, 16 GB RAM, Windows 10  and Matlab R2020b.

\subsection{Comparison of undersampling patterns}

\begin{figure}
	\begin{subfigure}{0.4\textwidth}
		\includegraphics[width = \linewidth]{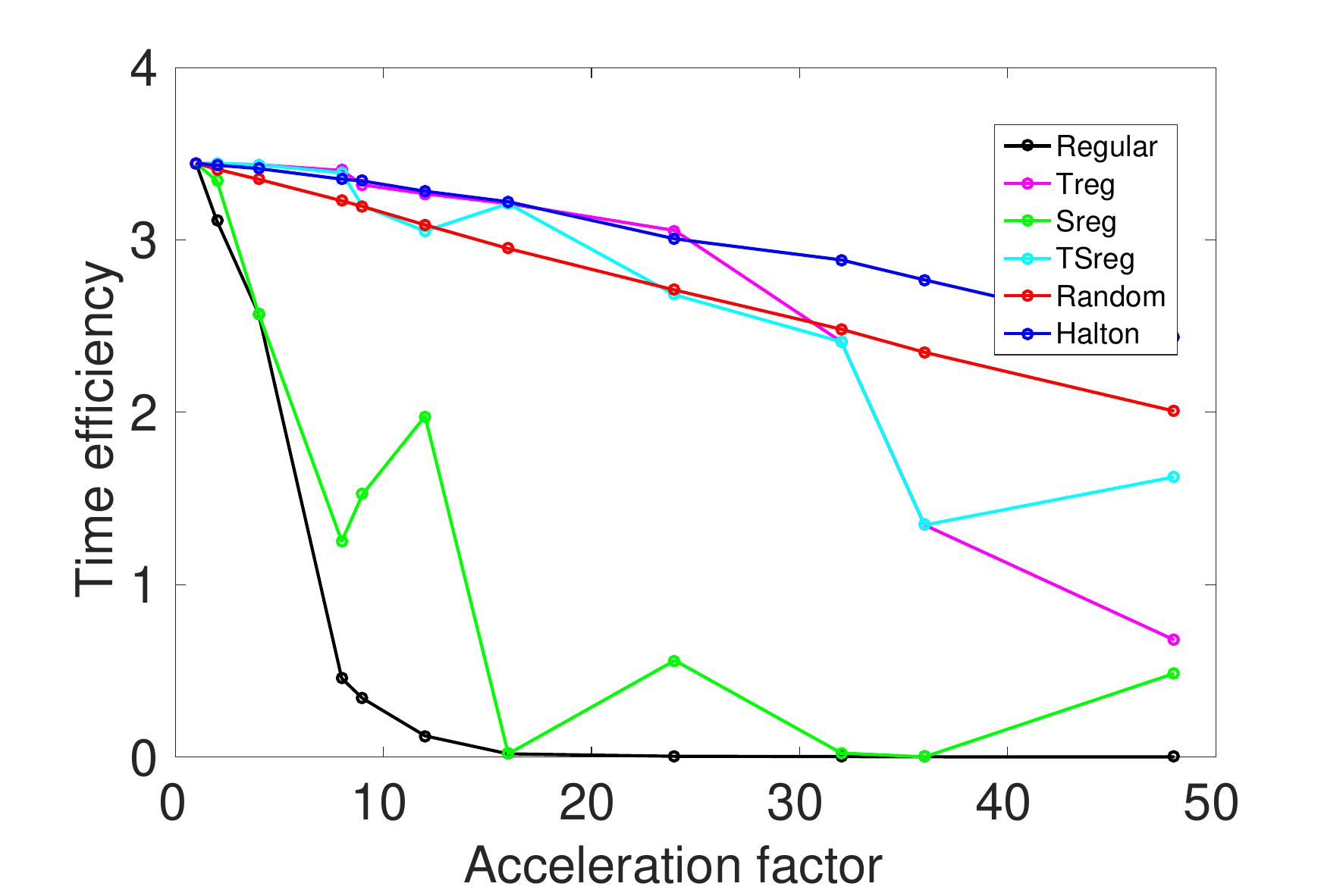}
		\subcaption{\centering	$\eta_{p}$ computed for $T_1$}
		\label{subfig: Discrepancy_real}
	\end{subfigure}
	\begin{subfigure}{0.4\textwidth}
		\includegraphics[width = \linewidth]{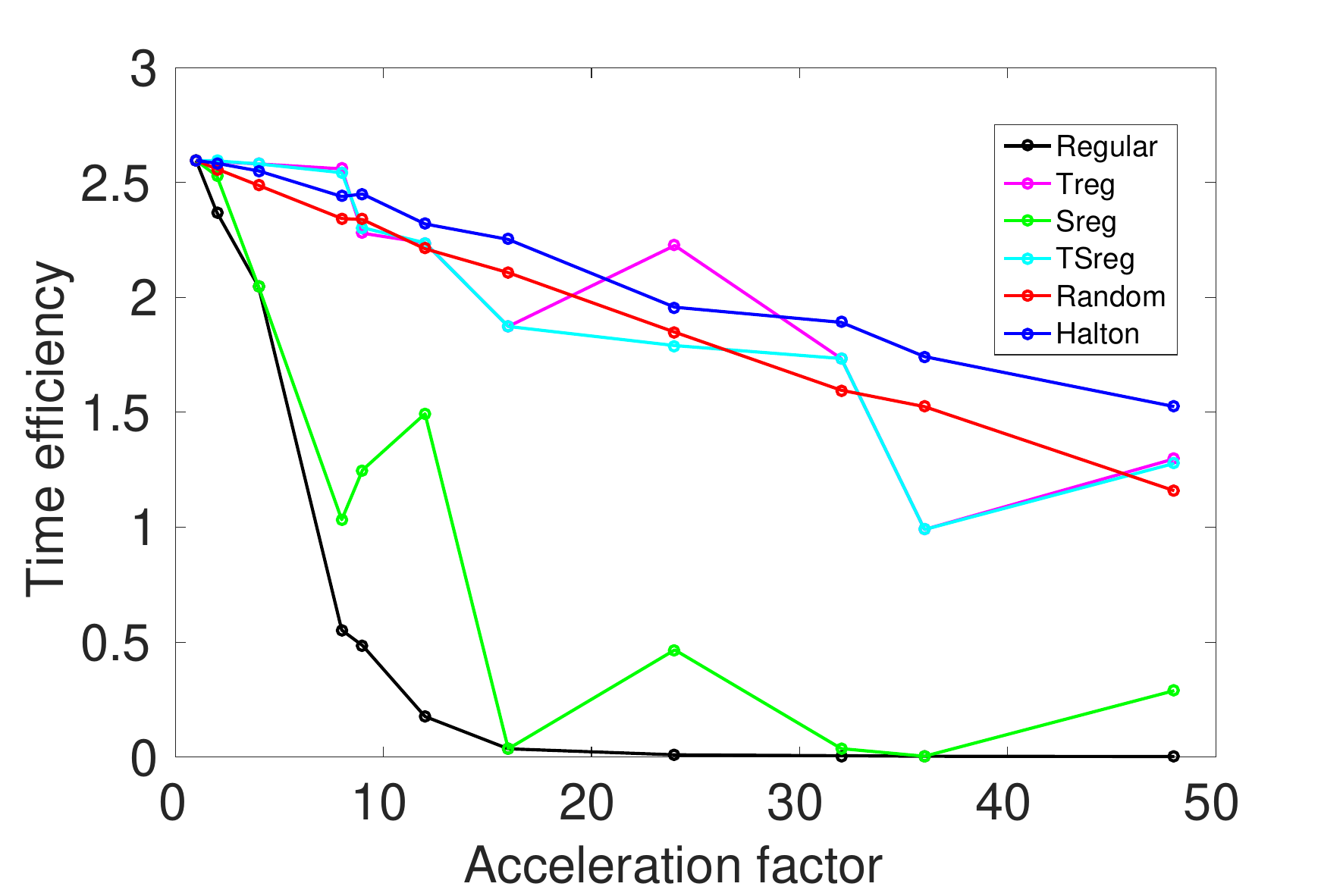}
		\subcaption{\centering $\eta_{p}$ computed for $T_2$}
		\label{subfig: SNK_real}
	\end{subfigure}
	\centering
	\begin{subfigure}{0.4\textwidth}
		\includegraphics[width = \linewidth]{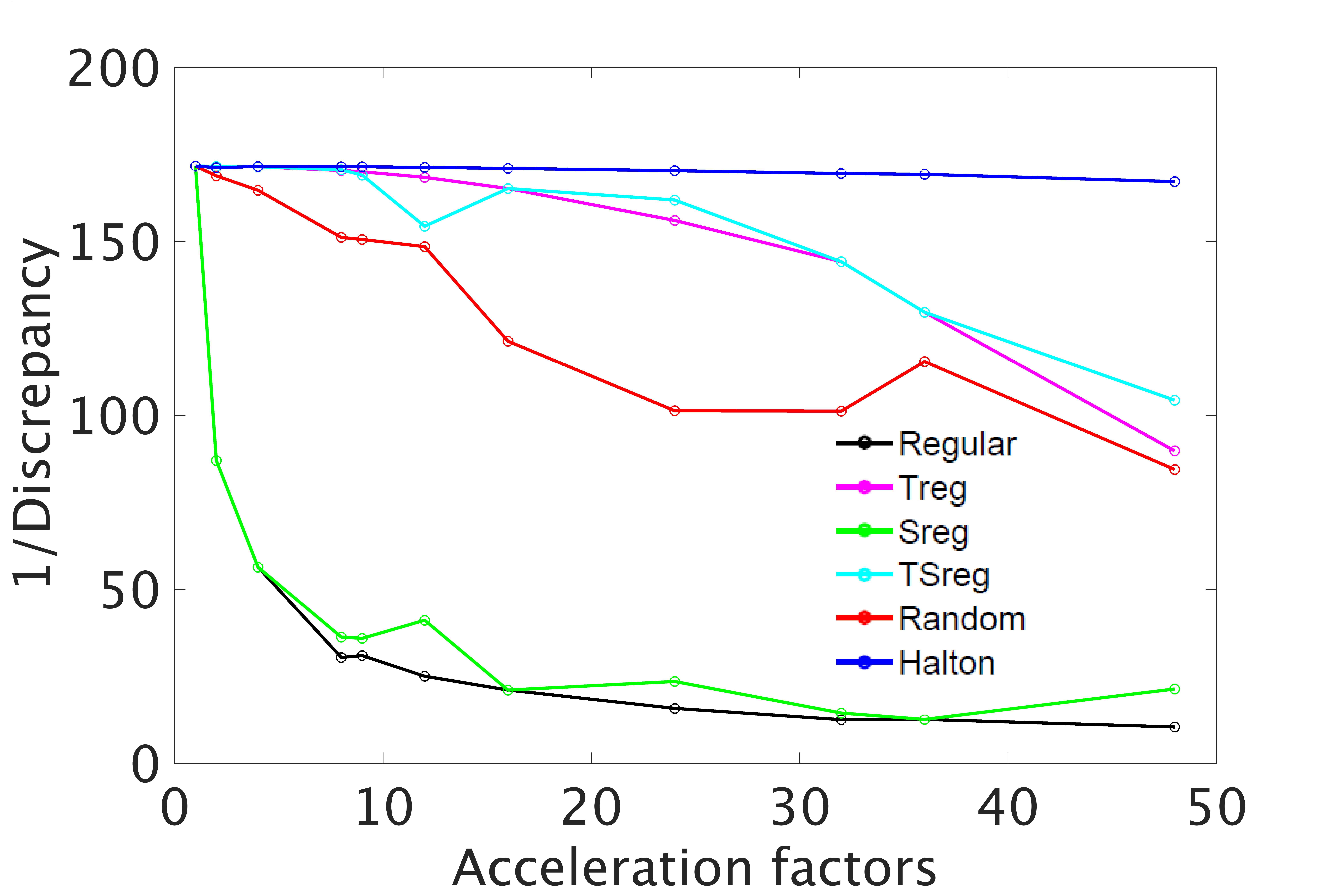}
		\subcaption{\centering Discrepancy}
		%\label{subfig: Discrepancy_small}
	\end{subfigure}
	\caption{{ (a and b) $\eta_{p}$ computed for $T_1$ and $T_2$ for all evaluated  undersampling patterns and acceleration factors.  (c) Discrepancy of downsampled undersampling patterns used for time efficiency computation.}}
	\label{Fig: Time_eff}
\end{figure}

Figure \ref{Fig: Time_eff} shows the time efficiency computed for all the considered undersampling patterns as a function of acceleration factors and reciprocal of Discrepancy. All pattern generation techniques showed a decline in time efficiency with increasing acceleration factor. Some patterns, such as Regular and Sreg, showed a steeper decline than others. Note that these undersampling patterns also showed a difference between $\eta^{MC}_{p}$ and $\eta_{p}$ in the Monte Carlo simulation. The Halton  pattern showed the lowest decline or the highest time efficiency with few exceptions.  

Comparing the Discrepancy with time efficiency, the patterns with high Discrepancy showed lower time efficiency. Regular and Sreg were two patterns that showed the highest Discrepancy, also showed the lowest time efficiency. The other four patterns have a similar level of Discrepancy and showed a similar level of time efficiency. From these, the Halton pattern had the lowest Discrepancy and the highest time efficiency. 

Given these results, the Halton pattern with acceleration factor 32 was chosen for the prospective acquisition, obtaining the best maps within 60 minutes of scan time. 

\subsection{Verification of TEUSQA with prospective acquisition}
\begin{figure}[h]
	\centering
	\includegraphics[width = \linewidth]{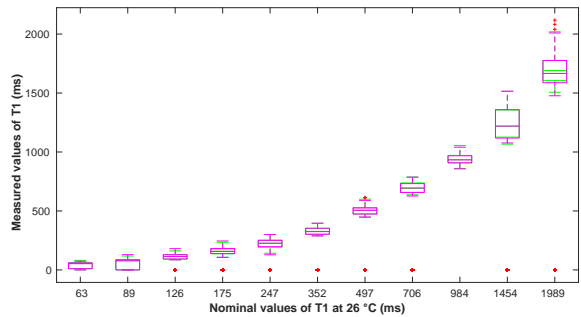}
	\caption{ $T_1$ measured in the selected spheres plotted against their nominal values. Green boxes indicate test acquisition and magenta boxes indicate retest.}
	\label{Fig: pros1}
\end{figure}

\begin{figure}
	\centering
	\includegraphics[width = \linewidth]{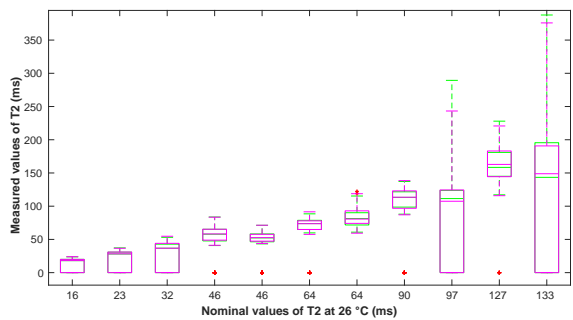}
	\caption{$T_2$ measured in the selected spheres plotted against their nominal values. Green boxes indicate test acquisition and magenta boxes indicate retest.}
	\label{Fig: pros2}
\end{figure}

\begin{figure}
	\centering
	\begin{subfigure}{0.22\textwidth}
		\includegraphics[width=\linewidth]{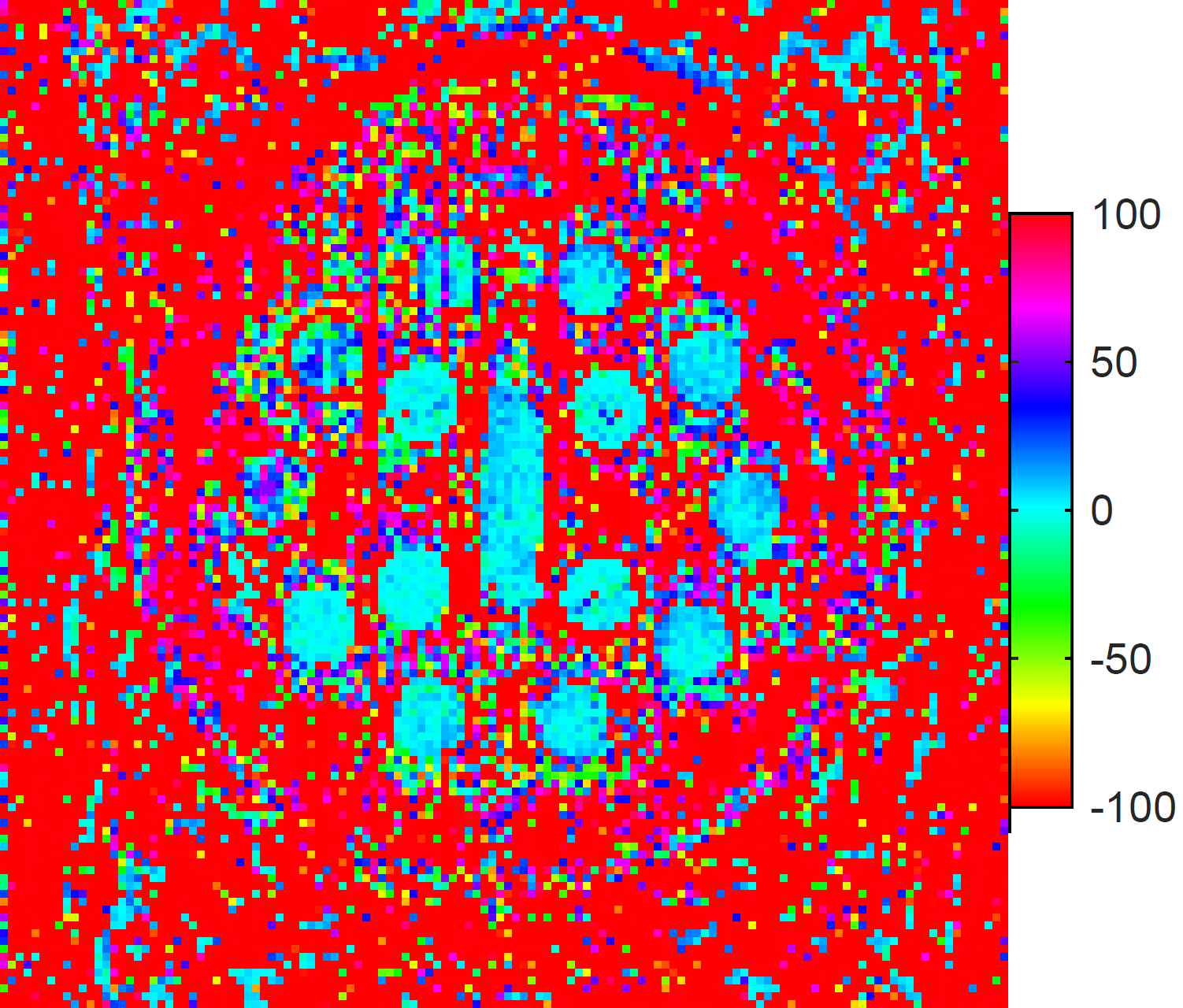}
		\caption{\centering {$T_1$ difference map from $T_1$ array }}
	\end{subfigure}
	%	\begin{subfigure}{0.03\textwidth}
	%		\includegraphics[scale=0.25]{t1_diff_legend}
	%	\end{subfigure}
	\begin{subfigure}{0.22\textwidth}
		\includegraphics[width=\linewidth]{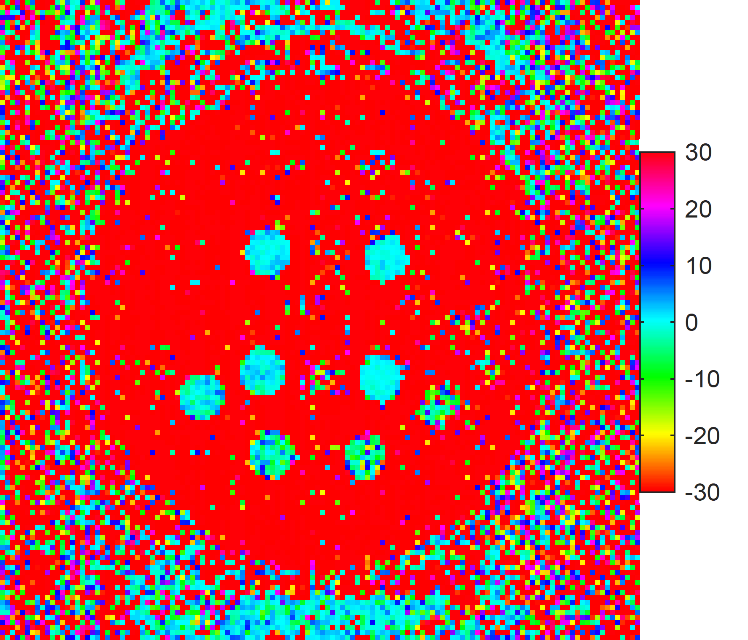}
		\caption{\centering {$T_2$ difference map from $T_1$ array}}
		\label{Fig: T1_pros2}
	\end{subfigure}
	%	\begin{subfigure}{0.03\textwidth}
	%		\includegraphics[scale=0.12]{T2_map_difflegend.eps}
	%	\end{subfigure}
	\begin{subfigure}{0.22\textwidth}
		\includegraphics[width=\linewidth]{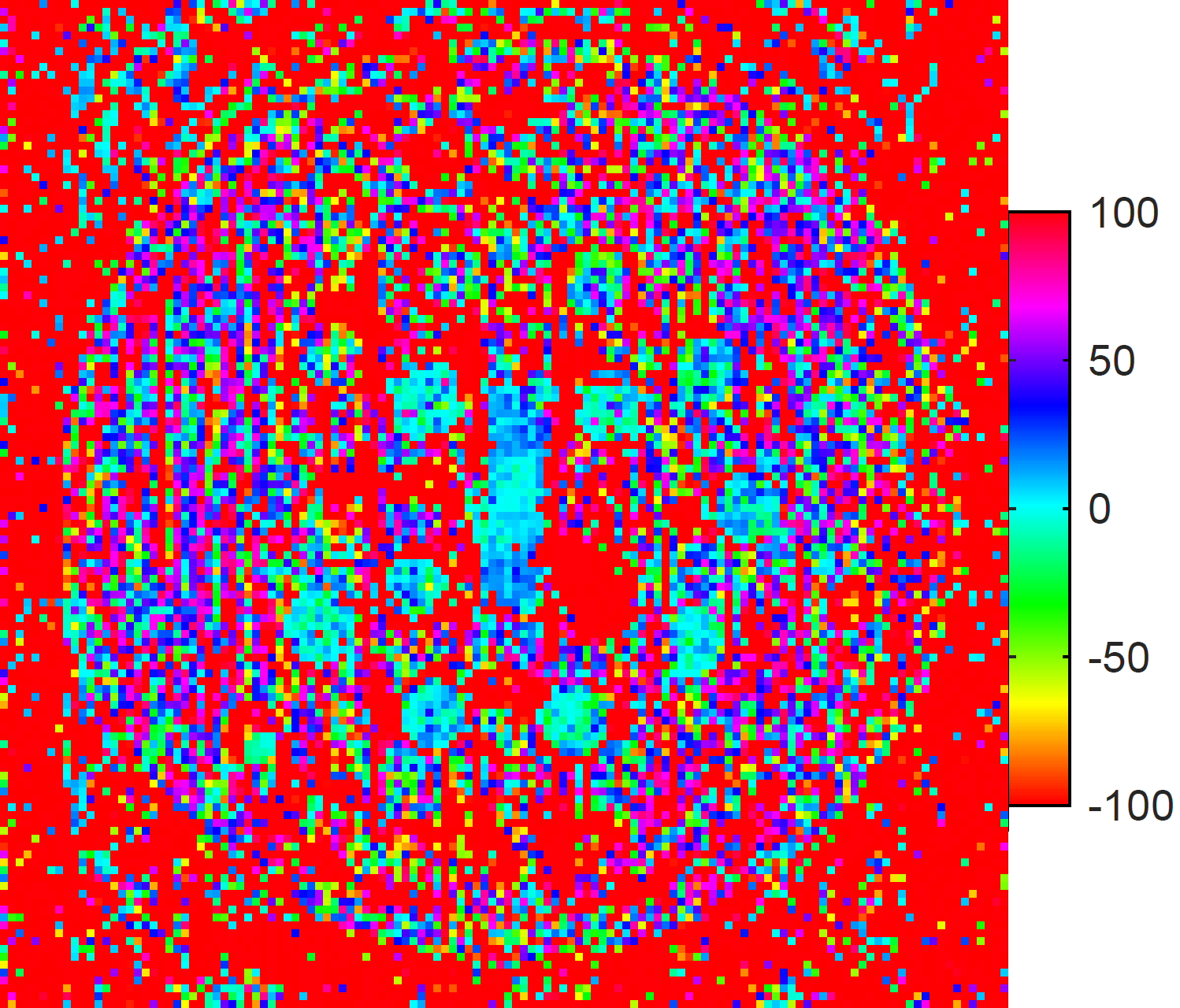}
		\caption{\centering {$T_1$ difference map  from $T_2$ array}}
	\end{subfigure}
	%	\begin{subfigure}{0.03\textwidth}
	%		\includegraphics[scale=0.15]{T1_map_difflegend.eps}
	%	\end{subfigure}
	\begin{subfigure}{0.22\textwidth}
		\includegraphics[width=\linewidth]{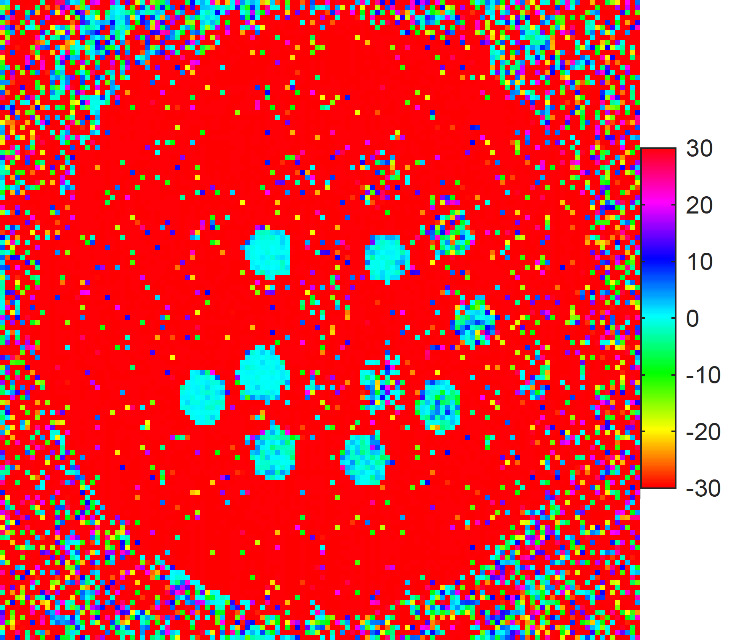}
		\caption{\centering {$T_2$ difference map from $T_2$ array}}
		\label{Fig: T1_pros2}
	\end{subfigure}
	%	\begin{subfigure}{0.03\textwidth}
	%		\includegraphics[scale=0.12]{T2_map_difflegend.eps}
	%	\end{subfigure}
	
	\caption{Test-retest difference maps for $T_1$ (left) and $T_2$ (right) estimates in the $T_1$ (top) and $T_2$ array (bottom) of the ISMRM model 130 phantom \cite{jiang_repeatability_2017} (units in ms). Scans were made using prospective undersampling using the Halton pattern with acceleration factor $R=32$.}
	\label{Fig: pros_result}
	
\end{figure}

%\ref{Fig: pros_result}
Figure \ref{Fig: pros_result} shows the difference in  estimated $T_1$ and $T_2$ maps for the test and retest acquisition. 
We select the spheres where the nominal $T_1$ and $T_2$ values are expected to be mapped correctly by the sequence settings. In these spheres the mean difference between maps from test and retest is close to zero. Outside of these spheres large differences can be observed.  
The bias in the $T_1$ and $T_2$ estimates compared to the nominal values was on average about $5 \%$ and $28\%$ respectively in the selected spheres. A detailed comparison with the nominal values is presented in the figures 	\ref{Fig: pros1} and \ref{Fig: pros2}.
Following this we computed  $\eta^{ACQ}_{p}$ over the selected spheres which was found to be $0.221$ with $95\%$ confidence bounds $[ 0.201, 0.241 ]$ for $T_1$ and $0.122$  with $[ 0.111, 0.134 ]$ $95\%$ confidence bounds for $T_2$. The predicted  $\eta_{p}$  for $T_1$ and $T_2$ were $0.254$ and $0.125$, respectively. The predicted $\eta_{p}$ for $T_1$ was within $12\%$ of the observed $\eta^{ACQ}_{p}$ while for $T_2$ it was within the $95\%$ confidence bounds of observed $\eta^{ACQ}_{p}$.

\begin{table*}[]
	\centering
		\caption{ The median values (ms) of $T_1$ and $T_2$ from test and re-test scans compared to nominal values of selected spheres.  }
	\label{Fig: results_tab}
	\begin{tabular}{|cccccc|}
		\hline
		Nominal value $T_1$ & Median $T_1$ test & Median $T_1$ re-test & Nominal value $T_2$ & Median $T_2$ test & Median $T_2$ re-test \\ \hline
		62.7 & 57.5 & 57 & 15.8 & 19.6 & 19.5 \\ 
		89 & 79.8 & 80.5 & 23 & 29.1 & 28.9 \\ 
		125.9 & 120.5 & 123.4 & 32 & 40 & 41.5 \\ 
		244.2 & 282.5 & 272.7 & 45.7 & 54.2 & 54.4 \\ 
		336.5 & 344.5 & 347.7 & 46.4 & 59.4 & 60.9 \\ 
		458.4 & 468.2 & 468.1 & 64 & 81.7 & 81 \\ 
		608.6 & 590.1 & 591.2 & 64.3 & 74.9 & 74.6 \\ 
		801.7 & 764.7 & 753.1 & 90.3 & 117.7 & 118.9 \\ 
		1044 & 923.3 & 934.4 & 96.9 & 113.8 & 113.2 \\ \hline
	\end{tabular}
\end{table*}

\subsection{In-vivo scan}
\begin{figure*}
	\centering
	\includegraphics[width=\linewidth]{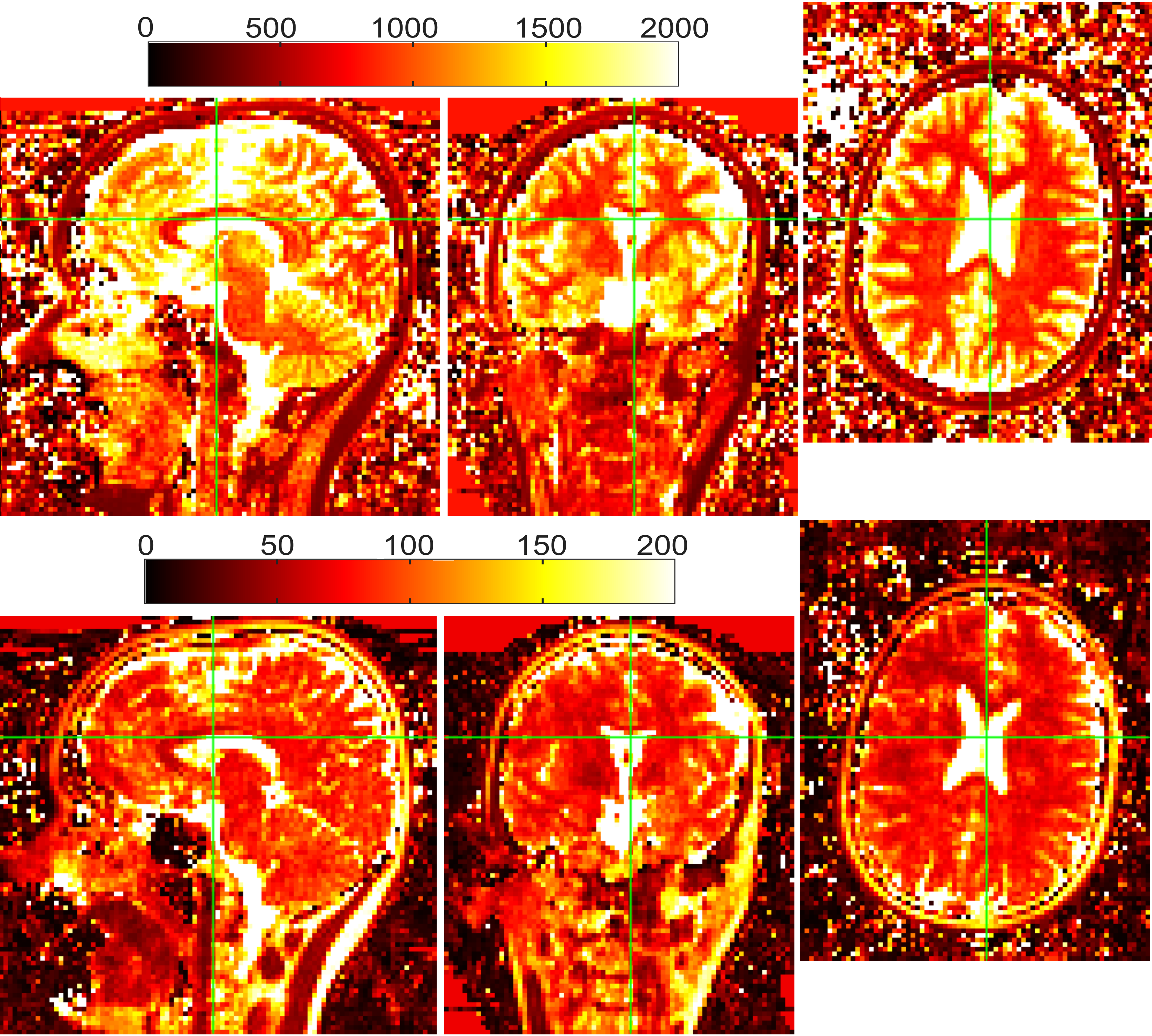}
\caption{Aagittal (left), coronal (center) and axial (right) view of $T_1$ map (Top) and $T_2$ map (Bottom) (ms) of a healthy volunteer obtained with an 32-fold accelerated 3D IP-FSE scan. The green line shows the slice selected in each of the three dimensions.The SI (FE) dimension is cropped to focus on the Subject's head.}
\label{Fig: invivo}
\end{figure*}

Figure\ref{Fig: invivo}  shows the $T_1$ and $T_2$ map of the in-vivo acquisition. The reconstruction was performed for each FE line along the SI orientation. All the FE lines were reconstructed successfully. There are no visible artefacts related to undersampling and $T_1$ and $T_2$ maps are in the range expected in the human brain.
%\begin{figure}
%	\centering
%		\begin{subfigure}[t]{ 0.2\textwidth}
%		\includegraphics[width=\linewidth]{axial_t112}
%		\caption{\centering { Axial view of $T_1$ map}}
%	\end{subfigure}
%	\begin{subfigure}[t]{ 0.2\textwidth}
%	\includegraphics[width=\linewidth]{saggital_t131}
%	\caption{\centering { Sagittal view of $T_1$ map}}
%\end{subfigure}
%	\begin{subfigure}[t]{ 0.2\textwidth}
%	\includegraphics[width=\linewidth]{back_head_t132}
%	\caption{\centering { Coronal view of $T_1$ map}}
%\end{subfigure}
%		\begin{subfigure}[t]{0.03\textwidth}
%		\includegraphics[scale=0.8]{axial_t1legend}
%	\end{subfigure}
%\caption{$T_1$ map of a healthy volunteer obtained in an accelerated 3D IP-FSE scan.}
%\label{Fig: T1_invivo}
%\end{figure}
%
%
%\begin{figure}
%	\centering
%	\begin{subfigure}[t]{ 0.2\textwidth}
%		\includegraphics[width=\linewidth]{axial12}
%		\caption{\centering { Axial view of $T_2$ map}}
%	\end{subfigure}
%	\begin{subfigure}[t]{ 0.2\textwidth}
%		\includegraphics[width=\linewidth]{saggital31}
%		\caption{\centering { Sagittal view of $T_2$ map}}
%	\end{subfigure}
%	\begin{subfigure}[t]{ 0.2\textwidth}
%		\includegraphics[width=\linewidth]{back_head32}
%		\caption{\centering { Coronal view of $T_2$ map}}
%	\end{subfigure}
%	\begin{subfigure}[t]{0.03\textwidth}
%		\includegraphics[scale=0.8]{axiallegend}
%	\end{subfigure}
%\caption{$T_1$ map of a healthy volunteer obtained in an accelerated 3D IP-FSE scan.}
%\label{Fig: T2_invivo}
%\end{figure}

\section{Discussion}
We derived TEUSQA as a function of the QMRI sequence settings, undersampling pattern, and  scan time. We verified the applicability of TEUSQA with Monte Carlo simulations and actual acquisitions. 

The results show that predictions from TEUSQA match with those observed in QMRI experiments. There was a difference in $\eta_{p}$ observed in the Monte Carlo simulations and acquisition for some undersampling patterns with low time efficiency. For such low-time efficiency patterns, the model's nonlinearity becomes relevant, and the estimator in the Monte Carlo simulation produces biased results. Hence the mismatch for these patterns is not unexpected. For patterns with high time efficiency, TEUSQA predicts $\eta^{MC}_{p}$ to within 15\%. Out of the mismatch,  $10\%$ can be attributed to standard error due to limited number of realisations in Monte Carlo simulation. The $\eta_{p}$ computed by TEUSQA from the downsampled map matched to those observed in actual acquisitions, $\eta^{ACQ}_{p}$, to within $12 \%$.  

We repeated the experiments for verifying TEUSQA with simulation and prospective acquisition for 3D GRASE sequence  (for $T_2$ and $B0$ mapping), which is summarized in \ref{A: 3D GRASE}.
Similar results were obtained in both cases, demonstrating the generic applicability of TEUSQA. 

The undersampling pattern influences the effect of the prior information on the estimates. Currently, TEUSQA is formulated with general multivariate Gaussian priors but evaluated with an independent prior per parameter. As this prior contains no spatial dependencies, the uniform undersampling provided by the evaluated schemes seems optimal. When TEUSQA could be extended to have sparsity constraints as prior  \cite{haldar_oedipus_2019}, variable density patterns might be more favorable. 

The comparison among the undersampling patterns for 3D IP-FSE and 3D GRASE shows that patterns with low-discrepancy have higher time efficiency. This was shown to be true also for sequences with less contrasts (32) which has been described in \ref{A: FSE}. Discrepancy quantifies the uniformity of the sampling pattern in the spatial dimensions as well as along contrast. So, uniformity of the pattern seems to be a desirable property. The metric proposed in \cite{levine_--fly_2018} can also take spatial as well as contrast dimensions into account. In the dynamic imaging experiment, where the temporal dimension was part of the pattern, the resulting pattern produced better results than pseudo-random patterns such as poison-disk and uniform patterns \cite{levine_--fly_2018}. The upper bound for the time efficiency at any acceleration factor is given by its value in a fully sampled scan. For the Halton patterns, these were within 25\% of this bound up to an acceleration factor of 32.

TEUSQA has been derived for 3D acquisitions, but it is also applicable to 2D acquisitions, assuming a 3D volume in which one of the dimensions has size 1. However, different k-space sampling patterns should be designed in that case. Furthermore, TEUSQA, although derived as general framework,  is limited to Cartesian acquisitions in the current work. For non-cartesian (spiral, radial) acquisitions, the frequency encoding dimension needs to be considered in the computation.

The prospectively undersampled phantom maps showed good repeatability, corresponding to the prediction of TEUSQA, within the range of $T_1$ and $T_2$ where the sequence was expected to map the values correctly. Large differences between the estimates were observed outside the selected spheres where the sequence is not accurate. Therefore, those regions were excluded from the evaluation. However, these regions with large differences did not prevent the estimation in the selected spheres to be accurate. 

We demonstrated $T_1$ and $T_2$ mapping on a healthy volunteer and a test-retest experiment using a high acceleration factor. Nevertheless, our primary focus was on the verification of TEUSQA. Further optimisation of sequence setting and targeting particular parameters, for example, $T_2$, could enable a shorter $TR$ and consequently scan time.

%and model-based reconstruction procedure with in-vivo scan and a test-retest experiment. This acceleration factor can be compared to the ETL in standard weighted FSE acquisitions, where all the echoes are combined into a single k-space, achieving an acceleration factor equal to the ETL. The artifacts due to the decay along the echo train expected in standard FSE scans are absent in our method as the differences among echoes are contained in the model and used to map $T_2$. Although we performed an accelerated scan,

In this work, we focused on finding good undersampling patterns using TEUSQA; however, TEUSQA can potentially be beneficial also to find optimal scan settings such as TE, TR, and flip angles similar to previously proposed metrics \cite{leitao_efficiency_2021,deoni_rapid_2003,crawley_comparison_1988,asslander_optimized_2019,zhao_optimal_2019,poot_optimal_2010,jones_optimal_1996,brihuega-moreno_optimization_2003}. Compared to these work the main benefit is the inclusion of the undersampling pattern  and hence that was the main focus of our investigation.

TEUSQA, in its present form, is only applicable for reconstruction techniques that directly estimate tissue parameters from undersampled
k-space measurements. Most of the approaches for time-resolved images for cardiac or perfusion acquisitions use a two step approach where contrast images are reconstructed followed by estimation of parameters
\cite{ahmad_variable_2015}. % if that reference is appropriate here. 
As such TEUSQA is not directly applicable to such dynamic acquisitions. However, given a forward model that relates perfusion or cardiac
parameters directly to under sampled kspace, TEUSQA would be applicable.

%The joint reconstruction of contrast images has been performed to reconstruct time-resolved images for cardiac or perfusion scans \cite{ahmad_variable_2015}. However, most of these works use two step approach where contrast images are reconstructed followed by estimation of parameters. TEUSQA, in its present form, is only applicable for reconstruction techniques that directly estimate tissue parameters from undersampled k-space measurements. Given a reconstruction technique that uses a forward model that can relate perfusion or cardiac parameters, such as blood flow directly from under sampled k-space, then TEUSQA would be applicable in such cases. 

\section{Conclusion}
\label{sec:conclusion}
The proposed metric takes into account essential aspects needed to accelerate Q-MRI scans, such as parallel imaging, sequence settings, and the k-space undersampling pattern. The metric can be used to inform sequence design and sample pattern optimisation in quantitative MRI studies, assuming a reconstruction technique is used that directly estimates tissue parameters from undersampled k-space measurements. We used the metric to evaluate undersampling patterns for multi-contrast QMRI acquisitions \textit{in silico}. With our metric we showed that low-discrepancy is a desirable design property when searching for a time efficient  undersampling pattern. Overall the patterns  produced with Halton sampling showed the best time efficiency. The accelerated acquisitions using 3D IP-FSE and 3D GRASE were reconstructed successfully and showed a time efficiency close to the value predicted with TEUSQA. In-vivo scan of a healthy volunteer with acceleration factor of 32 using 8-channel coil and $128$ contrasts produced artefact free $T_1$ and $T_2$ maps. As such TEUSQA could be useful for decreasing  scan time of other multi-contrast QMRI acquisitions in the future. 

\section*{Acknowledgments}
This work is part of the project B-Q MINDED which has received funding from the European Union’s Horizon 2020 research and innovation programme under the Marie Sklodowska-Curie grant agreement No 76451.

\appendix
%\section*{Appendix}

\section{Derivation of posterior distribution} 
\label{section: S1}
In this section we derive posterior distribution for the parameters with the addition of a prior distribution.  Using Bayes theorem, the posterior distribution is given by: 
\begin{equation}
	\label{step1}
	p( \bm{\theta} | {\bm{Z} ,\sigma)}  \propto \!p({\bm{Z} | \bm{\theta}  ,\sigma)}   .  p( \bm{ \bm{\theta}} | \overline{ \bm{\theta}}, \bm{\Gamma})\\ \! = \!\frac{e^{-\frac{1}{2 \sigma^2}||\bm{Z} - \m||^2}}{(2\pi{\sigma}^2)^{N}}     .     \frac{e^{-\frac{1}{2}( \bm{\theta} - \bar{ \bm{\theta}})^{T}\bm{\Gamma}^{-1} ( \bm{\theta} - \bar{ \bm{\theta}})}}{\sqrt{\left(2\pi\right)^{L} \det \bm{\Gamma}}}  
\end{equation}
Taking, only the arguments of the first exponential in Equation \ref{step1},
\begin{equation*}
	-\tfrac{1}{2 \sigma^2}{||\bm{Z} - \m||^2} = -\tfrac{1}{2 \sigma^2} [\z^{T}\z - \z^{T}\m \\ -\m^{T}\z + \m^{T}\m ]
\end{equation*}

Substituting the first order approximation of Taylor series $\bm{\mu}( \bm{\theta}) \approx \bm{\mu}(\ttil) + \jac(\ttil)  \left\{ \bm{\theta} - \ttil\right\}$ where $\ttil$ is the ground truth. In the rest of the section we denote  $\jac(\ttil)$  as $\jac$ and $\bm{\mu}(\ttil)$ as $\bm{\mu}$  for notational convenience: \\
$
%	\label{I}
	=  -\tfrac{1}{2 \sigma^2}   \left[ K +  \left(  - \z^{T}   \jac + \mt^{T} \jac -  \ttil^{T} \jac^{T} \jac  \right)\bm{\theta}  +   \bm{\theta}^{T} \jac^{T} \jac \bm{\theta} \right.\\  \left.   +  \bm{\theta}^{T}   \left( -  \jac^{T}\z +  \jac^{T}   \mt    -  \jac^{T} \jac \ttil   \right)    \right]
$
\\ where 
$  K   =   \z^{T}\z   -    \z^{T}\mt   +  \z^{T} \jac \ttil   -   \mt^{T}\z +  \ttil^{T} \jac^{T}\z  + \mt^{T}\mt    -   \mt^{T}\jac \ttil -  \ttil^{T} \jac^{T}  \mt  + \ttil^{T} \jac^{T}\jac \ttil $  contains all elements not dependent on $\bm{\theta}$.\\
Similarly taking the second exponential term from Equation \ref{step1} and  since  $  \bm{\theta}^{T} \bm{\Gamma}^{-1} \overline{ \bm{\theta}}$  =   $\overline{ \bm{\theta}}^{T}\bm{\Gamma}^{-1} \bm{\theta}  $,
\begin{equation}
	\label{II}
	\tfrac{1}{2}	(  \bm{\theta} - \overline{ \bm{\theta}})^{T}\bm{\Gamma}^{-1} (  \bm{\theta} - \overline{ \bm{\theta}})   = \tfrac{1}{2} \left( \bm{\theta}^{T} \bm{\Gamma}^{-1}  \bm{\theta} - 2\overline{ \bm{\theta}}^{T}\bm{\Gamma}^{-1} \bm{\theta}   +  \overline{ \bm{\theta}}^{T}\bm{\Gamma}^{-1}\overline{ \bm{\theta}}\right).
\end{equation}
Combining the two exponential terms of Equation \ref{step1} gives
\begin{equation}
	\label{form}
	A + B \bm{\theta} +  \bm{\theta}^{T} C  \bm{\theta},    
\end{equation} 
\begin{align*}
	\text{ where } A =& -\tfrac{1}{2 \sigma^2} K + -\tfrac{1}{2} \bar{ \bm{\theta}}^{-T}\bm{\Gamma}^{-1}\bar{ \bm{\theta}},\\ 
	B =& \tfrac{1}{ \sigma^2}  \left(   \z^{T}   \jac - \mt^{T} \jac +  \ttil^{T} \jac^{T} \jac  \right) + \overline{ \bm{\theta}}^{T}\bm{\Gamma}^{-1} \\ \text{ and }C =& -\tfrac{1}{2} \left( \tfrac{1}{\sigma^2}\jac^{T} \jac +\bm{\Gamma}^{-1}  \right).
\end{align*}

The derivative of Equation \ref{form} with respect to  $ \bm{\theta} $ can be equated to $0$ to  find its maximum value. Since our prior is conjugate for the likelihood function, the posterior distribution should also be normally distributed. Therefore, the location of the maximum is the mean $\breve{ \bm{\theta}}$ of the posterior distribution. That is,

\begin{align}
	\label{min}
	\breve{ \bm{\theta}} =& -\frac{1}{2} C^{-1} B^{T}\\
	=&    \left[   \tfrac{1}{\sigma^2}{\jac^{T} \jac} + \bm{\Gamma}^{-1} \right]^{-1}     \left[  \tfrac{1}{ \sigma^2}  \left(     \jac^{T}\z  - \jac^{T}\mt  +  \jac^{T}\jac\ttil    \right) + \bm{\Gamma}^{-1}\overline{ \bm{\theta}}\right]
	%\end{split}
\end{align}
Now the Equation of posterior distribution using the mean $\breve{ \bm{\theta}}$ and assuming the covariance matrix of this distribution to be $\breve{\bm{\Gamma}}$ is given by,
\begin{equation}
	\label{postprob2}
	p( \bm{\theta} | \breve{\bm{\theta}},\breve{\bm{\Gamma}}) = \frac{e^{-\frac{1}{2}( \bm{\theta} - \breve{ \bm{\theta}})^{T}\breve{\bm{\Gamma}}^{-1}( \bm{\theta} - \breve{ \bm{\theta}})}}{\sqrt{(2\pi)^{L} \det \breve{\bm{\Gamma}}}}
\end{equation}

Since Equation \ref{postprob2} and \ref{step1} give the same distribution, taking the argument of exponential in the numerator in Equation \ref{postprob2} and then arranging in the form shown in Equation \ref{form}, we get $C= -{\breve{\bm{\Gamma}}} / {2} $. Hence,
\begin{equation}
	\breve{\bm{\Gamma}}^{-1} =   \bm{\Gamma}^{-1} + \bm{I}(\ttil, \sigma)
\end{equation}
with $\bm{I}(\ttil, \sigma) = \tfrac{1}{\sigma^2}{\jac^{T}\jac}$ from  Equation (7) in the main manuscript.

\
\section{Pseudo code for Halton undersampling pattern}
\label{A: algo}
The pseudo code shows the implementation of the Halton undersampling pattern. The code generates a binary undersampling mask $Umask$ where sampled positions are $1$. The Halton sampling function used is based on the implementation of Wang et al.\cite{wang_randomized_2000}.
\begin{algorithm}[h]
	\SetAlgoLined
	$NS \leftarrow Q( k_1 \times k_2)/R_1 R_2 $\\
	$S_1 \leftarrow $\textbf{Halton sampling:} generate $NS$ samples $\in [0,1]$ with base 2 \\
	$S_2 \leftarrow $\textbf{Halton sampling:} generate $NS$ samples $\in [0,1]$ with base  3\\
	
	\For{q = 1 to Q}{
		$SamplesPerContrast \leftarrow( k_1 \times k_2)/R_1 R_2 $\\
		\While { $SamplesPerContrast > 0$}{ 
			$x1 = S1.next$,
			$x2 = S2.next$ \\
			\If{$Umask(\lfloor x1*k_1 \rfloor, \lfloor x2*k_2 \rfloor,q) == 0$}
			{$Umask(\lfloor x1*k_1 \rfloor, \lfloor x2*k_2 \rfloor,q) \leftarrow 1$}
			%			$no \leftarrow no +1$
			SamplePerContrast = SamplePerContrast - 1	}}
	\caption{Halton undersampling pattern, INPUT: $k_1$ (size of $PE_1$), $k_2$(size of $PE_2$), $Q$}
	\label{Algo: halton}
\end{algorithm}
\section{Verification of TEUSQA with 3D GRASE} 
\label{A: 3D GRASE}
	
\subsection{Sequence and estimator details}
\label{sec: sequence used2} 

To evaluate the generalisability of TESUQA, an additional evaluation was performed with a 3D GRASE sequence. 3D GRASE acquisitions can be used in a joint $T_2$ and $\Delta B0$ mapping QMRI protocol, by considering each echo as a different contrast, and fitting the model as described by \cite{jovicich_grase_1998}. To reduce the number of parameters, we assume $\Delta_t$ to be small, such that $T2'$ and $T_2$ decay between gradient and spin echo can be ignored.  Thus, the parameter vector we use for a single voxel is   $\bm{\theta_x} = \left[ \Re(M_0), \Im(M_0), \ln(T_2), \Delta B0 \right]$, where   $\Re(M_0)$, $\Im(M_0)$ are real and imaginary component of the complex valued apparent proton density $M_0$. Similar to the 3D IP-FSE,
the logarithm of $T_2$ was taken.  Due to unavailablity of nominal values of $\Delta B0$ we only present the evaluation of $T_2$.

We used sequence settings in Table \ref{tab: scan set}. The prediction function  $f_q( \bm{\theta}_{ \bm{x}})$ performs EPG simulation  \cite{busse_fast_2006}, additionally, the gradient echoes were adjusted according to the model described in \cite{jovicich_grase_1998}. In post processing the $\ln(T_2)$ maps were converted to  $T_2$ using principles of propagation of uncertainty. This conversion was also applied in the time efficiency analysis.  As prior we used: $\overline{ \bm{\theta}}_1 =[0,0, \ln(70), 0]$ and
$\bm{\Gamma_1} =
\left[ {\begin{array}{cccc}
		20^2 & 0 & 0 & 0\\
		0 & 20^2 & 0 & 0\\
		0 & 0 & \ln(7)^2 & 0\\
		0 & 0 & 0 & (\pi)^2\\
\end{array} } \right]$.

%\begin{table}[h]
%	\centering
%	\caption{Acquisition settings for 3D GRASE.}
%	\begin{tabular}{|c|c|}
%		\hline
%		\textbf{Sequence settings} & \textbf{Values} \\ \hline
%		Delay between Spin echo and Gradient echo ($\Delta_t$) & 2 ms \\ 
%		Repetition time (TR) & \begin{tabular}[c]{@{}c@{}}1800  ms\end{tabular} \\ 
%		\begin{tabular}[c]{@{}c@{}}Echo train length   (ETL)\end{tabular} & 32 \\ 
%		Echo spacing (ESP) & 10 ms \\ 
%		Flip angles (FA) & $180^{\circ}$ \\
%		Contrasts (${Q}$)	& 96 \\ \hline
%	\end{tabular}
%	\label{table: tabs2}
%\end{table}

\subsection{Verification of TEUSQA with numerical simulation}
\label{sec: A_MC}

The ground truth parameter maps obtained using a  fully sampled scan of the ISMRM model 130 phantom \cite{jiang_repeatability_2017} with sequence settings described in Section \ref{sec: sequence used2} and a $3.0$ T  clinical scanner (Discovery MR750, GE Healthcare, Waukesha, WI) using a 32-channel head coil. The acquisitions were performed with a reduced acquisition matrix of size  $ 128 \times 84 \times 128$ in $PE_2 (SI) \times PE_1 (AP) \times FE (LR)$. The coil maps were computed using ESPIRIT technique \cite{uecker_espirit-eigenvalue_2014} and BART toolbox \cite{uecker_generalized_2016}. 

\begin{figure*}
	\begin{subfigure}{\textwidth}
		\centering
		\includegraphics[width = 0.5\linewidth]{legend_expt1_new}
		\label{Fig: expt_1_legend_t1ag}
	\end{subfigure}
	\begin{subfigure}{\textwidth}
		\centering
		\includegraphics[width = \linewidth]{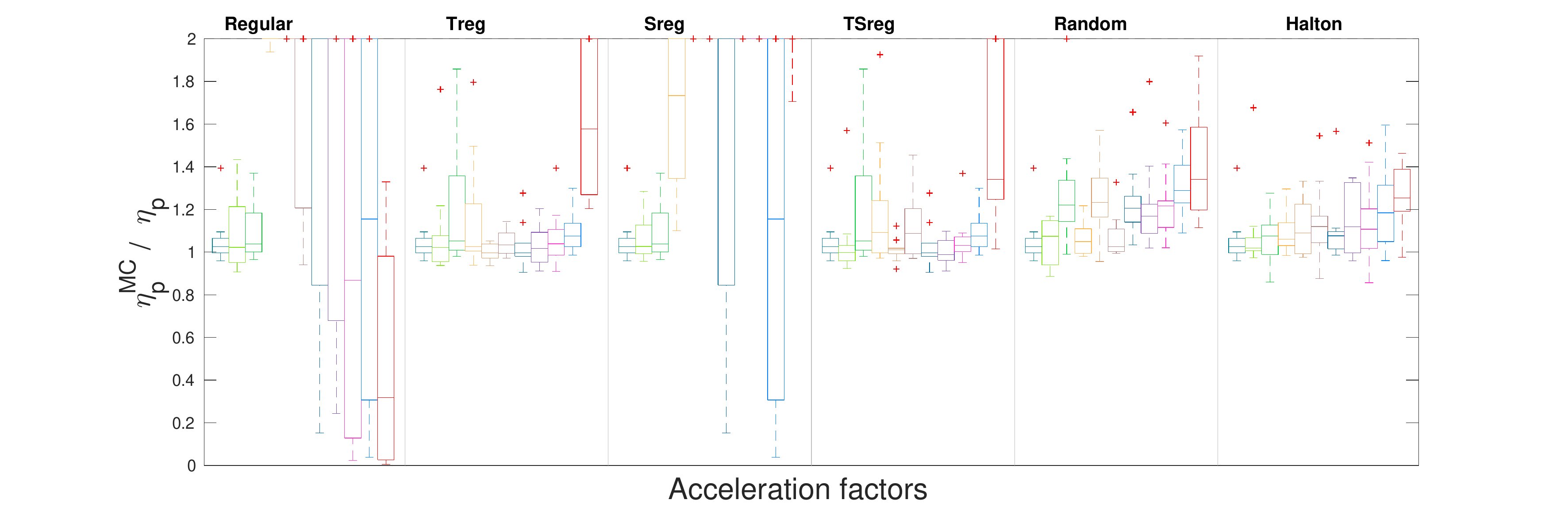}
		%		\subcaption{	\centering Result for $T_2$}
		%		\label{Fig: expt_1_t2s}
	\end{subfigure}
	\caption{ Results from Monte Carlo simulation for evaluation of TEUSQA for $T_2$ using 3D GRASE. The  box plots are grouped together according to the undersampling patterns shown by the vertical line separating the figure. The  box plots are colored according to the respective acceleration factor shown in the legend. Each  box plot represents the distribution of $\eta^{MC}_{p}/\eta_{p}$ over voxels within the ROI of the phantom for a particular acceleration factor. The ratio is shown for range $[0, 2]$.} 
	\label{Fig: expt_1_t2_grase}
\end{figure*}

The ratio of $\eta^{MC}_{p}$ to  $\eta_{p}$ were computed in similar way as described for 3D IP-FSE in Section \ref{sec: MC}.  
The box plot showing the ratio of $\eta^{MC}_{p}$ to  $\eta_{p}$ is shown in Figure \ref{Fig: expt_1_t2_grase}.

The results show that the ratio of $\eta^{MC}_{p}$ to  $\eta_{p}$ is higher than those observed for 3D IP-FSE in some cases. First, for undersampling patterns that have low time efficiencies, such as Regular and Sreg. Second, for acceleration factors greater than $R = [4, 4]$.  Apart from these, the Random undersampling pattern also shows a slightly higher ratio for 3D GRASE than in the case of 3D IP-FSE. These can be because of the lower SNR used for the simulation and the greater degree of model mismatch. 

For time-efficient undersampling patterns and to an acceleration factors that have sufficient measurements, TEUSQA can predict time efficiency of 3D GRASE scans for mapping $T_2$ and $\Delta B0$.

\subsection{Selection of undersampling pattern}
\label{sec: A_MC2}
\begin{figure}
	\begin{subfigure}{0.4\textwidth}
		\includegraphics[width = \linewidth]{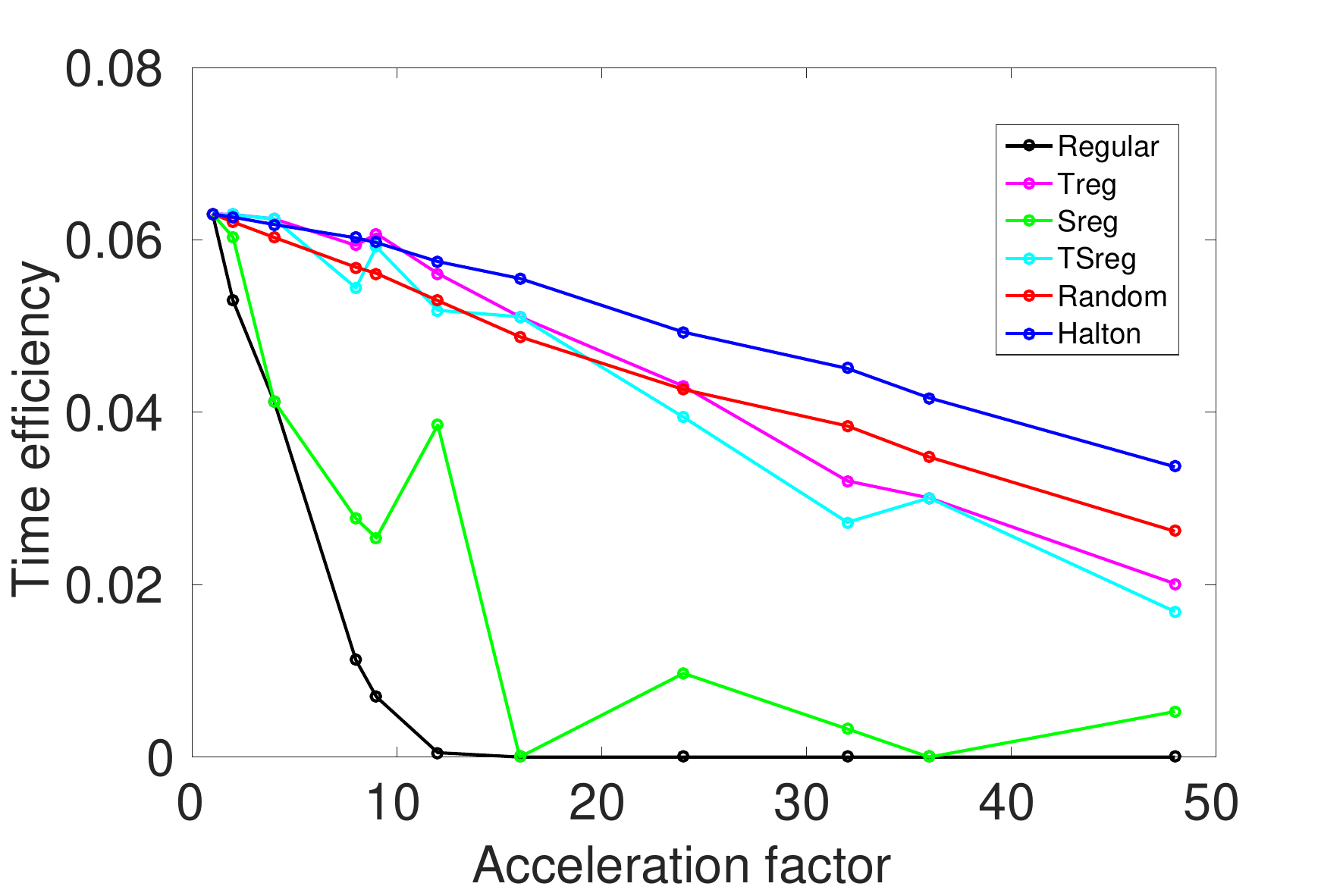}
		\subcaption{\centering $\eta_{p}$ computed for $T_2$}
		\label{subfig: SNK_real21}
	\end{subfigure}
	\centering
	\begin{subfigure}{0.4\textwidth}
		\includegraphics[width = \linewidth]{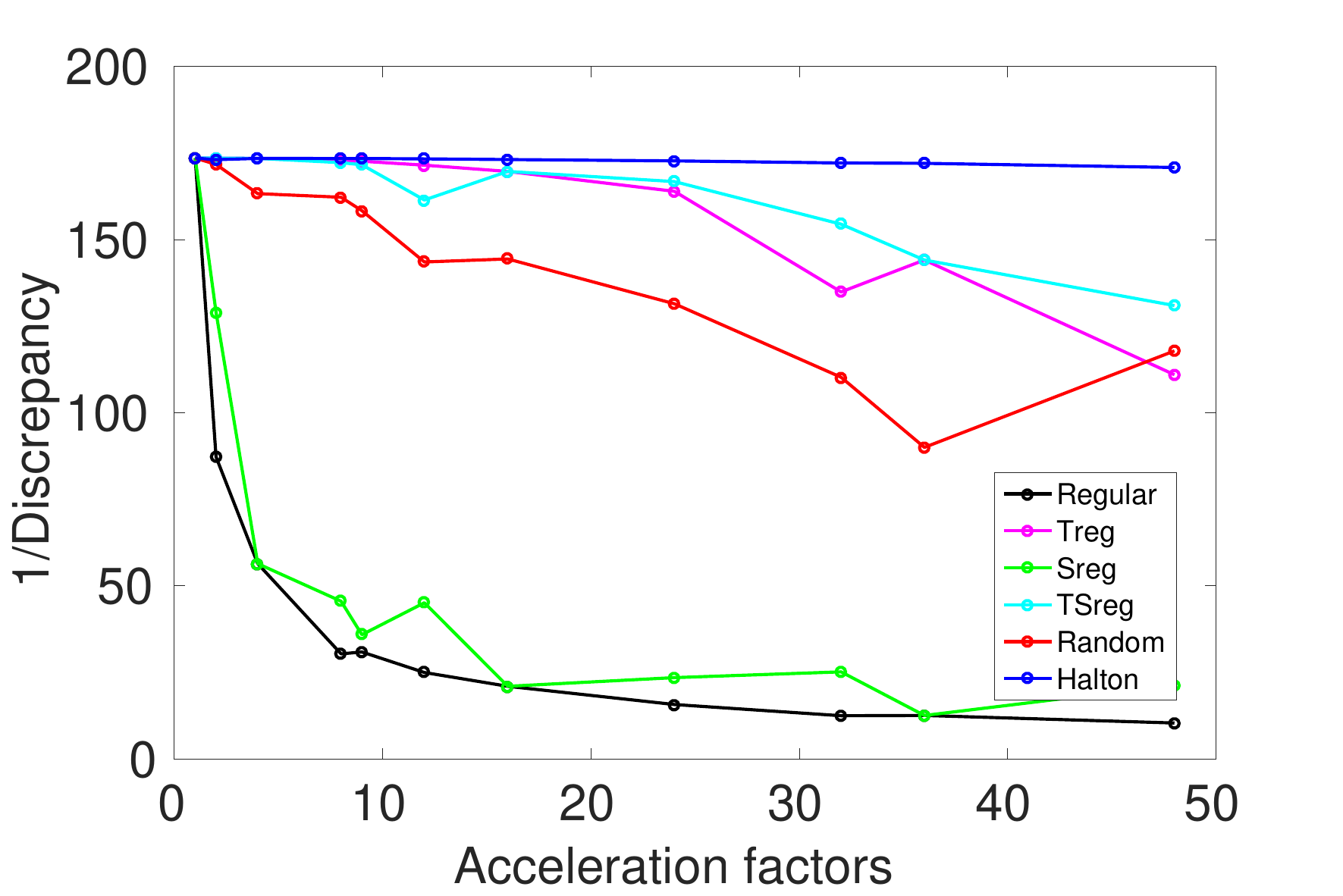}
		\subcaption{\centering Discrepancy}
		\label{subfig: Discrepancy_small21}
	\end{subfigure}
	\caption{{ (a) $\eta_{p}$ computed for  $T_2$ for all evaluated  undersampling patterns and acceleration factors.  (b) Reciprocal of Discrepancy of undersampling patterns.}}
	\label{Fig: Time_eff12}
\end{figure}

Comparision of different undersampling patterns was done for various acceleration factors similarly as described for 3D IP-FSE in Section \ref{sec: pattern}.  Result from the comparison are shown in Figure \ref{Fig: Time_eff12} where \ref{subfig: SNK_real21} shows the TEUSQA and \ref{subfig: Discrepancy_small21} shows the reciprocal of Discrepancy.

The Halton undersampling pattern with low-discrepancy showed the best time efficiency similar to what was observed for 3D IP-FSE.

\subsection{Verification with prospective scan}
With the seclected Halton undersampling pattern, a test-retest scan was performed on a ISMRM model 130 phantom, with the same scan settings as shown in Table \ref{tab: scan set} and acquisition settings described in  \ref{sec: sequence used2}, but with acceleration factor of 16.

\begin{figure}
	\centering
	\begin{subfigure}{0.2\textwidth}
		\includegraphics[width=\linewidth]{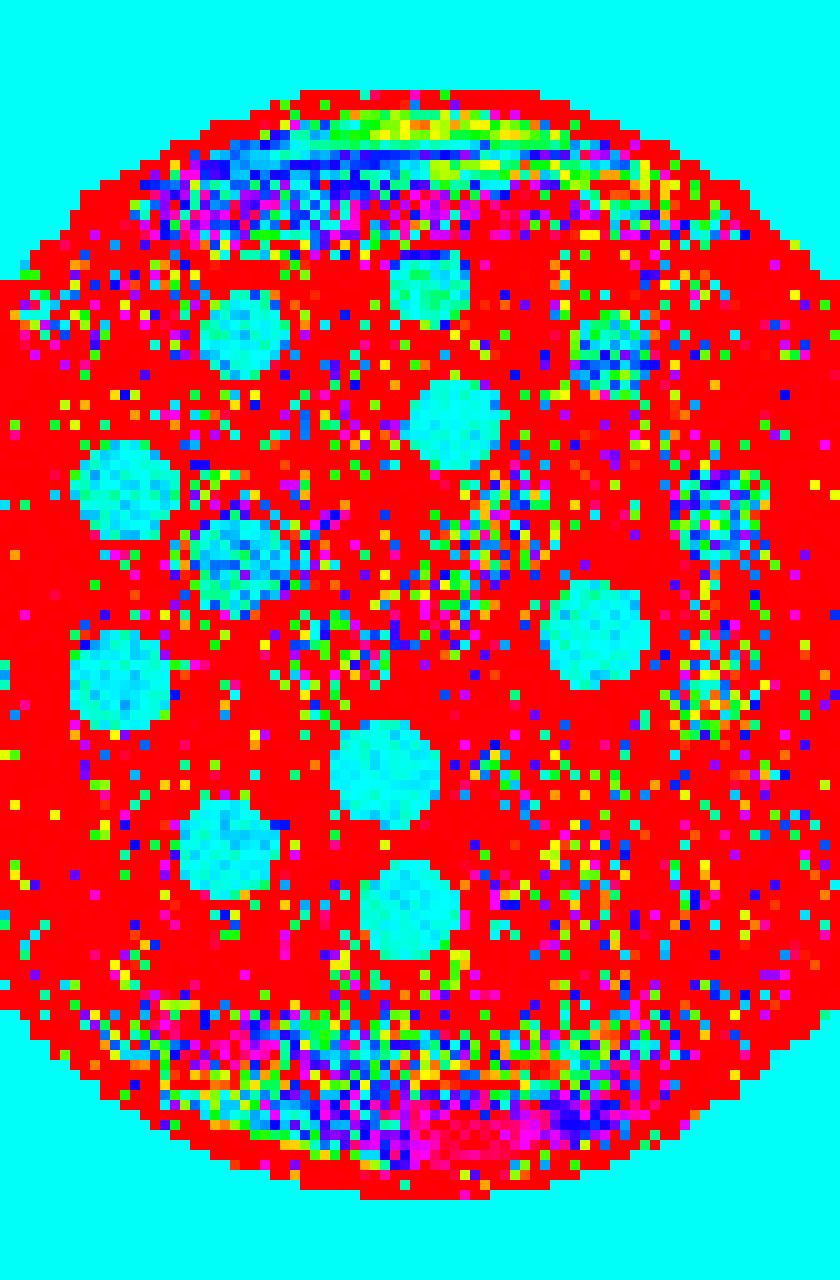}
		\caption{\centering {$T_2$ difference map from $T_1$ array }}
	\end{subfigure}
		\begin{subfigure}{0.03\textwidth}
	\includegraphics[scale=0.5]{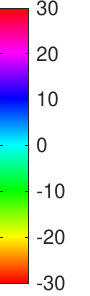}
\end{subfigure}
	\begin{subfigure}{0.2\textwidth}
		\includegraphics[width=\linewidth]{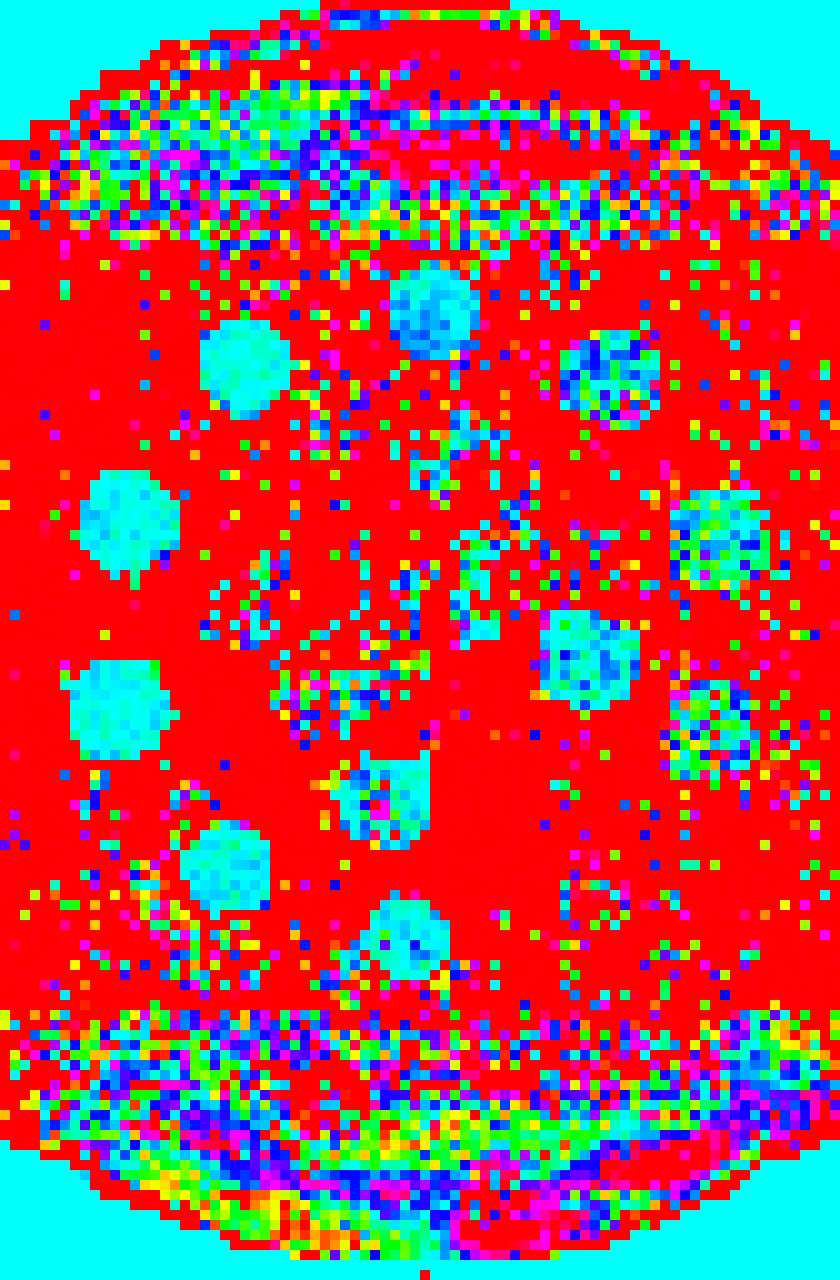}
		\caption{\centering {$T_2$ difference map from $T_2$ array}}
		\label{Fig: T1_pros22}
	\end{subfigure}
		\begin{subfigure}{0.03\textwidth}
			\includegraphics[scale=0.5]{T2_slicelegend}
		\end{subfigure}
	\begin{subfigure}{0.2\textwidth}
		\includegraphics[width=\linewidth]{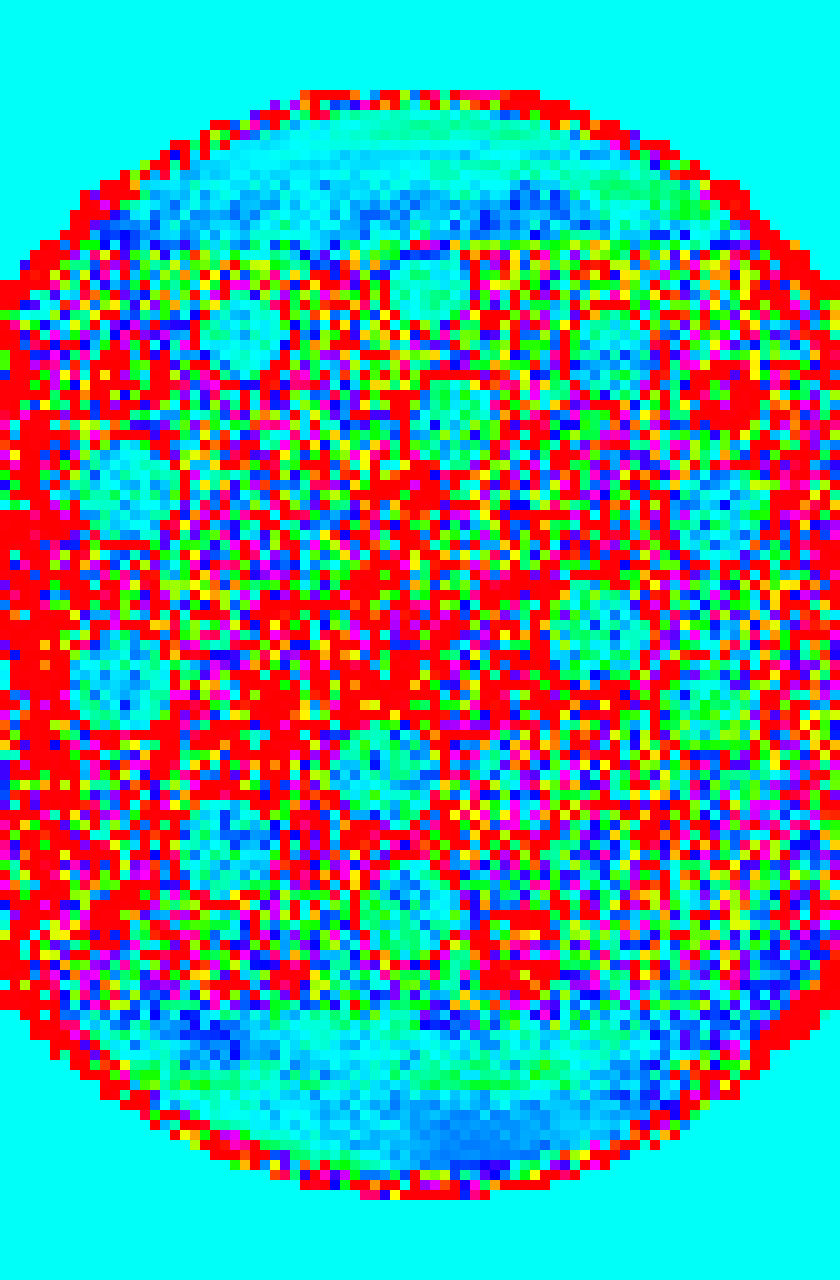}
		\caption{\centering {$\Delta B0$ difference map  from $T_2$ array}}
	\end{subfigure}
			\begin{subfigure}{0.03\textwidth}
		\includegraphics[scale=0.5]{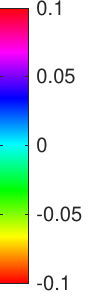}
	\end{subfigure}
	\begin{subfigure}{0.2\textwidth}
		\includegraphics[width=\linewidth]{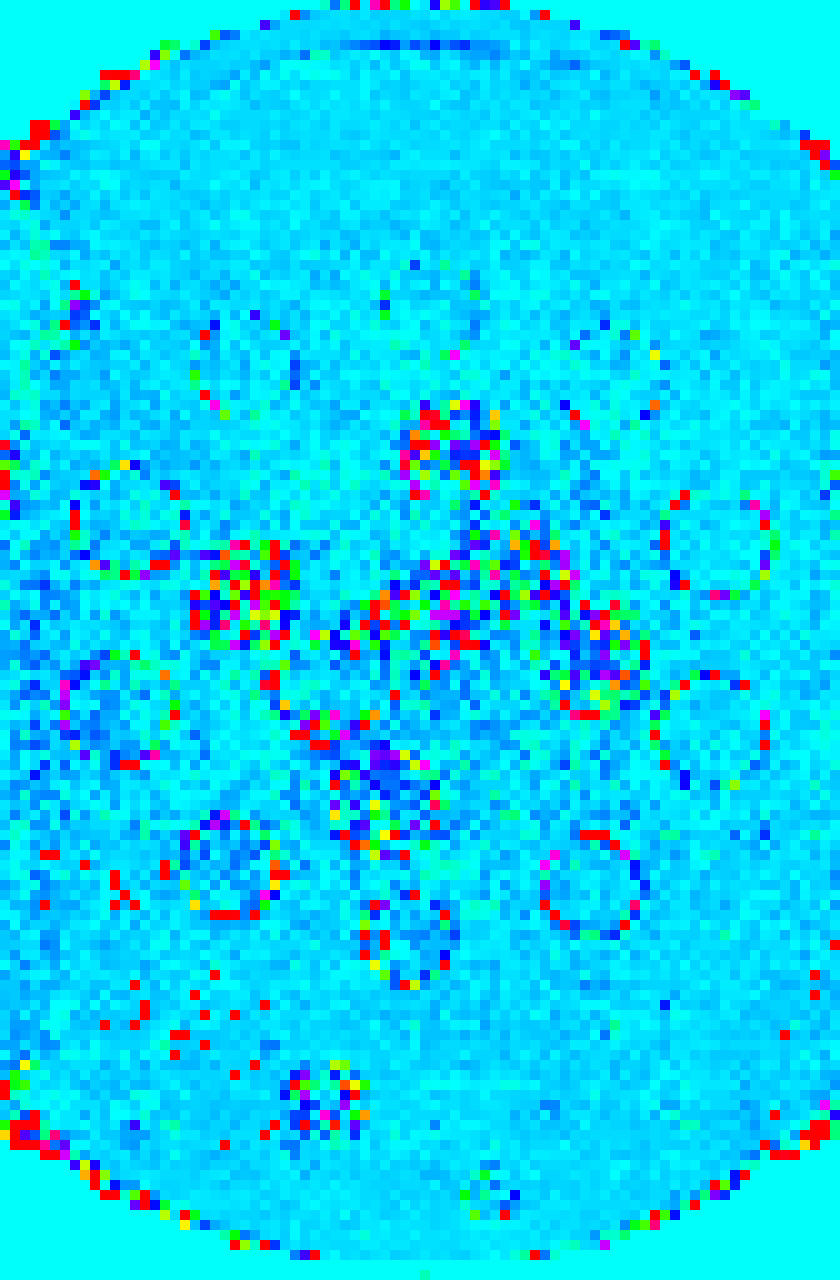}
		\caption{\centering {$\Delta B0$ difference map from $T_2$ array}}
		\label{Fig: T1_pros22}
	\end{subfigure}
		\begin{subfigure}{0.03\textwidth}
			\includegraphics[scale=0.5]{T2_slice_b0legend}
		\end{subfigure}
	
	\caption{Test-retest difference maps for $T_2$ (left) and $\Delta B0$ (right) estimates in the $T_1$ (top) and $T_2$ array (bottom) of the ISMRM model 130 phantom \cite{jiang_repeatability_2017} (units in ms). Scans were made using prospective undersampling using the Halton pattern with acceleration factor $R=16$.}
	\label{Fig: pros_result2}
	
\end{figure}

Figure \ref{Fig: pros_result2} shows the difference in  estimated $T_2$ and $\Delta B0$ maps for the test and retest acquisition. 
Similar to the case of 3D IP-FSE, the  spheres where nominal values of $T_2$ are within the range [second TE, last TE]  the mean difference between maps from test and retest is close to zero. Outside of these spheres large differences can be observed.  

\begin{figure}[h]
	\centering
	\includegraphics[width = \linewidth]{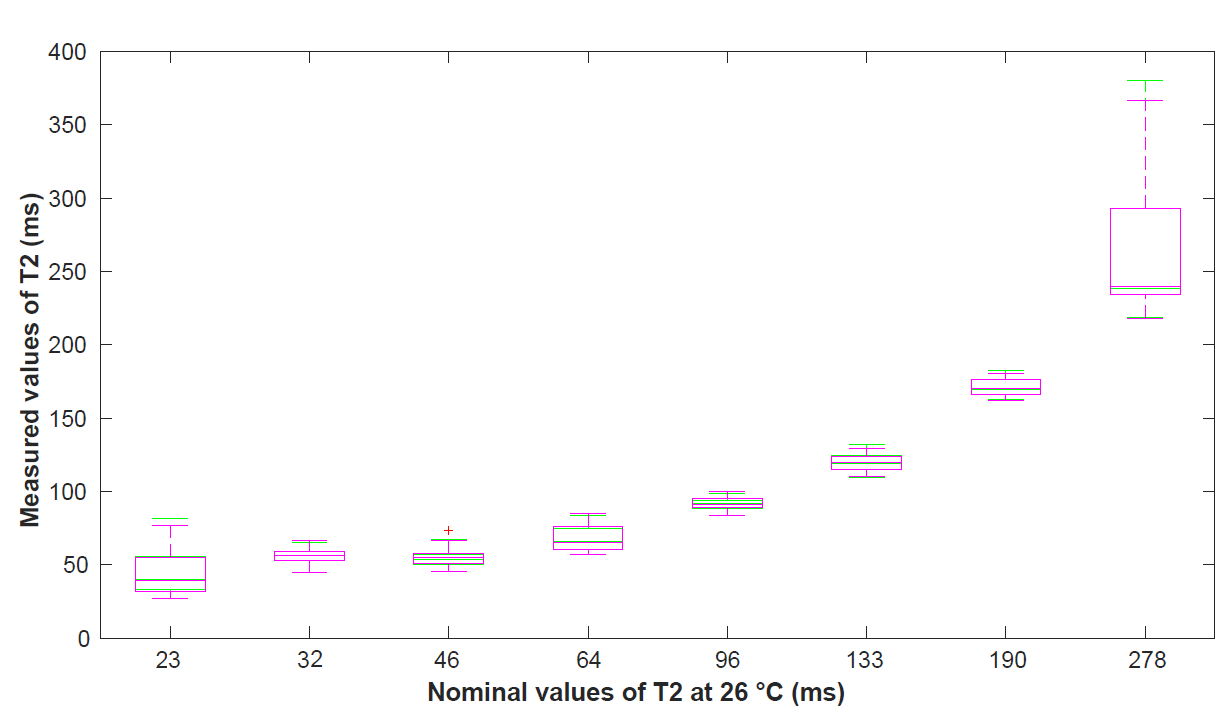}
	\caption{ $T_2$ measured in the selected spheres in $T_1$ array plotted against their nominal values. Green boxes indicate test acquisition and magenta boxes indicate retest.}
	\label{Fig: pros1g}
\end{figure}

\begin{figure}
	\centering
	\includegraphics[width = \linewidth]{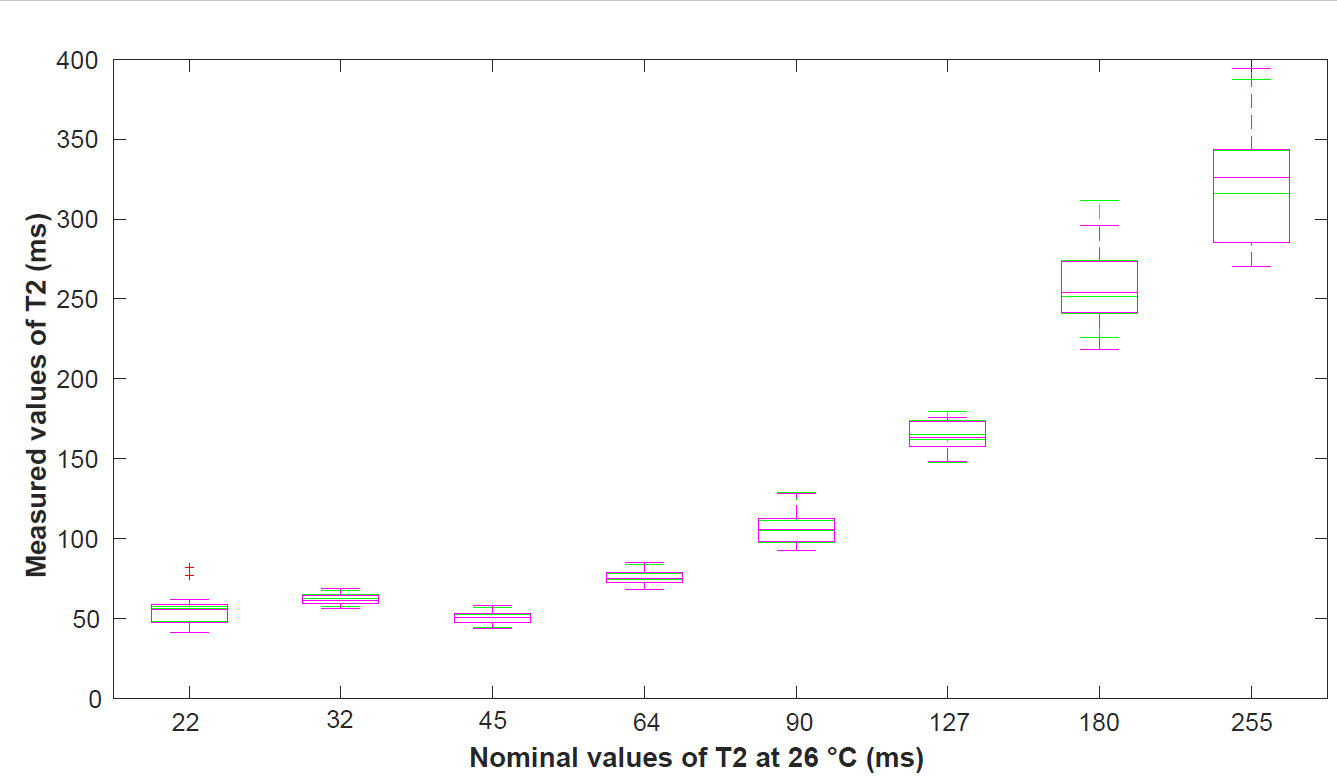}
	\caption{$T_2$ measured in the selected spheres in $T_2$ array plotted against their nominal values. Green boxes indicate test acquisition and magenta boxes indicate retest.}
	\label{Fig: pros2g}
\end{figure}

The bias in the $T_2$ estimates compared to the nominal values was on average about  $17\%$  in the selected spheres. A detailed comparison with the nominal values is presented in the Figures 	\ref{Fig: pros1g} and \ref{Fig: pros2g}.
Following this we computed  $\eta^{ACQ}_{p}$ over the selected spheres which was found to be $0.0196$ with $95\%$ confidence bounds $[ 0.018, 0.021 ]$ for $T_2$. The predicted  $\eta_{p}$ for $T_2$ was $0.020$. The prediction  was within the $95\%$ confidence bounds of observed $\eta^{ACQ}_{p}$.

\begin{table*}[]
	\centering
	\caption{ The median values (ms) of $T_2$ from test and re-test scans compared to nominal values of selected spheres.  }
	\label{Fig: results_tab_grase}
	\begin{tabular}{|ccc|}
		\hline
	 Nominal value $T_2$ & Median $T_2$ test & Median $T_2$ re-test \\ \hline
		 22 & 19.4 & 19.1 \\ 
		 32 & 53.1 & 53.5 \\ 
		 46 & 47.7 & 48.4 \\ 
		 64 & 72.8 & 72.6 \\ 
		 97 & 95.4 & 95.3 \\ 
		 133 & 163.2 & 159.5 \\ 
		 190 & 284.9 & 282.8 \\
		  278 & 325.34 & 34.8 \\ \hline
	\end{tabular}
\end{table*}

\section{Comparison of undersampling patterns for FSE with 32 echoes}
\label{A: FSE}
In this experiment, we compare the undersampling pattern generation techniques for a sequence with fewer contrasts than the 3D IP-FSE and 3D GRASE sequence and check if the low discrepancy is still desirable for such shorter sequences.  For this purpose, we select an FSE sequence with 32 echoes and echo spacing of 10 ms and $TR = 1800$. The ground truth for this experiment was a checkered board pattern that had $T_2$ values of 50 ms and 100 ms.  The coil maps were taken from acquisition described in Section \ref{sec: MC_1}. We used  acceleration factors upto 24.

\begin{figure}
	\begin{subfigure}{0.4\textwidth}
		\includegraphics[width = \linewidth]{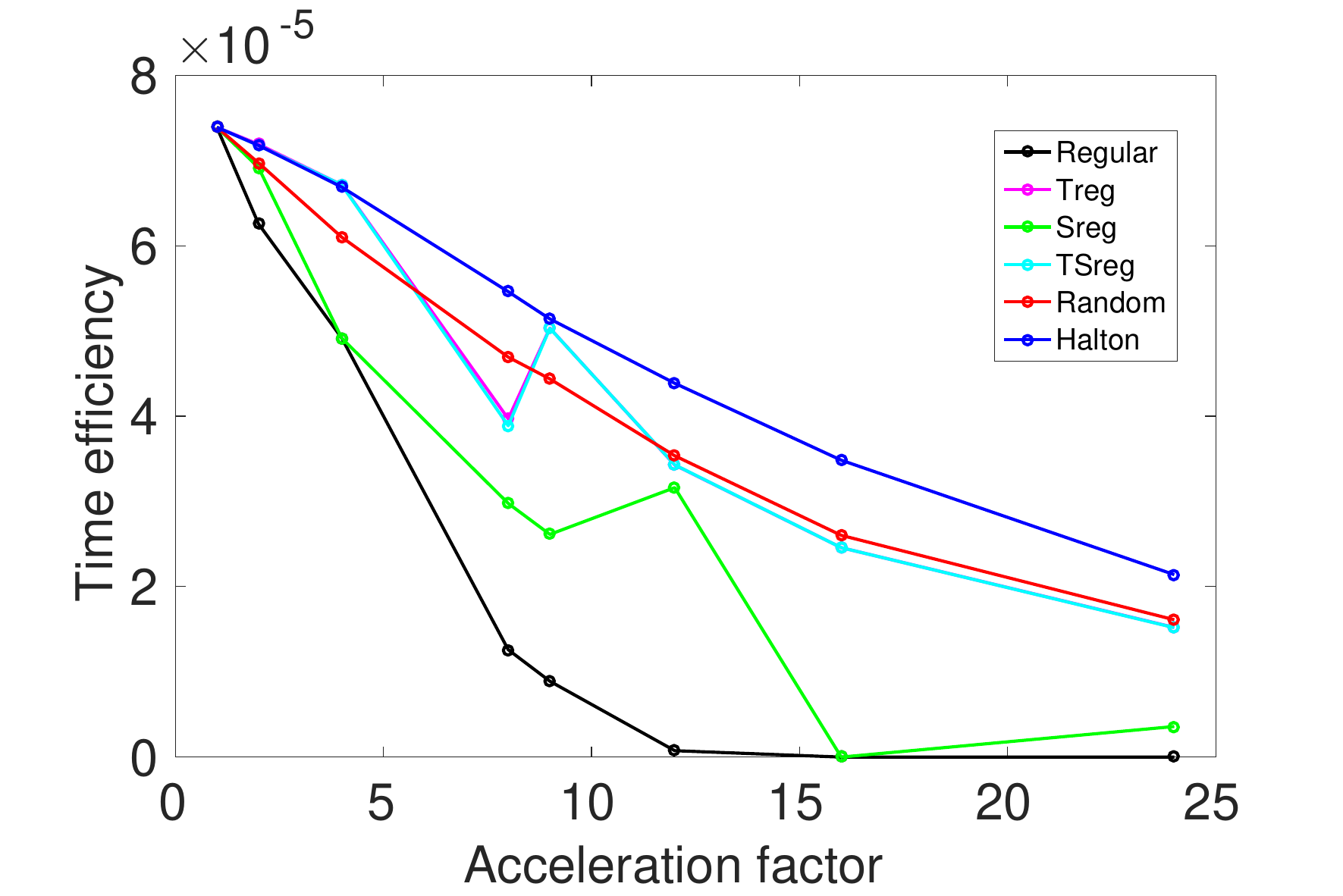}
		\subcaption{\centering $\eta_{p}$ computed for $T_2$}
		\label{subfig: SNK_real2}
	\end{subfigure}
	\centering
	\begin{subfigure}{0.4\textwidth}
		\includegraphics[width = \linewidth]{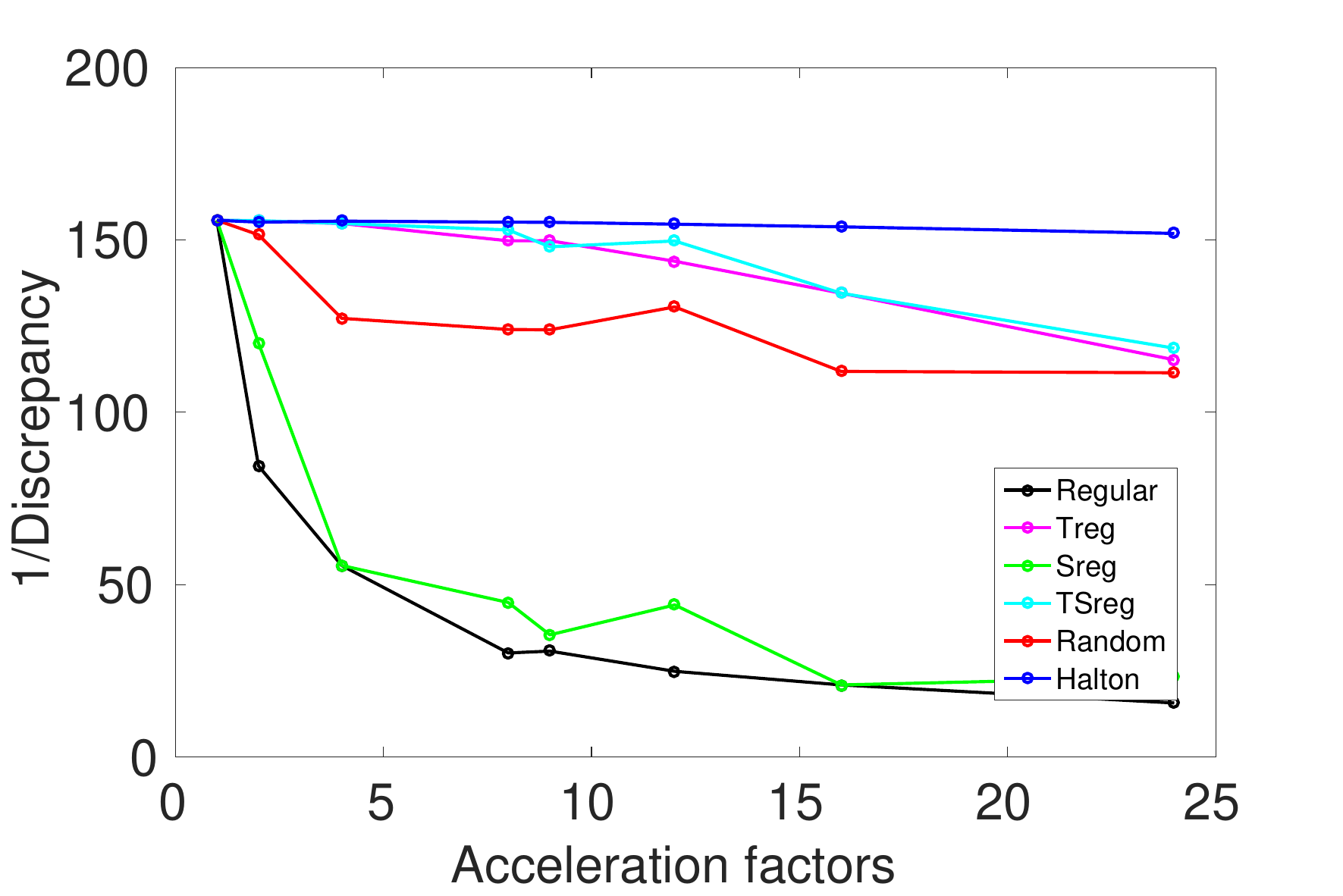}
		\subcaption{\centering Discrepancy}
		\label{subfig: Discrepancy_small2}
	\end{subfigure}
	\caption{{ (a) $\eta_{p}$ computed for  $T_2$ for all evaluated  undersampling patterns and acceleration factors.  (b) Reciprocal of Discrepancy of undersampling patterns.}}
	\label{Fig: Time_eff2}
\end{figure}

Figure \ref{subfig: SNK_real2} shows the TEUSQA, and Figure \ref{subfig: Discrepancy_small2} shows the reciprocal of Discrepancy computed for each undersampling pattern and undersampling factors from 1 to 24. Patterns generated using Halton have the highest TEUSQA score and lowest Discrepancy.

We conclude from this experiment that the low-discrepancy is still desireable for lower number of contrasts.

%\section{Choice of echo for calibration region}
%
%Supplementary material that may be helpful in the review process should
%be prepared and provided as a separate electronic file. That file can
%then be transformed into PDF format and submitted along with the
%manuscript and graphic files to the appropriate editorial office.
%%Harvard
\bibliographystyle{model2-names.bst}\biboptions{authoryear}
\bibliography{papers3.bib}
\end{document}

\verso{Riwaj Byanju  \textit{et~al.}}
	
\begin{frontmatter}
		
		\title{ Supplementary material for Time efficiency analysis for undersampled quantitative MRI acquisitions}%
		\tnotetext[tnote1]{This is an example for title footnote coding.}
		
		\author[1]{Riwaj \snm{Byanju}\corref{cor1}}
		\cortext[cor1]{Corresponding author: 
			Tel.: +31-068-391-2364;}
		\ead{r.byanju@erasmusmc.nl}
		\author[1]{Stefan \snm{Klein}}
		\author[1]{Alexandra \snm{Cristobal-Huerta}}
		\author[1]{Juan A. \snm{Hernandez-Tamames}}
		\author[1]{Dirk H. J. \snm{Poot}}
		
		\address[1]{Department of Radiology and Nuclear Medicine, Erasmus MC,  Rotterdam, 3015 GE, The Netherlands}
		%\address[2]{Affiliation 2, Address, City and Postal Code, Country}
		
		\received{1 May 2013}
		\finalform{10 May 2013}
		\accepted{13 May 2013}
		\availableonline{15 May 2013}
		\communicated{S. Sarkar}

\begin{abstract}
	%%%
	Undersampled scans supported by conventional acceleration techniques such as parallel imaging are inadequate to reduce the scan time for clinical use of Quantitative MRI (QMRI). Further acceleration is possible using Model-based reconstruction. To inform sequence design and sample pattern optimization, we propose a theoretical metric called TEUSQA: Time Efficiency for UnderSampled QMRI Acquisitions. TEUSQA is designed for a particular class of reconstruction techniques that directly estimate tissue parameters, possibly using prior information to regularize the estimation. TEUSQA can be used to evaluate undersampling patterns for multi-contrast QMRI sequences targeting any tissue parameter. To verify the time efficiency predicted by TEUSQA, we performed Monte Carlo simulations and a 32-fold accelerated joint T1 \& T2 mapping experiment. Using TEUSQA, we assessed several ways to generate undersampling patterns \textit{in silico}, providing insight into the relation between sample distribution and time efficiency for different acceleration factors.The time efficiency predicted by TEUSQA was within 15\% of that observed in the Monte Carlo simulations and the prospective acquisition experiment. The assessment of undersampling patterns showed that a class of good patterns could be obtained by low-discrepancy sampling. We believe that TEUSQA offers a valuable instrument for developers of novel QMRI sequences pushing the boundaries of acceleration to achieve clinically feasible protocols.
\end{abstract}
\end{frontmatter}
\section{Derivation of posterior distribution} 
		\label{section: S1}
		In this section we derive posterior distribution for the parameters with the addition of a prior distribution.  Using Bayes theorem, the posterior distribution is given by: 
		\begin{equation}
			\label{step1}
			p( \bm{\theta} | {\bm{Z} ,\sigma)} \propto p({\bm{Z} | \bm{\theta}  ,\sigma)}   \times  p( \bm{ \bm{\theta}} | \overline{ \bm{\theta}}, \bm{\Gamma})\\  =  \frac{e^{-\frac{1}{2 \sigma^2}||\bm{Z} - \m||^2}}{(2\pi{\sigma}^2)^{N}}     \times     \frac{e^{-\frac{1}{2}( \bm{\theta} - \bar{ \bm{\theta}})^{T}\bm{\Gamma}^{-1} ( \bm{\theta} - \bar{ \bm{\theta}})}}{\sqrt{\left(2\pi\right)^{L} \det \bm{\Gamma}}}  
		\end{equation}
		Taking, only the arguments of the first exponential in equation \ref{step1},
		\begin{equation*}
			-\tfrac{1}{2 \sigma^2}{||\bm{Z} - \m||^2} = -\tfrac{1}{2 \sigma^2} [\z^{T}\z - \z^{T}\m \\ -\m^{T}\z + \m^{T}\m ]
		\end{equation*}
		
		Substituting the first order approximation of Taylor series $\bm{\mu}( \bm{\theta}) \approx \bm{\mu}(\ttil) + \jac(\ttil)  \left\{ \bm{\theta} - \ttil\right\}$ where $\ttil$ is the ground truth. In the rest of the section we denote  $\jac(\ttil)$  as $\jac$ and $\bm{\mu}(\ttil)$ as $\bm{\mu}$  for notational convenience:
		\begin{equation}
			\label{I}
			=  -\tfrac{1}{2 \sigma^2}   \left[ K +  \left(  - \z^{T}   \jac + \mt^{T} \jac -  \ttil^{T} \jac^{T} \jac  \right)\bm{\theta}  +   \bm{\theta}^{T} \jac^{T} \jac \bm{\theta} \right.\\  \left.   +  \bm{\theta}^{T}   \left( -  \jac^{T}\z +  \jac^{T}   \mt    -  \jac^{T} \jac \ttil   \right)    \right]
		\end{equation}
		where 
		$  K   =   \z^{T}\z   -    \z^{T}\mt   +  \z^{T} \jac \ttil   -   \mt^{T}\z +  \ttil^{T} \jac^{T}\z  + \mt^{T}\mt    -   \mt^{T}\jac \ttil -  \ttil^{T} \jac^{T}  \mt  + \ttil^{T} \jac^{T}\jac \ttil $  contains all elements not dependent on $\bm{\theta}$.\\
		Similarly taking the second exponential term from equation \ref{step1} and  since  $  \bm{\theta}^{T} \bm{\Gamma}^{-1} \overline{ \bm{\theta}}$  =   $\overline{ \bm{\theta}}^{T}\bm{\Gamma}^{-1} \bm{\theta}  $,
		\begin{equation}
			\label{II}
			\tfrac{1}{2}	(  \bm{\theta} - \overline{ \bm{\theta}})^{T}\bm{\Gamma}^{-1} (  \bm{\theta} - \overline{ \bm{\theta}})   = \tfrac{1}{2} \left( \bm{\theta}^{T} \bm{\Gamma}^{-1}  \bm{\theta} - 2\overline{ \bm{\theta}}^{T}\bm{\Gamma}^{-1} \bm{\theta}   +  \overline{ \bm{\theta}}^{T}\bm{\Gamma}^{-1}\overline{ \bm{\theta}}\right).
		\end{equation}
		Combining the two exponential terms of equation \ref{step1} gives
		\begin{equation}
			\label{form}
			A + B \bm{\theta} +  \bm{\theta}^{T} C  \bm{\theta},    
		\end{equation} 
		\begin{align*}
			\text{ where } A =& -\tfrac{1}{2 \sigma^2} K + -\tfrac{1}{2} \bar{ \bm{\theta}}^{-T}\bm{\Gamma}^{-1}\bar{ \bm{\theta}},\\ 
			B =& \tfrac{1}{ \sigma^2}  \left(   \z^{T}   \jac - \mt^{T} \jac +  \ttil^{T} \jac^{T} \jac  \right) + \overline{ \bm{\theta}}^{T}\bm{\Gamma}^{-1} \\ \text{ and }C =& -\tfrac{1}{2} \left( \tfrac{1}{\sigma^2}\jac^{T} \jac +\bm{\Gamma}^{-1}  \right).
		\end{align*}
		
		The derivative of equation \ref{form} with respect to  $ \bm{\theta} $ can be equated to $0$ to  find its maximum value. Since our prior is conjugate for the likelihood function, the posterior distribution should also be normally distributed. Therefore, the location of the maximum is the mean $\breve{ \bm{\theta}}$ of the posterior distribution. That is,
		
		\begin{align}
			\label{min}
			\breve{ \bm{\theta}} =& -\frac{1}{2} C^{-1} B^{T}\\
			=&    \left[   \tfrac{1}{\sigma^2}{\jac^{T} \jac} + \bm{\Gamma}^{-1} \right]^{-1}     \left[  \tfrac{1}{ \sigma^2}  \left(     \jac^{T}\z  - \jac^{T}\mt  +  \jac^{T}\jac\ttil    \right) + \bm{\Gamma}^{-1}\overline{ \bm{\theta}}\right]
			%\end{split}
		\end{align}
		Now the equation of posterior distribution using the mean $\breve{ \bm{\theta}}$ and assuming the covariance matrix of this distribution to be $\breve{\bm{\Gamma}}$ is given by,
		\begin{equation}
			\label{postprob2}
			p( \bm{\theta} | \breve{\bm{\theta}},\breve{\bm{\Gamma}}) = \frac{e^{-\frac{1}{2}( \bm{\theta} - \breve{ \bm{\theta}})^{T}\breve{\bm{\Gamma}}^{-1}( \bm{\theta} - \breve{ \bm{\theta}})}}{\sqrt{(2\pi)^{L} \det \breve{\bm{\Gamma}}}}
		\end{equation}
		
		Since equation \ref{postprob2} and \ref{step1} give the same distribution, taking the argument of exponential in the numerator in equation \ref{postprob2} and then arranging in the form shown in equation \ref{form}, we get $C= -{\breve{\bm{\Gamma}}} / {2} $. Hence,
		\begin{equation}
			\breve{\bm{\Gamma}}^{-1} =   \bm{\Gamma}^{-1} + \bm{I}(\ttil, \sigma)
		\end{equation}
		with $\bm{I}(\ttil, \sigma) = \tfrac{1}{\sigma^2}{\jac^{T}\jac}$ from  equation (7) in the main manuscript.
		
		\
		\section{Pseudo code for Halton undersampling pattern}
		The pseudo code shows the implementation of the Halton undersampling pattern. The code generates a binary undersampling mask $Umask$ where sampled positions are $1$. The Halton sampling function used is based on the implementation of Wang et al.\cite{wang_randomized_2000}.
		\begin{algorithm}[h]
			\SetAlgoLined
			$NS \leftarrow Q( k_1 \times k_2)/R_1 R_2 $\\
			$S_1 \leftarrow $\textbf{Halton sampling:} generate $NS$ samples $\in [0,1]$ with base 2 \\
			$S_2 \leftarrow $\textbf{Halton sampling:} generate $NS$ samples $\in [0,1]$ with base  3\\
			
			\For{q = 1 to Q}{
				$SamplesPerContrast \leftarrow( k_1 \times k_2)/R_1 R_2 $\\
				\While { $SamplesPerContrast > 0$}{ 
					$x1 = S1.next$,
					$x2 = S2.next$ \\
					\If{$Umask(\lfloor x1*k_1 \rfloor, \lfloor x2*k_2 \rfloor,q) == 0$}
					{$Umask(\lfloor x1*k_1 \rfloor, \lfloor x2*k_2 \rfloor,q) \leftarrow 1$}
					%			$no \leftarrow no +1$
					SamplePerContrast = SamplePerContrast - 1	}}
			\caption{Halton undersampling pattern, INPUT: $k_1$ (size of PE1), $k_2$(size of PE2), $Q$}
			\label{Algo: halton}
		\end{algorithm}

\bibliographystyle{model2-names.bst}\biboptions{authoryear}
\bibliography{papers3.bib}